\documentclass{aa}

\usepackage{natbib}
\usepackage{txfonts}
\usepackage{graphicx}
\usepackage[colorlinks=true,linkcolor=blue,citecolor=blue]{hyperref}

\usepackage{multirow}
\newcommand{\RomanNumeralCaps}[1]
    {\MakeUppercase{\romannumeral #1}}
\bibpunct{(}{)}{;}{a}{}{,}

\begin{document}

%\title{BINARY POPULATION SYNTHESIS STUDY OF MERGING NEUTRON STAR BLACK HOLE BINARIES AT SOLAR METALLICITY, USING DETAILED BINARY EVOLUTION MODELS}
  \title{From ZAMS to Merger: Detailed Binary Evolution Models of Coalescing Neutron Star-Black Hole Systems at Solar Metallicity}

  \author{Zepei\,Xing
          \inst{1,2}\fnmsep\thanks{e-mail: Zepei.Xing@unige.ch},
          Simone\,S.\,Bavera
          \inst{1,2},
          Tassos\,Fragos
          \inst{1,2},
          Matthias\,U.\,Kruckow
          \inst{1,2},
          Jaim\,Román-Garza
          \inst{1,2}
          Jeff\, J.\,Andrews
          \inst{3,4},
          Aaron\,Dotter
          \inst{3},
          Konstantinos\,Kovlakas
          \inst{1,5,6},
          Devina\,Misra
          \inst{1,7},
          Philipp\,M.\,Srivastava
          \inst{3,8},
          Kyle\,A.\,Rocha
          \inst{3,8},
          Meng\,Sun
          \inst{3},
          Emmanouil\,Zapartas
          \inst{9,10}
          }
  \authorrunning{Xing et al.}
    \titlerunning{Neutron Star-Black Hole Binaries}
  \institute{
    Département d’Astronomie, Université de Genève, Chemin Pegasi 51, CH-1290 Versoix, Switzerland
   \and
   Gravitational Wave Science Center (GWSC), Université de Genève, CH1211 Geneva, Switzerland
   \and
    Center for Interdisciplinary Exploration and Research in Astrophysics (CIERA), Northwestern University, 1800 Sherman Ave, Evanston, IL 60201, USA
   \and 
   Department of Physics, University of Florida, 2001 Museum Rd, Gainesville, FL 32611, USA
   \and
   Institute of Space Sciences (ICE, CSIC), Campus UAB, Carrer de Magrans, 08193 Barcelona, Spain
   \and
   Institut d’Estudis Espacials de Catalunya (IEEC), Carrer Gran Capit\`a, 08034 Barcelona, Spain
   \and 
   Institutt for Fysikk, Norwegian University of Science and Technology, Trondheim, Norway
   \and 
   Electrical and Computer Engineering, Northwestern University, 2145 Sheridan Road, Evanston, IL 60208, USA
   \and
   Institute of Astrophysics, FORTH, N. Plastira 100,  Heraklion, 70013, Greece
   \and 
   IAASARS, National Observatory of Athens, Vas. Pavlou and I. Metaxa, Penteli, 15236, Greece
     }
   \date{Received XXYYZZ; Accepted XXYYZZ}

\abstract{Neutron star -- black hole (NSBH) merger events bring us new opportunities to constrain theories of stellar and binary evolution, and understand the nature of compact objects. In this work, we investigate the formation of merging NSBH binaries at solar metallicity by performing a binary population synthesis study of merging NSBH binaries with the newly developed code \texttt{POSYDON}. The latter incorporates extensive grids of detailed single and binary evolution models, covering the entire evolution of a double compact object progenitor. We explore the evolution of NSBHs originating from different formation channels, which in some cases differ from earlier studies performed with rapid binary population synthesis codes. Then, we present the population properties of merging NSBH systems and their progenitors such as component masses, orbital features, and BH spins, and investigate the model uncertainties in our treatment of common envelope (CE) evolution and  core-collapse process. We find that at solar metallicity, under the default model assumptions, most of the merging NSBHs have BH masses in a range of $3-11\,M{_\sun}$ and chirp masses within $1.5-4\,M{_\sun}$. Independently of our model variations, the BH always forms first with dimensionless spin parameter $\lesssim 0.2$, which is correlated to the initial binary orbital period. Some BHs can subsequently spin up moderately ($\chi_{\rm BH} \lesssim 0.4$) due to mass transfer, which we assume to be Eddington limited. Binaries that experienced CE evolution rarely demonstrate large tilt angles. Conversely, approximately $40\%$ of the binaries that undergo only stable mass transfer without CE evolution contain an anti-aligned BH. Finally, accounting for uncertainties in both the population modeling and the NS equation of state, we find that $0-18.6\%$ of NSBH mergers may be accompanied by an electromagnetic counterpart.}
%The binaries that experienced common envelope evolution have effective inspiral spins highly centered around zero and the binaries that went through only stable mass transfer have a flatter distribution within $\sim -0.2$ and $\sim 0.2$. %The difference is due to the different distributions of the BH spin and the orbital tilt. 
\maketitle

\section{Introduction} \label{sec:intro}
After the detection of binary black hole (BH) and binary neutron star (NS) mergers \citep{2016PhRvL.116f1102A,2017PhRvL.119p1101A} by the ground-based gravitational-wave (GW) observatories LIGO \citep{2015CQGra..32g4001L} and Virgo \citep{2015CQGra..32b4001A}, NS -- BH (NSBH) systems had remained elusive until the detection of the NSBH merger events GW200115 and GW200105 \citep{2021ApJ...915L...5A}. For GW200105, with a high-spin prior for the NS, the BH and NS masses are estimated to be $M_{\rm BH} = 8.9^{+1.2}_{-1.5}\,M_{\sun}$ and $M_{\rm NS} = 1.9^{+0.3}_{-0.2}\,M_{\sun}$, respectively. The BH spin magnitude is constrained as $\chi_{\rm BH} = 0.08^{+0.22}_{-0.08}$ and its effective inspiral spin parameter $\chi_{\rm eff} = -0.01^{+0.11}_{-0.15}$. For GW200115, the binary mass components are estimated to be $M_{\rm BH} = 5.7^{+1.8}_{-2.1}\,M_{\sun}$ and $M_{\rm NS} = 1.5^{+0.7}_{-0.3}\,M_{\sun}$. The primary spin magnitude, $\chi_{\rm BH} = 0.33^{+0.48}_{-0.29}$, is not well constrained, while the effective inspiral spin parameter is found to be $-0.19^{+0.23}_{-0.35}$. The latter indicates a possible misalignment between the BH spin and the orbital angular momentum. The misalignment angle is estimated to be $2.30^{+0.59}_{-1.18}\,\rm{rad}$ \citep{2021ApJ...918L..38F}. \citet{2021ApJ...922L..14M} then reanalyzed the GW200115 signal with their astrophysically motivated priors and suggested a non-spinning BH for GW200115. 

With the release of the third Gravitational-Wave Transient Catalog (GWTC-3), in addition to GW200105 and GW200115, NSBH merger candidates include GW190426 \citep{2021PhRvX..11b1053A}, GW190917 \citep{2021arXiv210801045T} and GW191219 \citep{2021arXiv211103606T}. GW190426 and GW190917 are marginal candidates because their probability of astrophysical origin, $p_{\rm astro}$, is below $0.5$ and GW191219 exhibits large model-dependent uncertainties in $p_{\rm astro}$. In the GWTC-3 analysis, GW200105 was also listed as a marginal candidate because its $p_{\rm astro}$ is below $0.5$. Nevertheless, it is still an interesting NSBH candidate with a clear outlier from the noise background \citep{2021arXiv211103606T}. The properties of these NSBH candidates are listed in Table \ref{tab:NSBH}. Apart from these events, GW190814 \citep{2020ApJ...896L..44A,2020ApJ...904...39H,2021ApJ...910...62Z} and GW200210 \citep{2021arXiv211103606T} are suspected to be NSBH mergers. Both systems possess a compact object that is within the mass range of $2-3\,M_{\sun}$ and close to the boundary between the heaviest NS and the lightest BH. In the near future, more NSBHs are expected to be detected in the fourth (O4) and fifth (O5) observing runs of ground-based GW detectors at improved sensitivities \citep{2018LRR....21....3A}, as well as, in the longer term,  the next generation ground-based observatories, such as the Einstein Telescope \citep{2010CQGra..27s4002P,2011CQGra..28i4013H} and the Cosmic Explorer \citep{2017CQGra..34d4001A,2019BAAS...51g..35R} and future space-based observatory, such as the Laser interferometer Space Antenna \citep[LISA,][]{2017arXiv170200786A,2019arXiv190706482B}, TianQin \citep{2016CQGra..33c5010L,2021PTEP.2021eA107M}, and Taiji \citep{2020IJMPA..3550075R}.

NSBH inspiral can present different GW signals, depending on whether the NS is tidally disrupted and whether the mass shedding happens before the NS crosses the innermost-stable circular orbit (ISCO) \citep{2010PhRvD..82d4049K,2011PhRvD..84f4018K,2013PhRvD..87h4006F,2021PhRvD.104l3024T}. If the NS is not plunging as a whole into the BH but is tidally disrupted outside ISCO, the NSBH mergers may have potential electromagnetic counterparts (EMCs) such as kilonovae \citep{1998ApJ...507L..59L,2020ApJ...893..153K,2021ApJ...917...24Z,2021ApJ...915...69D}, short gamma-ray burst emission \citep{2007PhR...442..166N,2014ApJ...791L...7P,2020ApJ...895...58G}, long gamma-ray burst emission \citep{2023arXiv230900038G}, and radio emission \citep{2013MNRAS.430.2121P,2016ApJ...831..190H}. Generally, stiff NS equation of states corresponding to large NS radii, low-mass BHs, and high BH spins, which affect the position of the ISCO, increase the possibility of generating EMCs \citep{2011ApJ...727...95P,2018PhRvD..98h1501F,2019MNRAS.486.5289B,2021ApJ...912L..23R}. The recent study of \citet{2021ApJ...923L...2F} showed that only if the BHs are spinning fast and NSs have stiff equation of states one would expect a significant fraction of NSBH mergers to be associated with EMCs. However, in the classic isolated binary formation channel, it is commonly thought that if the BH forms first, it is likely to have a small spin \citep{2015ApJ...800...17F,2018A&A...616A..28Q,2019ApJ...881L...1F,2023NatAs.tmp..142B} because the angular momentum of the core can be efficiently transported to the outer layer, which would be removed by mass transfer or winds. Alternatively, NSBHs where the NS forms first, which allow for a tidally span-up BH progenitor, become candidates of EMC sources \citep{2021ApJ...912L..23R,2022ApJ...928..163H}. Despite the possibility, this corresponds to a very small portion of the whole NSBH population \citep{2021ApJ...920...81S,2021MNRAS.508.5028B,2022MNRAS.513.5780C}.

Regarding the formation mechanisms of merging NSBH binaries, multiple channels have been proposed. These channels include the evolution of isolated field binaries; dynamical interactions in global clusters \citep{2013MNRAS.428.3618C,2020ApJ...888L..10Y}, nuclear star clusters \citep{2017ApJ...846..146P,2020CmPhy...3...43A,2021ApJ...917...76W}, and young star clusters \citep{2014MNRAS.441.3703Z,2020ApJ...901L..16F,2020MNRAS.497.1563R,2020ApJ...898..152S,2021ApJ...908L..38A}; evolution from triples \citep{2019MNRAS.490.4991F,2022MNRAS.511.1362T}; and chemical homogeneous evolution \citep{2017A&A...604A..55M}. The isolated binary evolution channel is widely explored through binary population synthesis (BPS) methods \citep{1993MNRAS.260..675T,1999ApJ...526..152F,2003MNRAS.342.1169V,2015ApJ...806..263D,2015ApJ...814...58D,2018ApJ...866..151A,2018MNRAS.480.2011G,2018MNRAS.481.1908K,2019MNRAS.490.3740N,2020A&A...636A.104B,2020MNRAS.493L...6T,2020ApJ...899L...1Z,2021MNRAS.508.5028B,2021ApJ...920...81S}. These methods estimate a merger rate density of NSBHs consistent with the observations, given appropriate physical assumptions. However, relying solely on the merger rate density is not sufficient for constraining model uncertainties. The merger rate density is subject to many uncertainties in binary and stellar evolution such as common envelope (CE) evolution, mass transfer process, supernova (SN) mechanisms, and natal kicks as well as the metallicity-specific star formation rate density and its redshift evolution \citep[see, e.g.,][for a review]{2021MNRAS.508.5028B}. Varying model assumptions can make the rate predictions plausible but alone is not enough to obtain decisive constraints on the uncertain physical processes without detailed modeling and systematic joint investigations of all GW observables. As a result, the relatively well measured GW observables like chirp masses and effective spins which include the information of individual compact object masses, BH spins, and spin-orbit tilt angles, provide valuable constraints to the astrophysical models. 

In this work, we present for the first time detailed binary evolution models that follow the entire evolution of binaries from zero-age main sequence (ZAMS) until the formation of merging NSBH. We use the population synthesis framework \texttt{POSYDON} \citep{2023ApJS..264...45F} to carry out a BPS study of NSBH systems that will merge within a Hubble time. We focus on NSBH populations at solar metallicity, and investigate their formation channels in detail. With \texttt{POSYDON}, we can evolve binaries using extensive simulation grids of stellar structure and binary evolution simulations computed with the stellar evolution code Modules for Experiments in Stellar Astrophysics \citep[MESA,][]{2011ApJS..192....3P,2013ApJS..208....4P,2015ApJS..220...15P,2018ApJS..234...34P,2019ApJS..243...10P, 2023ApJS..265...15J}. The detailed simulations allow us to track the changes of the internal structure of the stars, taking into account both single-star evolution and binary interactions, as well as the angular momentum transport in the interior of the stars and between binary components. This enables us to make a self-consistent estimation of mass-transfer rate and thus an accurate appraisal of the mass transfer stability. Furthermore, the feedback of mass changes from stellar winds or mass transfer on the stellar structure is modeled self-consistently \citep[see, e.g.][]{2023NatAs.tmp..142B}. Such model advancement allows us to infer the end of the mass-transfer phase more accurately, resolving potentially partially stripped envelopes \citep{2022A&A...662A..56K}. When the binary ensues into a CE phase, we are able to calculate the binding energy self-consistently from the internal structure of the actual donor star model. Finally, we can estimate the properties of compact objects such as masses and BH spins based on the internal stellar-structure profiles of the immediate progenitor stars.
%We will display the evolutionary path of individual NSBH mergers and show the population properties of merging NSBHs and their progenitors, exploring the implications for GW observations and binary evolution theories.

This paper is organized as follows. In section \ref{sec:method}, we briefly introduce \texttt{POSYDON} and describe the most important input physics in the code. We present individual examples of binaries leading to the formation of NSBH, as well as the population properties of merging NSBH in section \ref{sec:res}, and show the impact of a set of model uncertainties in CE evolution and core-collapse prescriptions in section \ref{sec:var}. Finally, we we discuss our findings in section \ref{sec:discussion} and summarise them in section \ref{sec:sum}.

%\begin{deluxetable*}{cccccccc}
\begin{table*}[]
\centering
\renewcommand{\arraystretch}{1.5}
\caption{Median and $90\%$ symmetric credible intervals on component masses, chirp masses, effective inspiral spins for the NSBH merger event candidates. \label{tab:NSBH}}
\begin{tabular}{l|cccc}
\hline\hline
Event&$M_{\rm BH}\,[M_{\sun}]$& $M_{\rm NS}\,[M_{\sun}]$&$M_{\rm chirp}\,[M_{\sun}]$& $\chi_{\rm eff}$ \\
\hline
GW200115\tablefootmark{1} & $5.7^{+1.8}_{-2.1}$ & $1.5^{+0.7}_{-0.3}$ & $2.42^{+0.05}_{-0.07}$ & $-0.19^{+0.23}_{-0.35}$ \\
GW200105\tablefootmark{1} & $8.9^{+1.2}_{-1.5}$ & $1.9^{+0.3}_{-0.2}$ & $3.41^{+0.08}_{-0.07}$ & $-0.01^{+0.11}_{-0.15}$ \\
GW190426\tablefootmark{2} & $5.7^{+3.9}_{-2.3}$ & $1.5^{+0.8}_{-0.5}$ & $2.41^{+0.08}_{-0.08}$ & $-0.03^{+0.32}_{-0.30}$ \\
GW190917\tablefootmark{3} & $9.7^{+3.4}_{-3.9}$ & $2.1^{+1.1}_{-0.4}$ & $3.7^{+0.2}_{-0.2}$ & $-0.08^{+0.21}_{-0.43}$ \\
GW191219\tablefootmark{4} & $31.1^{+2.2}_{-2.8}$ & $1.17^{+0.07}_{-0.06}$ & $4.32^{+0.12}_{-0.17}$ & $-0.00^{+0.07}_{-0.09}$ \\
\hline
\end{tabular}

\tablebib{
\tablefoottext{1}{\citet{2021ApJ...915L...5A}}
\tablefoottext{2}{\citet{2021PhRvX..11b1053A}}
\tablefoottext{3}{\citet{2021arXiv210801045T}}
\tablefoottext{4}{\citet{2021arXiv211103606T}}
}
\end{table*}

\section{The binary population synthesis code - {\tt POSYDON}} \label{sec:method}

We use the newly developed BPS code \texttt{POSYDON} to study merging NSBHs. In contrast to most rapid BPS codes, \texttt{POSYDON} evolves binaries based on detailed stellar and binary models. For a detailed description of the code see \citet{2023ApJS..264...45F}. In what follows, we only summarize some of its key aspects.
%The evolution of stars is treated self-consistently with the input stellar physics as well as binary interactions. The feedback of mass changes from stellar winds or mass transfer is modeled self-consistently \citep[see, e.g.][for ]{} and the angular momentum transport in the interior of the stars and between stars in binaries is taken account of. The inner profiles of the stars can be traced during binary evolution. Hence, we are able to model the SN process with on-the-fly calculations, using the final actual pre-supernova structure of the modeled . In addition, the profile of stellar rotation enables us to estimate the birth spin of BHs, while the information of internal stellar structure allows us to treat CEE in a self-consistent way without using fitted parameters. 

\texttt{POSYDON} contains five pre-calculated stellar and binary evolution grids including two single-star grids: hydrogen (H)-rich star grid and helium (He)-rich star grid, as well as three binary grids in circular orbits with various initial orbital periods, primary masses, and mass ratios: hydrogen ZAMS binary grid; a H-rich star at the onset of Roche-lobe overflow (RLOF) and a compact object grid; a He-rich ZAMS and a compact object (CO-HeMS) grid. The single-star grids are used to treat the non-interacting binaries in eccentric orbits \citep[see Section 8.1 in][]{2023ApJS..264...45F}. 
%We interpolate each non-degenerate star in the binary within the appropriate single-star grids and take into account tides, gravitational radiation, stellar winds, and magnetic breaking along with the binary evolution. 
The binary grids are first postprocessed, so that the data size of the original binary evolution calculations is reduced and additional quantities used in BPS are computed. We then apply classification and interpolation algorithms on the outputs of the three grids, which enables us to effectively interpolate between the pre-computed binary evolution simulations and estimate the evolution of any arbitrary binary. These algorithms currently allow for the interpolation between the initial and the final state of a binary evolution calculation. As a simpler alternative, we also provide functionality to evolve individual binaries using nearest-neighbor matching. While the former approach is the default in population synthesis calculations, nearest-neighbor matching can be used in situations where for a specific binary we need to demonstrate its detailed time evolution (see, e.g., Figures~\ref{fig:ce2} and \ref{fig:smt}).

%after a series of postprocessing steps of the raw data coming from \texttt{POSYDON} allows applying interpolation algorithms to process the outputs of the grids with classification methods to ensure accuracy. For a set of input initial binary parameters, we can implement the 'initial-final' interpolation that generates binary output from the initial conditions based on the interpolation algorithms trained on the detailed binary grids. Alternatively, we can adopt a nearest-neighbor matching scheme that matches the binary with the closest one in the grids. At the onsets of critical evolutionary phases such as SN and CEE, \texttt{POSYDON} allows using profiles of the stars and dynamical information of the binaries to do on-the-fly calculations. The 'initial-final' interpolation can also be applied to obtain the compact object properties as the output of SN step. In this study, we adopt the 'initial-final' algorithms to get the population properties of merging NSBHs.

\subsection{Stellar physics}

We summarize the most important stellar physics model assumptions including stellar winds and mixing adopted in our stellar models. Regarding stellar winds, we use the {\tt Dutch} scheme in \texttt{MESA} for stars initially more massive than $8 M_\odot$. Specifically, we adopt wind prescription of \citet{1988A&AS...72..259D} for cool stars with effective temperature below $10\,000\,\rm{K}$ and \citet{2001A&A...369..574V} for hot stars with effective temperature above $11\,000\,\rm{K}$. Between these two temperatures, we linearly interpolate the wind loss rates. If the surface H fraction of the hot star is below $0.4$, we employ the Wolf–Rayet wind mass loss rates of \citet{2000A&A...360..227N}. For stars initially below $8\,M_\odot$, the {\tt Dutch} scheme is used again when their effective temperature exceeds $12\,000\,\rm{K}$. Whereas for stars cooler than $8\,000\,\rm{K}$, \citet{1975psae.book..229R} wind with scaling factor $\eta_{R} = 0.1$ is adopted for those on the red giant branch and the \citet{1995A&A...297..727B} wind with scaling factor $\eta_{B} = 0.1$ is adopted for those in the thermally-pulsating phase. Between $8\,000\,\rm{K}$ and $12\,000\,\rm{K}$, we linearly interpolate the wind loss rates between the cool and hot wind. 

To treat the convective mixing, we adopt the mixing-length theory \citep[MLT,][]{1958ZA.....46..108B} with a mixing length parameter of $\alpha = 1.93$. For the overshooting mixing, we use the exponential decay formalism \citep{2000A&A...360..952H,2011ApJS..192....3P} with the free parameter $f_{\rm ov} = 0.016$ for stars below $4\, M_\odot$ \citep{2016ApJ...823..102C} and $f_{\rm ov} = 0.0415$ for stars more massive than $8\,M_\odot$. The latter is converted from the result of \citet{2011A&A...530A.115B} that is $\alpha_{\rm ov} = 0.335$ via the rough ratio $\sim 10$ between the step overshoot formalism parameter $\alpha_{\rm ov}$ and the exponential decay formalism parameter $f_{\rm ov}$ \citep{2017ApJ...849...18C}. Note that we back up $0.008$ times the scale height from the edge of the convective zone when assigning $f_{\rm ov}$. We make a smooth transition between the two values of $f_{\rm ov}$ for stars within $4-8\, M_\odot$. Additionally, we adopt the Spruit–Tayler dynamo \citep{2002A&A...381..923S} to account for the efficient coupling between the stellar core and envelope. We treat rotation mixing and angular momentum transport following the MIST project \citep{2016ApJ...823..102C}.

\subsection{Tidal interactions and binary mass transfer}

Tidal interactions in a binary system affect the orbital evolution and the rotational properties of the stars. In a close detached binary, tidal forces efficiently circularise the orbit and lead to the synchronization between the stars' rotation rates and the orbit. We follow \citet{1981A&A....99..126H} to calculate the evolution of binary separation, eccentricity, and star rotation, considering the difference between convective and radiative envelope \citep{2002MNRAS.329..897H,2018A&A...616A..28Q}. For the initial state of the binary models, we assume the spins of two ZAMS stars are synchronized with the orbits.

By virtue of detailed simulations, we are able to model the mass transfer in a self-consistent way, considering the feedback of mass loss or gain along with the angular momentum change for both donors and acrretors. For a main-sequence star, we adopt the \texttt{contact} scheme in \texttt{MESA} to calculate the mass transfer rate during RLOF. In contrast, for an evolved star, when its center H abundance is below $10^{-6}$, we use the \texttt{Kolb} scheme \citep{1990A&A...236..385K} allowing for a better treatment of the extended envelope. Based on the nature of the accretor, we treat the mass accretion process differently. In the case of accretion onto a compact object, we limit the accretion at the rate corresponding to the Eddington luminosity $L_{\rm Edd}$, which can be written as:
\begin{equation}\label{Ledd}
L_{\rm Edd} = \frac{4\pi G M_{\rm acc} c }{\kappa},
\end{equation}
where $G$ is the gravitational constant, $c$ is the speed of light, $M_{\rm acc}$ is the mass of the accretor, and $\kappa$ is the opacity. The excess mass transferred is assumed to be lost as isotropic wind, carrying with it the specific orbital angular momentum of the accretor. In the case of accretion onto a non-degenerate star, we consider a rotation-dependent accretion. The transferred material carrying angular momentum will spin up the outer layer of the accretor, considering both ballistic and Keplerian disk mass transfer \citep{2013ApJ...764..166D}. The accretor's rotation will reduce the accretion rate by enhancing its stellar wind \citep{1998A&A...329..551L}. Most importantly, accretion is ceased when the accretor reaches critical rotation, implicitly boosting its winds as much as needed in order to not exceed $\omega / \omega_{\rm crit}\sim 1$. 
%and the rotation will reduce the accretion rate in the form of enhanced stellar wind \citep{1998A&A...329..551L}:
%\begin{equation}\label{ml}
%\centering
%\dot{M}(\omega)= \dot{M}(\omega = 0)\left(\frac{1}{1-\omega_{\rm s}/\omega_{\rm s,crit}}\right)^\xi,
%\end{equation}
%where $\xi$ is a constant of $0.43$, $\dot{M}$ is the accretion rate, $\omega$ is the surface angular velocity of the accretor, and $\omega_{\rm s,crit} = \sqrt{(1- L/L_{\rm Edd})GM/R^3}$ is the critical surface angular velocity, in which $L$, $M$, $R$ represent the accretor's luminosity, mass, and radius. The specific angular momentum of the transferred material is computed by

\subsection{Common-envelope evolution}

CE evolution is crucial for the formation of close double compact objects. If the binary experiences unstable mass transfer and the donor is not a main-sequence star or a stripped He star, the binary will enter a CE phase. Then, the companion transports orbital energy and angular momentum to the CE, causing dramatic orbital shrinkage. If the CE is successfully ejected during this process, the binary will avoid merger and go into a detached phase with the core of the donor and the companion. In our detailed binary evolution simulations, we consider the following conditions as the criterion for the onset of unstable mass transfer. Firstly, whenever the mass transfer rate exceeds $0.1\,M_{\odot}\,\rm{yr^{-1}}$, we assume it becomes unstable as the mass transfer is only expected to be more rapid after reaching this limit. Secondly, we check if the star has mass loss from the second Lagrangian point (L2). The mass loss from the L2 point is considered to take away a significant amount of angular momentum from the binary leading to a rapid orbital shrinkage and triggering CE evolution \citep{2011A&A...528A.114T,2014ApJ...786...39N}. For the case where the donor is a post main-sequence star, we check L2 RLOF following the implementation of \citet{2020A&A...642A.174M}, while for the case of two main sequence stars, we calculate the L2 radius using the prescription from \citet{2016A&A...588A..50M}. Thirdly, for a compact object accretor, if its photon trapping radius \citep{1979MNRAS.187..237B} extends beyond the Roche-lobe radius, the binary would enter CE evolution. Finally, if both binary components fill their Roche lobe , i.e. the binary is in a contact phase, and at least one of the stars has evolved beyond its main sequence, we assume the binary enters a CE phase.

To treat CE evolution, we use the $\alpha-\lambda$ formalism \citep{1984ApJ...277..355W,1988ApJ...329..764L}. The value of $\lambda_{\rm CE}$ depends on the mass and evolutionary stage of the stars, affected by the internal structure and internal energy of the star \citep{1990ApJ...358..189D,2000A&A...360.1043D}. At the onset of the CE phase, we calculate the value of $\lambda$ self-consistently using the stellar profile. To encompass the model uncertainty on the boundary of the core and the envelope, we implemented three hydrogen abundances for the definitions of the He core-envelope boundary, which are $X_{\rm H}$ = $0.01, 0.1$ and $0.3$. We take the value $0.1$ as the default. The $\alpha_{\rm CE}$ parameter determines how efficient the orbital energy can be converted to eject the envelope. The value of $\alpha_{\rm CE}$ is still uncertain and not necessarily constant for stars with different properties. We take $\alpha_{\rm CE} = 1.0$ as the default value. With these two parameters, we can calculate the binding energy of the envelope and the post-CE orbit where the transferred orbital energy is enough to disperse the envelope. If in the post-CE orbit computed by the $\alpha-\lambda$ formalism, the core of the donor and the companion are not filling their Roche lobe, we consider the CE phase to be successful. 

\subsection{Core-collapse physics and compact-object formation}

We adopt the \citet{2012ApJ...749...91F} delayed core-collapse prescription by default to calculate the remnant baryonic mass of core-collapse SN (CCSN). The pre-SN carbon-oxygen core mass $M_{\rm CO,core}$ determines whether the star explodes as well as the fall back fraction $f_{\rm fb}$ of the ejected mass.
In addition, we explore other core-collapse prescriptions including the \citet{2012ApJ...749...91F} rapid prescription and the \citet{2020MNRAS.499.2803P} prescription. The latter determines the explodability of a star based on the $M_{\rm CO,core}$ and the carbon mass fraction of the stellar core at the point of carbon ignition. For electron-capture SN (ECSN), we consider stars with the final He core masses in the range of $1.4-2.5\,M_{\odot}$ \citep{2004ApJ...612.1044P} as their progenitors. 
%We use the fit to the simulations of \citet{2019ApJ...882...36M}  \citep[see Eq. 4 in][]{2020ApJ...898...71B} to map CO-core masses to BH masses. 
To covert the remnant baryonic mass to gravitational mass for the compact objects, we follow the method of \citet{2020ApJ...899L...1Z}. Furthermore, we assume the maximum mass for a NS to be $2.5\,M_{\odot}$ \citep[see discussion in][and references therein]{2020ApJ...896L..44A}. Finally, we estimate the birth BH spins based on the final stellar profiles in our grids, following the method described in Appendix D of \citet{2021A&A...647A.153B}, while we assume the spins of NSs are zero for simplicity. %In our simulation, the BH masses and spins are obtained directly with 'initial-final' interpolation based on the calculation with the stellar profile.% The NS masses are calculated using the interpolated core masses because we only care about the NS masses and do not need other information from the profile of the progenitors before core collapse. 

A newly-born compact object may receive a SN natal kick because of asymmetric mass ejection and neutrino mass loss that can tilt the orbit, alter the orbital separation and eccentricity, or even disrupt the binary. By default, we adopt a natal kick velocity following a Maxwellian distribution with a velocity dispersion of $\sigma_{\rm CCSN} = 265\,\rm{km\,s^{-1}}$ \citep{2005MNRAS.360..974H} for CCSN and $\sigma_{\rm ECSN} = 20\,\rm{km\,s^{-1}}$ for ECSN \citep{2019MNRAS.482.2234G}. The BH kick velocities follow the CCSN distribution, rescaled by a factor of $1.4\,M_{\odot}/\,M_{\rm BH}$. We calculate the orbital tilt $\theta$ introduced by the natal kick using equation (5) in \citet{1996ApJ...471..352K}, which is defined as the angle between the angular momentum of the post-SN orbit and the spin vector of the compact object. 

After the formation of the BH, the secondary star may initiate a mass-transfer phase. In that case, we assume that the secondary star's spin would always be aligned with the first post-SN orbit. The accretion process onto BHs is expected to affect both the magnitude and the direction of the BH spins \citep{2010ApJ...719L..79F}. In addition, the accretion disc may become a warped disc due to the Lensing-Thirring effect \citep{1996MNRAS.282..291S}, which would accelerate the alignment between the BH spin and the orbit. However, in the case of binaries with stellar-mass BHs, the effect can be very weak \citep{2005MNRAS.363...49K}. We assume that the transferred material has the specific angular momentum at ISCO. The BH's angular momentum $J$ can be divided into two components, along ($J_{\rm{y}}$) and perpendicular ($J_{\rm{x}}$) to the angular momentum of the orbit. For simplicity, we assume the BH gains all the mass at once and the transferred angular momentum contribute to changing $J_{\rm{y}}$. After accretion, the total angular momentum of the BH is equal to $\sqrt{J_{\rm{y,acc}}^2 + J_{\rm{x}}^2}$, the tilt angle from the first SN, $\theta_{1}$, becomes $\arccos(J_{\rm{y,acc}}/J)$, and the updated BH spin $\chi_{\rm BH}$ is determined by $J$. The simplified assumption might lead to underestimation of the change on BH spin and tilt because the specific angular momentum of the ISCO is constantly changing during the accretion phase. However, we find that this has only a marginal impact on the majority of the binaries, as discussed in the next section. 

When the second SN happens, the orbit is tilted again. The final tilt angle of the BH can be calculated from the two tilt angles $\theta_{1}$ and $\theta_{2}$:
\begin{equation}\label{tilt}
\cos{\theta} = \cos{\theta_{1}}\cos{\theta_{2}} \pm \sin{\theta_{1}}\sin{\theta_{2}}\cos{\alpha},
\end{equation}
where $\alpha \in [0,\pi]$ is the angle of the relative position of two SN in the orbit, which we choose randomly as we attribute random mean anomaly for the two SN kicks. When $\alpha = 0$ or $\pi$, the three vectors\textemdash the BH spin, the first post-SN orbital plane, and the second post-SN orbital plane, all lie in the same plane. This occurs either when both SN take place at the same relative position within the orbit or at opposite sides of the orbit. In such case, the final tilt is the addition or subtraction of the two tilt angles, which is also determined by whether the tilts are in the same or opposite directions. 

With the information of component spins and orbital tilt angles, we can calculate the effective inspiral spin defined as:
\begin{equation}\label{eff}
\chi_{\rm eff} = \frac{M_{\rm BH}\chi_{\rm BH}\cos{\theta_{\rm BH}} + M_{\rm NS}\chi_{\rm NS}\cos{\theta_{\rm NS}}}{M_{\rm BH}+M_{\rm NS}},
\end{equation}
where $\theta_{\rm BH}$ and $\theta_{\rm NS}$ are the misalignment angles between spin vectors of compact objects and the vector along the final orbital angular momentum. $\theta_{\rm BH}$ is equal to the final tilt angle $\theta$ in Equation~\ref{tilt}. Because we assume $\chi_{\rm NS} = 0$, $\chi_{\rm eff}$ is determined by the component masses, BH spin and the final tilt angle.

\subsection{Initial binary properties}

For each of our populations synthesis models, we evolve $10^{7}$ binaries, starting from two H-rich ZAMS stars. We randomly sample the initial masses of the primary stars, $M_{1,i}$ in the range $7-120\,M_{\odot}$, following a distribution of the initial mass function from \citet{2001MNRAS.322..231K}. The secondary masses, $M_{2,i}$ are in the range $0.35-120\,M_{\odot}$ following a uniform distribution between the minimum and $M_{1,i}$ \citep{2007ApJ...670..747K}. The distribution of the initial orbital separation is logarithmically flat between $1$ and $10^{5}\,R_{\odot}$ and we assume zero eccentricity for the primordial binaries. To display the population properties of merging NSBHs, we apply a burst star formation history, which assumes all stars form at the same time.

\section{Results} \label{sec:res}
\subsection{Formation channels}
In our simulation, we found BH always form first in merging NSBH systems. We identify two main formation channels for the merging NSBH systems. One involves stable mass transfer before the formation of the first-born compact object and CE evolution between a BH and a non-degenerate star ($\sim70\%$, the percentage refers to the number of merging NSBHs systems per unit of simulated stellar mass) and the other one involves only stable mass transfer episode(s) during the evolution ($\sim30\%$). Each channel can be further divided into two sub-channels based on the details of the mass transfer episodes, as we will discuss in the following sections. For each channel, we illustrate the evolutionary path of a typical binary for the dominant sub-channel from a \texttt{POSYDON} simulation as an example. 

\subsubsection{Stable mass transfer and common envelope evolution channel}
\begin{figure*}\center
\includegraphics[width=\textwidth]{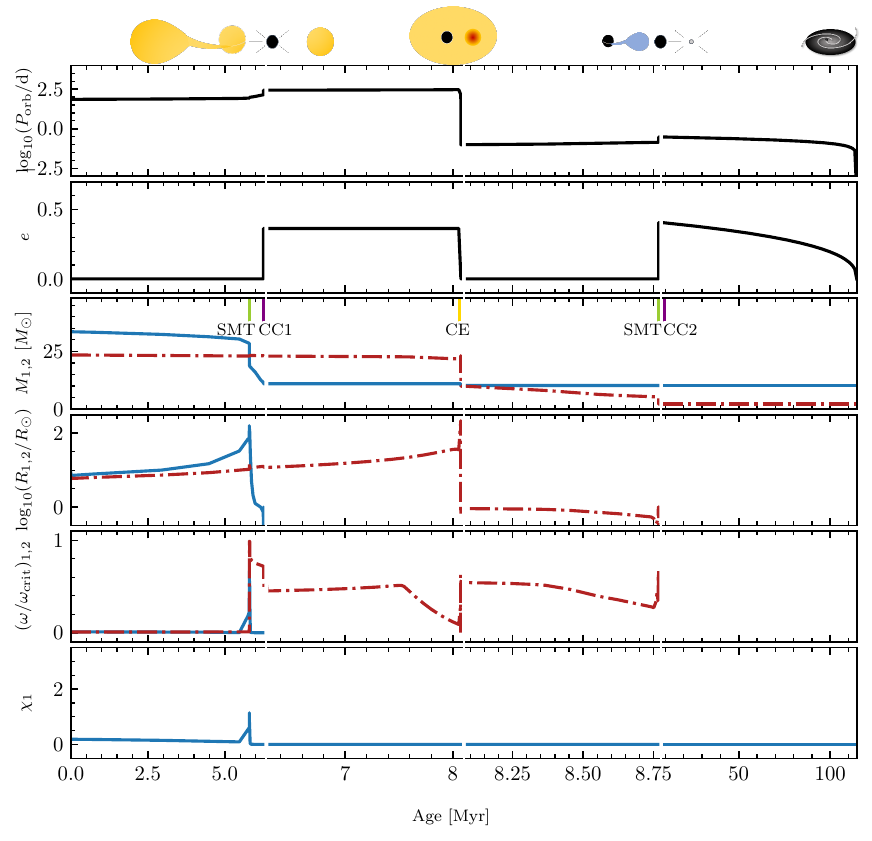}\\
\caption{Evolutionary path, from ZAMS to the merger, of a NSBH binary from channel \RomanNumeralCaps 1a. From the top to the bottom we show the evolution of the orbital period $P_{\rm orb}$, eccentricity $e$, component masses $M_{1,2}$, star radii $R_{1,2}$, surface angular velocity over the critical value $(\omega/\omega_{\rm crit})_{1,2}$, and the spin of the primary $\chi_{1}$. The blue solid line represents the star that evolves to form a BH while the red dashed-dotted line represents the star that becomes a NS. In the panel of stellar mass evolution, we indicate with the short vertical lines the onset of the stable mass transfer (SMT), the first core collapse (CC1), the common envelope (CE) phase, and the second core collapse (CC2). 
}
\label{fig:ce2}
\end{figure*}

We refer to the channel involving stable mass transfer between two non-degenerate stars and a CE phase after the BH formation as channel \RomanNumeralCaps 1. 
%Binaries from this channel typically have wide initial orbits ($P_{\rm orb} \gtrsim 50\,\rm d$). 
The first stable mass-transfer phase occurs when the primary star begins expanding during the H shell-burning stage or the He core-burning stage. After the mass-transfer phase, the primary star loses its H-rich envelope through stellar winds. The stripped star then undergoes CCSN forming a BH. After the first core collapse, the binary remains detached until the secondary star enters post main-sequence phase and fills the Roche lobe. The second mass-transfer phase is unstable, leading to a CE phase. If the CE evolution is successful, the secondary star would lose the H-rich envelope with the ejection of the CE, leaving behind a He star. About $94\%$ of the binaries undergo a third mass transfer episode, a case-BB stable mass-transfer phase, between the BH and the He star, followed by the secondary star exploding and forming a NS. We refer to this channel as channel \RomanNumeralCaps{1}a. Alternatively, the remaining binaries evolve without undergoing any further mass transfer episode after the CE phase. This is regarded as channel \RomanNumeralCaps 1b. An illustration of the evolutionary phases are indicated at the top of Figure \ref{fig:ce2}, in which we show the evolutionary path of a typical NSBH merger evolving through channel \RomanNumeralCaps 1a. 

The binary in Figure \ref{fig:ce2} has initially a primary star of $\sim33.50\,M_{\sun}$, a secondary star of $\sim23.45\,M_{\sun}$, and an orbital period of $72.0\,\rm{d}$. The primary loses mass through stellar winds down to $\sim 28.50\,M_{\sun}$ before the mass-transfer phase initiates. The stable mass transfer starts at roughly $5.795\,\rm{Myr}$ after the birth of the binary system, lasting for only $\sim 15\,000\, \rm{yr}$. The secondary is spun up quickly due to the accretion, gaining about $0.4\,M_{\sun}$ in total. The primary experiences CCSN at $\sim 6.257\,\rm{Myr}$ forming a BH of $10.98\,M_{\sun}$ with a negligible spin of $\chi_{\rm BH} = 0.005$, and the binary goes into the detached phase. As the secondary evolves, the mass transfer from the secondary onto the BH starts, shrinking the orbit. Soon after the onset of the mass transfer, the secondary radius exceeds the L2 Roche lobe radius, leading to a CE phase at $\sim 8.073\,\rm{Myr}$. When the CE is ejected, the secondary is a stripped He-rich star of $\sim 9.97\,M_{\sun}$ and it expands at the end of He-burning phase, leading to a case-BB mass-transfer phase. Eventually, the secondary collapses to a $2.20\,M_{\sun}$ NS at $\sim 8.766\,\rm{Myr}$, and the binary merges at $\sim 552.274\,\rm{Myr}$ due to energy and angular momentum loss from GW radiation.

\subsubsection{Stable mass transfer channel}
We refer to the channel that involves only stable mass transfer episode(s) as channel \RomanNumeralCaps 2. %The binaries from this channel have relatively short initial orbital periods in general ($P_{\rm orb} \lesssim 50\,\rm d$). 
About $96.4\%$ of them undergo two stable mass-transfer phase. The binary first undergoes a stable case-A or case-B mass-transfer phase, resulting into the donors being partially stripped at the end of the main sequence. Consequently, the primary stars do not enter a giant phase as in channel \RomanNumeralCaps 1, but gradually shrink after evolving off the main sequence. After the formation of the BH, the binary experiences stable case-A or case-B mass transfer and some of them may experience both case-A and case-B mass transfer at the second mass-transfer phase. Finally, the secondary star explodes forming a NS. We refer to this channel as channel \RomanNumeralCaps 2a. In cases where the primary stars are more massive than $\sim 60\,M_{\sun}$, the binaries may avoid the first mass transfer. This is because at solar metallicity, our \texttt{POSYDON} stellar models predict that the stars above $\sim 60\,M_{\sun}$ do not expand to the red supergiant state with radii of $\sim 1000\,R_{\sun}$ \citep{2023NatAs.tmp..142B}. The mass loss through stellar winds from the stars keeps widening the orbits until the first core collapse. Subsequently, the binaries undergo case-A or case-B stable mass transfer from the secondary onto the BH and the secondary stars become NSs in the end. This sub-channel accounts for about $4.6\%$ of channel \RomanNumeralCaps 2 and we refer to it as channel \RomanNumeralCaps 2b.  Without an efficient orbital shrinkage mechanism, at the time of NS formation, binaries from channel \RomanNumeralCaps 2 have relatively wide orbits compared with channel \RomanNumeralCaps 1. Consequently, only the NSBHs formed with eccentric orbits (caused by SN natal kicks) result in delay time smaller than a Hubble time. 

\begin{figure*}\center
\includegraphics[width=\textwidth]{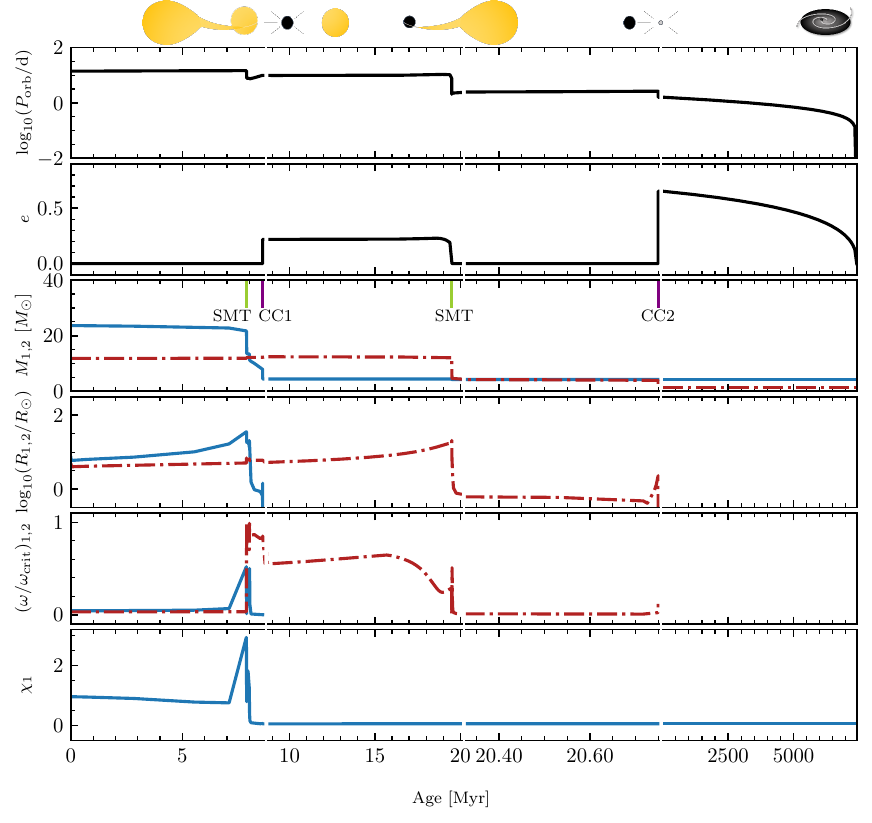}\\
\caption{Same as Figure \ref{fig:ce2} but for a NSBH binary formed through channel \RomanNumeralCaps 2a.
}
\label{fig:smt}
\end{figure*}

Figure \ref{fig:smt} shows a a typical binary evolving through the channel \RomanNumeralCaps 2a, which initially has a $\sim 23.66\,M_{\sun}$ primary, a $\sim 11.83\,M_{\sun}$ secondary, and an orbital period of $13.90\,\rm{d}$. The first stable mass-transfer phase starts at $\sim 7.867\,\rm{Myr}$ and ends at $\sim 7.882\,\rm{Myr}$ when the primary is still a main-sequence star. The secondary star reaches near-critical rotation during the mass-transfer phase, with the rate as high as $\sim 10^{-2.5}\,M_{\sun}\,\rm{yr^{-1}}$. After a short time of being detached from the secondary, the primary evolves off the main sequence and the mass transfer starts again at $\sim 8.003\,\rm{Myr}$ and ends at $\sim 8.012\,\rm{Myr}$ with a peak of $\sim 10^{-3.3}\,M_{\sun}\,\rm{yr^{-1}}$. During the mass-transfer phase, the secondary star accretes $\sim 0.40\,M_{\sun}$ in total. In the mean time, the primary loses its envelope through mass transfer and winds. At $\sim 8.600\,\rm{Myr}$, the primary explodes to form a BH of $4.40\,M_{\sun}$ with a dimensionless spin parameter of $\chi_{\rm BH} = 0.05$. Then, a stable mass-transfer phase ensues from the secondary onto the BH. The secondary later explodes to form an NS of $1.28\,M_{\sun}$ at $\sim 21.738\,\rm{Myr}$. Finally, the binary merges at $\sim 6115.647\,\rm{Myr}$.

\subsection{Population properties of merging NSBHs}

In this section, we present the population properties of merging NSBHs and their progenitors, including component masses, orbital properties, BH spins, final tilt angles and effective spins. We select all the NSBH systems that merge within a Hubble time and distinguish them by the formation channels\footnote{For $\lesssim$2\% of the total number of simulated NSBH binaries, specifically those originating from areas where the parameter space is close to the boundary between stable and unstable mass transfer, our classifiers fail to classify the type of the resulting compact object. While we are still able to correctly infer the correct masses and orbital properties for these binary compact objects, we cannot properly estimate the BH spins. Given the small number of systems that encounter this issue, we decided to exclude them from our analysis. This known technical issue of {\tt POSYDON} v1 will be addressed in future releases.}
%We exclude potential NSBHs near the boundary of different binary types separated by the mass transfer scenario, where the 'initial-final' classifier failed to obtain the correct compact object type. These systems account for less than two percent of the total number of merging NSBHs.} 

\subsubsection{Progenitors of merging NSBHs}

\begin{figure*}\center
\includegraphics[width=0.96\textwidth]{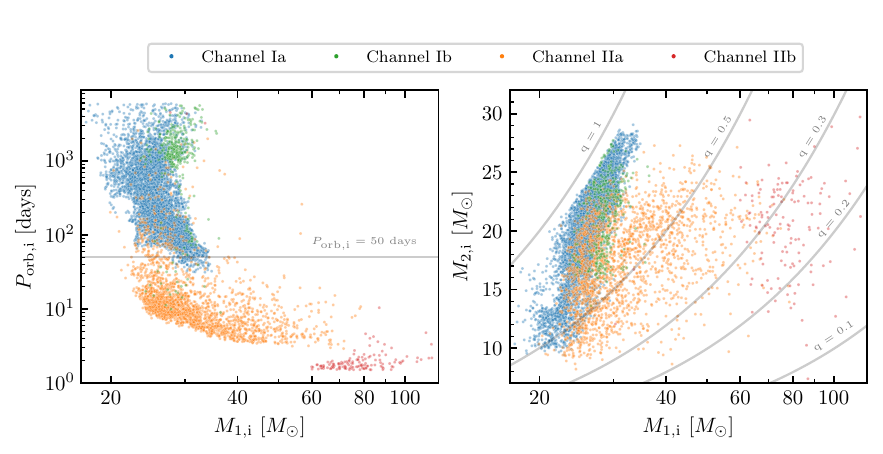}\\
\caption{Initial properties of the progenitors of merging NSBHs. The blue, green, orange and red dots represent binaries from channel \RomanNumeralCaps 1a, \RomanNumeralCaps 1b, \RomanNumeralCaps 2a and \RomanNumeralCaps 2b, respectively. The left panel shows the primary masses versus initial binary orbital periods. The gray line indicates $50\,\rm{days}$ of orbital period for reference. The right panel displays the component masses with the gray lines indicating different mass ratios $q$, defined as $M_{\rm{2,i}} / M_{\rm{1,i}}$.
}
\label{fig:3}
\end{figure*}

On the left panel of Figure \ref{fig:3}, we show the initial orbital periods $P_{\rm orb,i}$ versus the initial primary masses $M_{\rm 1,i}$ of merging NSBHs. The majority of the binaries from channel \RomanNumeralCaps 1 have $P_{\rm orb,i}$ above $\sim 50\,\rm d$ and most of the binaries evolving through channel \RomanNumeralCaps 2 have closer orbits with $P_{\rm orb,i}$ in the range from a few days to a few tens of days. Binaries with initial primary masses high enough to produce a BH and initial periods $\lesssim 50\,\rm d$ are more likely to results in post-BH-formation orbits that subsequently lead to stable case-A or case-B mass transfer, from the secondary star onto the BH this time. Even if the mass-ratio is such that the mass transfer is unstable, the CE evolution will result in a merger, as the donor star is still very compact. For binaries with initial periods $\gtrsim 50\,\rm d$, even when they result in post-BH formation binaries that undergo stable mass transfer,the final orbital periods at the formation of NSBH that are too wide to lead to a merger within a Hubble time. In contrast, the progenitor binaries from channel \RomanNumeralCaps 1, must have initial periods $\gtrsim 50\,\rm d$, such that the post-BH formation orbital periods are wide enough to not only lead to unstable mass transfer, but also have the secondary star sufficiently evolved at the onset of the unstable mass transfer so that CE evolution can lead to the successful ejection of the envelope. The interplay of all these processes determine the relatively sharp boundary in initial orbital periods that separate channel \RomanNumeralCaps 1 and channel \RomanNumeralCaps 2.

%Most systems formed with orbital periods less than $\sim 50\,\rm d$ are more likely to undergo stable case-A or case-AB mass transfer after the BH formation. Even if they experience unstable mass transfer, they merge during the CEE because of the close orbital separation. The binaries going through only stable mass transfer are not able to form NSBHs merging within a Hubble time if they have initial orbital periods above $\sim 50\,\rm d$, unless they acquire large eccentricities and tighter orbits from the SN kick.
%In channel \RomanNumeralCaps 1, most systems formed with orbital periods less than 50 d experiencing unstable mass transfer merge during the CEE because of their close orbital separation. In channel \RomanNumeralCaps 2, most systems forming with oribtal periods above 50 d merge only if they acquire large eccentricities fro the SN kick at NS formation. 
In channel \RomanNumeralCaps 2b, the binaries with the most massive primary stars have the shortest initial orbital periods of a few days. The strong winds of the massive stars will keep widening the orbit during the early evolution, resulting in orbital periods of $\sim 10\,\rm{d}$ in the end. These binaries experience no RLOF mass transfer before the BH formation because the massive primary stars do not expand to become supergiant stars.

The ZAMS masses of the progenitor stars of merging NSBHs are shown in the right panel of Figure \ref{fig:3}. The primary stars that evolve to BHs have initial masses $M_{\rm{1,i}}$ varying from $\sim 17$ to $\sim 120\,M_{\sun}$ with the majority of them in the range of $\sim 20$ to $\sim 40\,M_{\sun}$. The secondary stars that form NSs have initial masses $M_{\rm{i,2}}$ lying within the range of $\sim 8$ to $\sim 30\,M_{\sun}$. Most of the binaries evolving through channel \RomanNumeralCaps 1 exhibit initial mass ratios $q = M_{\rm{i,2}} / M_{\rm{i,1}}$ greater than $0.5$, while binaries originating from channel \RomanNumeralCaps 2 display a broader range of mass ratios. Notably, binaries from Channel \RomanNumeralCaps 2b have small mass ratios and can reach below $0.2$. The right panel of Figure \ref{fig:3} also shows the scarcity of massive ($\gtrsim 40\,M_{\sun}$) primary stars in channel \RomanNumeralCaps 1. Given the adopted stellar physics parameters in \texttt{POSYDON}, a $40\,M_{\sun}$ can lead to the formation of BH as massive as $18\,M_{\sun}$. Such massive BHs, when paired with secondary stars that are potential NS progenitors ($\lesssim 28\,M_{\sun}$), they experience stable mass transfer independently of the post-BH formation orbital periods and thus cannot be progenitors of merging NSBH from channel \RomanNumeralCaps 1. 

%According to our \texttt{POSYDON} single stellar models at solar metallicity, BHs can form with masses up to $\sim 35\,M_{\sun}$. This means that primary stars more massive than $40\,M_{\sun}$ would produce even more massive BHs, which are likely to lead to stable mass transfer instead of a CE phase.

\subsubsection{Compact objects in merging NSBHs}

\begin{figure*}\center
\includegraphics[width=0.96\textwidth]{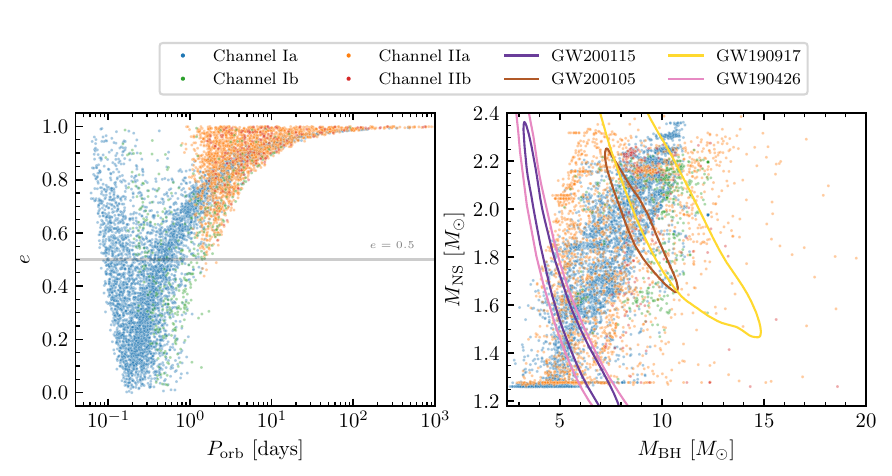}\\
\caption{Properties of merging NSBHs at the formation of the double compact object system. The left panel shows the orbital periods and eccentricities. The right panel shows the BH and NS masses and the $90\%$ confidence intervals of GW200115, GW200105, GW190917 and GW190426.}
\label{fig:NSBH}
\end{figure*}

The left panel of Figure \ref{fig:NSBH} shows the orbital periods and eccentricities of the merging NSBHs at the time of double compact object formation. Channel \RomanNumeralCaps 1 is capable of forming merging NSBH systems in close obits $\lesssim 0.1\,\rm{d}$ due to CE evolution, whereas channel \RomanNumeralCaps 2 can only form them in orbits $\gtrsim 1\,\rm{d}$. A large number of the NSBH systems forming without undergoing a CE phase have too long delay times to merge within a Hubble time, except for those that obtain a high eccentricity from natal kick at the second SN. The merging NSBHs from channel \RomanNumeralCaps 2 have eccentricities greater than $\sim 0.5$.

In Figure \ref{fig:NSBH} on the right, we show the BH and NS masses of merging NSBHs as well as the $90\%$ confidence intervals of NSBH merger candidates GW200115, GW200105, GW190917 and GW190426. GW191219 is not included because its BH mass is much larger than the maximal values in our population. The median of the component masses of the four candiates are within the mass range of merging NSBHs from our model. However, we cannot rule out the possibility that they were formed at lower-metallicity environment.
%For the other possible NSBH candidates, we do not show them in the plot as they are outside the mass range from our simulation. 
Overall, the NS masses are covering a mass range of $\sim 1.26$ to $\sim 2.4\,M_{\sun}$ and the BH masses are from $\sim 2.5$ to $\sim 20\,M_{\sun}$. Channel \RomanNumeralCaps 1 forms BHs less massive than $\sim 12\,M_{\sun}$, while channel \RomanNumeralCaps 2 is able to produce more massive BHs but only a small fraction of them are above $15\,M_{\sun}$.
%%A large group of NSs have masses more massive than $\sim 1.75\,M_{\sun}$. 
The \citet{2012ApJ...749...91F} delayed mechanism we adopt for calculating the compact object mass from CCSN does not produce a BH mass gap between $2.5-5\,M_{\sun}$. The BHs in the mass gap account for $16.2\%$ of the total NSBH population. In the \citet{2012ApJ...749...91F} delayed prescription, the discontinuity of calculating proto-compact object mass at $M_{\rm CO,core}$ of $3.5\,M_{\sun}$ introduces a NS mass gap of $\sim 0.1\,M_{\sun}$ around $\sim 1.7\,M_{\sun}$. However, the existence of this gap is not observed in practice. In our simulation, the interpolation applied to obtain the NS masses smooths out the discontinuity. In the \citet{2012ApJ...749...91F} delayed prescription, if the final $M_{\rm CO,core}$ of the compact object progenitor is less massive than $2.5\,M_{\sun}$, a constant NS mass of $\sim 1.27\,M_{\sun}$ is predicted. Accordingly, we can see the concentration at that mass especially for channel \RomanNumeralCaps 2. In channel \RomanNumeralCaps 1, the NS progenitors are more likely to have lower $M_{\rm He,core}$ in the range of $1.4-2.5\,M_{\sun}$, leading to ECSN and the formation of NSs at $\sim 1.26\,M_{\sun}$.   %And we can see that within the mass gap $\lesssim 5\,M_{\sun}$, the BHs are likely to have low mass NS companions.

\begin{figure*}\center
\includegraphics[width=0.96\textwidth]{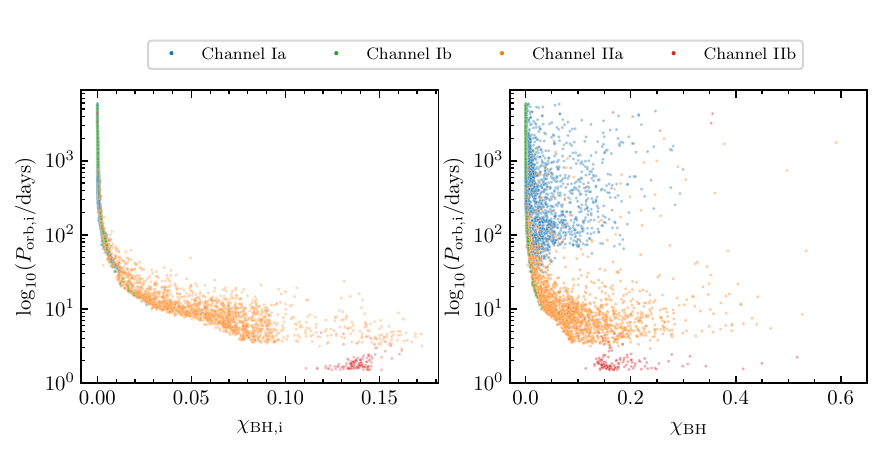}\\
\caption{BH spins versus initial orbital periods of merging NSBHs at BH birth (left) and after the NS formation (right).
}
\label{fig:spin}
\end{figure*}

We calculate the BH spins using the stellar profiles of the BH progenitors at the onset of the SN event \citep[see Sec.8.3.4 in][]{2023ApJS..264...45F}. The BH spins versus initial orbital periods is shown in Figure \ref{fig:spin}, where the left panel shows the BH spins at birth and the right panel shows the final BH spins after accretion. From the left panel, we can see a significant correlation between the BH spins and the initial orbital periods where shorter orbital periods correspond to higher BH spins. The prevalence of short initial orbital periods among the binaries in channel \RomanNumeralCaps 2 results in a substantial number of BHs with non-negligible spins. In our models, the BHs always form from the primary stars. As we assume efficient angular momentum transport inside stars in our models, the primary stars are expected to lose most of their angular momentum through mass transfer and winds, forming BHs with negligible spins. However, the orbits of the initially close binaries can remain tight before the first core collapse, and tides could help spin up the progenitor stars. In this case, although the stars lose the envelope with nearly all the stellar angular momentum, the stripped primary stars preserve a portion of angular momentum contributing to the BH spins. Nevertheless, neither channel produces BHs with spins larger than $\sim 0.2$ at BH birth, consistent with \citep{2018A&A...616A..28Q,2019ApJ...881L...1F,2020A&A...636A.104B}.
%In addition, for channel \RomanNumeralCaps 2b, the primary stars, which are in the closest orbits, would experience no mass transfer and lose angular momentum only by the winds, leading to spinning BHs. 
From the right panel, we can see that BHs can be moderately spun up by accretion. For channel \RomanNumeralCaps 1, the BHs gain spins less than $\sim 0.3$ through case-BB stable mass transfer. For channel \RomanNumeralCaps 2, the BH spins increase mainly due to case-A stable mass transfer. Most of the BH spins still stay below $0.2$, $8.4\%$ of them are within $0.2-0.4$ and only $0.7\%$ of them exceed $0.4$. In channel \RomanNumeralCaps 2, the inclination angle between the BH spin and the orbit is typically small because the pre-SN orbit is tightly bound and hard to be tilted significantly by natal kicks. In channel \RomanNumeralCaps 1, some binaries possess large tilt angles or are even anti-aligned after the BH formation. However, the BH spins are close to zero. As a result, the impact of the simplified assumption on the spin evolution due to accretion is negligible for most of the binaries. 

\subsubsection{Spin-orbit tilt and effective inspiral spin}\label{sec:tilt}
\begin{figure*}\center
\includegraphics[width=0.96\textwidth]{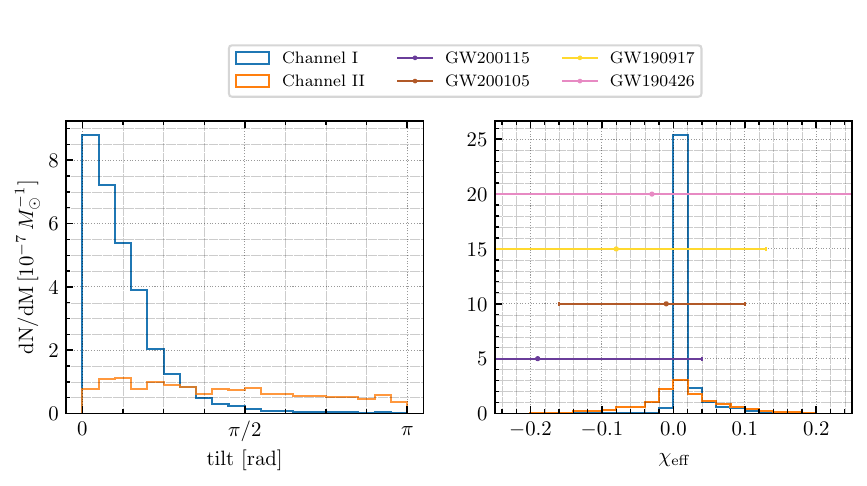}\\
\caption{Distribution of final tilt angles and effective inspiral spins for channel \RomanNumeralCaps 1 (blue) and \RomanNumeralCaps 2 (orange). The colored circles and corresponding error bars in the right panel show the median and $90\%$ credible intervals of effective inspiral spins for the NSBH merger event candidates. For some of the observed systems, the error bars extend outside the range of the figure (see also Table~\ref{tab:NSBH}).
%For the observed systems with less accurately determined effective spins, the error bars extend beyond the range of the figure.
}
\label{fig:tilt}
\end{figure*}

In Figure \ref{fig:tilt}, we show the distribution of the tilt angles between the BH spin and the orbital angular momentum, and the effective inspiral spins for the two channels. The numbers shown in the figure are normalized with the total underlying simulated mass of the population in units of $10^{-7}\,M_{\sun}^{-1}$. A large fraction of binaries from channel \RomanNumeralCaps 1 have small tilt angles and only about $2\%$ of them end up with anti-alignment between the BH spin vector and orbital angular momentum vector. In contrast, for channel \RomanNumeralCaps 2, about $40\%$ of the binaries have final tilt angles $>\pi/2\,\rm{rad}$. 
%This distribution could result from that channel \RomanNumeralCaps 2 favors a natal kick direction that makes small projection onto the original orbital plane. 

The difference for the two channels is attributed to the different orbital velocity at the time of the second SN. In channel \RomanNumeralCaps{2}, binaries experience a natal kick with a velocity comparable to the orbital velocity to provide high eccentricity to the orbit. Hence, the kick can tilt the orbit easily. In channel \RomanNumeralCaps 1, most binaries are in tightly bound orbits with high orbital velocities, thus natal kicks in excess of $\gtrsim 1000\,\rm km/s$ would be required to tilt significantly the orbit.
%meaning that it is demanding to tilt the orbit significantly by natal kicks.
With the final tilt angles and BH spins, we calculate the effective spins of NSBH systems. The right panel of Figure \ref{fig:tilt} shows the distribution of the effective inspiral spins for the two channels as well as the median and cut-off $90\%$ credible intervals of effective inspiral spins for the NSBH merger event candidates. Most of the binaries from channel \RomanNumeralCaps 1 possess BHs with nearly zero spins and small tilt angles, so the distribution of the effective spins peaks sharply at zero. A small fraction of BHs that have spun up due to accretion, contribute to the presence of systems with positive $\chi_{\rm eff}\gtrsim 0.02$. Channel \RomanNumeralCaps 2 produces a larger fraction of spinning BHs and anti-aligned systems compared with channel \RomanNumeralCaps 1, leading to a flatter distribution of effective spins in the range of $\sim -0.2$ to $\sim 0.2$. 
%Nevertheless, the value of $\chi_{\rm eff}$ is limited by the BH spin magnitude. 
The number of the binaries exhibiting non-negligible positive $\chi_{\rm eff}$, e.g. $\chi_{\rm eff}\gtrsim 0.05$ is comparable between the two channels. In contrast, it is only  channel \RomanNumeralCaps 2 that can produce NSBH mergers with $\chi_{\rm eff}\lesssim -0.02$. Binaries with $\chi_{\rm eff}$ below $-0.1$ and above $0.1$ account for only $\sim 2\%$ for channel \RomanNumeralCaps 1 and $\sim 14\%$ for channel \RomanNumeralCaps 2. 
%Regarding the simplified model we adopt for the tilt evolution due to accretion, the small BH spins in channel \RomanNumeralCaps 2 diminish its effect on the effective spins. In channel \RomanNumeralCaps 2, the binaries get their tilt angles mainly due to the second SN, resulting in an insignificant impact on both tilt and effective spin.

\subsubsection{Associated electromagnetic counterparts}

Observations of EMCs associated with NSBH mergers can provide unique insights into the GW physics as well as put constraints on the NS equation of state. To date, no EMC has been associated with identified NSBH merger candidates. To predict the fractions of NSBH mergers associated with an EMC in our simulations, we follow the method presented in \citet{2018PhRvD..98h1501F} which involves calculating the remnant mass outside the ISCO. We consider three scenarios for the NS radii of $11\,\rm{km}$, $12\,\rm{km}$ and $13\,\rm{km}$. In the calculations, we also take into account the misaligned angles between the BH spins and the orbits, using the component of the BH spin that is aligned with the orbital angular momentum.

The first line of table \ref{tab:emc} shows the fractions of NSBH mergers that are associated with an EMC under the different assumptions of NS radii. When assuming a NS radius of $R_{\rm NS} = 11\,\rm{km}$, the fraction of NSBH systems associated with an EMC for channel \RomanNumeralCaps 1 is $1.69\%$ and only $0.62\%$ for channel \RomanNumeralCaps 2. In total, the fraction is $1.35\%$. These NSBHs feature light NSs $<1.29\,M_{\sun}$ and light BHs $<3.54\,M_{\sun}$. Even when assuming a NS radius of $R_{\rm NS} = 12\,\rm{km}$, the BH masses in these NSBHs remain below $4.77\,M_{\sun}$. This suggests that the NSBH mergers associated with EMCs involve a BH that falls within the mass gap of $2.5-5\,M_{\sun}$. The largest NS radius assumption of $R_{\rm NS} = 13\,\rm{km}$ predicts a larger fraction of $15.80\%$, $9.83\%$, and $13.92\%$ for channel \RomanNumeralCaps 1, channel \RomanNumeralCaps 2, and the total, respectively. The maximal BH masses of the binaries with EMCs reach up to $5.87\,M_{\sun}$ and the NS masses extend to $1.59\,M_{\sun}$. In this scenario, approximately $88\%$, $75\%$, and $85\%$ of BHs in NSBH mergers associated with EMCs are within the mass gap for channel \RomanNumeralCaps 1, channel \RomanNumeralCaps 2, and the total, respectively.
%in channel \RomanNumeralCaps 2, the BH mass range extends to $\sim 9\,M_{\sun}$ because of their higher spins of $\sim 0.4$ that make the ISCO move inwards, increasing the possibility of disrupting the NSs outside the ISCO.

%Table \ref{tab:1} shows the fractions of EMC in all NSBH systems under different assumptions of the uncertain NS radius. When assuming a small NS radius, $R_{\rm NS} = 11\,\rm{km} $, the overall fraction of NSBH mergers that disrupt the NS outside the ISCO is less than $1\%$. This is because the NS is compact and, hence, it is hard to be disrupted outside the ISCO. The largest NS radius assumption predicts an overall fraction of $\sim 7.2\%$. These NSBHs with potential EMC have NSs with masses $< 1.36\,M_{\sun}$ and BHs with masses $<5.8\,M_{\sun}$. Channel \RomanNumeralCaps 2 has a larger fraction of EMC as there are more BHs with non-zero BH spins than Channel \RomanNumeralCaps 1. Higher spins can make the ISCO move inwards and increase the possibility of generating EMC.

Previous studies \citep{2021ApJ...912L..23R,2022ApJ...928..163H} showed that if the NS forms first in a NSBH system, 
%and the progenitor of the BH evolves through a CE phase ending up in a tight orbit
the BH progenitor can be spun up efficiently through tidal interactions with the NS, and results in a BH with high birth spin. The latter increases the possibility of having EMC. However, in our simulation, NSBHs with NS forming first account for a negligible fraction ($<0.01\%$). This is different from the results in \citet{2021ApJ...912L..23R} where the fraction of NSBHs with NS formed first can be higher than $10\%$. We attribute the difference to the modeling of accretion onto non-degenerate stars. While, in \citet{2021ApJ...912L..23R}, the rapid BPS code \texttt{COSMIC} \citep{2020ApJ...898...71B} was used to simulate the binaries from two ZAMS stars until the formation of compact object--stripped He star binaries, in this study, we use the detailed simulations throughout the evolution of the binaries. The mass accretion model adopted for non-degenerate accretors in \citet{2020ApJ...898...71B}, which only limits the accretion when it exceeds $10$ times the thermal-timescale mass-transfer rate of the accretor, is significantly more efficient compared with the rotation-dependent model \citep{2013ApJ...764..166D} adopted in \texttt{POSYDON}'s detailed binary evolution calculations. In this model, the accreted material carries with it the specific angular momentum of a Keplerian accretion disk whose inner edge is at the surface of the accreting star. As a star is accreting mass, it is being spun up due to the high specific angular momentum of the accreted material and it typically reaches critical rotation after accreting only a fraction of a solar mass. After the accreting star reaches critical rotation at its surface, we assume that further accretion of material ceases. Overall, this process results in very inefficient mass transfer ($\lesssim 10\%$), except in case-A mass transfer where tidal spin down from the companion star counteracts the spin up due to accretion and accretion efficiencies of $\sim 30\%$ can be reached \citep[see also][]{2020A&A...638A..39L}.

The amount of mass that can be accreted onto the secondary, during the initial mass-transfer phase between two non-degenerate stars, plays a crucial role in determining whether the formation of the NS can precede the formation of the BH \citep{2004MNRAS.354L..49S, 2021MNRAS.508.5028B}. The accretion model adopted in \texttt{POSYDON}'s detailed binary evolution calculations results in low accretion efficiency and thus does not lead to enough mass accreted by the secondary stars to form BHs, while the primary stripped stars forms a NS first. \citet{2021ApJ...920...81S} also predicted a very low fraction of NSBHs where the NS formed first, when using a similar rotation-dependent model. An additional factor that may lead to an increased fraction of first-born NSs in NSBH populations modeled by rapid codes is that case-A mass transfer, which is most often involved in the formation of NSBH binaries with first-born NSs, is not accurately modeled in rapid BPS codes \citep[especially those using Hurley's fitting formulae;][]{2000MNRAS.315..543H}. The primary reasons for this is that the formulae do not track the mass of the developing He core of the main sequence star and the radius evolution of the star used to infer the end of the mass transfer does not take into account altered structure of the stripped donor star. Both these factors can lead to more extended case-A mass-transfer cases where a larger fraction of the donor star is transferred onto the accretor \citep[see][for an extended discussion]{2022arXiv220708837D}.

In our detailed binary evolution calculations, the accretion efficiency in case-A mass-transfer phases between two non-degenerate stars can reach up to $\sim 30\%$, which is in principle sufficient to result in a first-born NS and a second-born BH. However, the orbits of those binaries after the formation of the NS are still compact, and the subsequent mass transfer from the secondary star onto the NS, which is unstable due to the high mass ratio, leads to a merger during the CE evolution.

%Besides, our treatment of CEE using the dynamic stellar information predicts a high merger rate with two non-degenerate stars at solar metallicity. The initially close binaries that may experience a highly conservative case-A mass-transfer phase, being able to reverse in masses, are likely to merge before the formation of NSs.

\section{Model Variations} \label{sec:var}
Binary evolution modelling is subject to various uncertainties that can affect the BPS results significantly. Thus, it is always instructive to investigate the effects of model variations. Because \texttt{POSYDON} allows on-the-fly calculations at the stage of core collapse and CE evolution, we are able to apply different models for core collapse and CE evolution, showing their impact on the population properties of merging NSBHs. In contrast, changing physical assumptions such as the accretion efficiency, as discussed in the previous section (see also Section~\ref{sec:discussion:uncertainties}), would require to rerun the entire library of pre-calculated binary evolution models, and thus it is outside the scope of this work.

\subsection{Variations in common envelope prescriptions}

In CE evolution, we consider the variations of the core-envelope boundary and $\alpha_{\rm CE}$ parameters. We explore two more values of the H abundance, which are $0.01$ and $0.3$ (the default value is $0.1$), to define the core-envelope boundary. Different core-envelope boundaries will directly affect the binding energy of the envelope. With lower H abundance for core-envelope definition, the boundary will be deeper inwards the star and the binding energy of the envelope that needed to be overcome will increase, leading to a smaller $\lambda_{\rm CE}$. In addition, as the default $\alpha_{\rm CE}=1.0$, we apply a smaller $\alpha_{\rm CE}$ parameter of $0.5$ as a model variation for comparison. 

\begin{figure*}\center
\includegraphics[width=\textwidth]{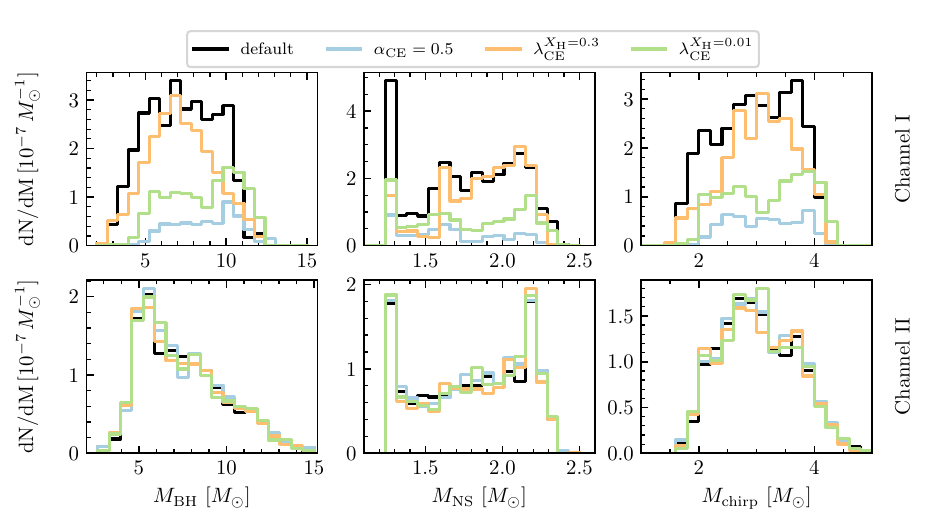}\\
\caption{Distributions of BH masses, NS masses, and chirp masses of merging NSBHs, scaled by the stellar mass of the parent population, for different parameters associated with CE evolution. The top and bottom correspond to channel \RomanNumeralCaps 1 and \RomanNumeralCaps 2, respectively. The default model corresponds to $\alpha_{\rm CE} = 1.0$ and adopting $\lambda_{\rm CE}^{X_{\rm H}=0.1}$.
}
\label{fig:CE_mass}
\end{figure*}

In Figure \ref{fig:CE_mass}, we show the distributions of NS masses, BH masses, and chirp masses for merging NSBHs under different parameters associated with CE evolution. The variations in CE evolution do not affect the stable mass transfer process and thus overall distributions have no differences for channel \RomanNumeralCaps 2. We can see that the $\lambda_{\rm CE}^{X_{\rm H}=0.3}$ model produces BHs peaking at around $7\,M_{\sun}$ while the other two variations have a larger fraction of BHs around $10\,M_{\sun}$. The $\lambda_{\rm CE}^{X_{\rm H}=0.01}$ model leads to a smaller core of the donor star and larger binding energy of the envelope, which makes the envelope harder to be expelled. A smaller value of $\alpha_{\rm CE}$ increases the energy needed to expel the donor's envelope. Consequently, a reduced $\alpha_{\rm CE}$ effectively decreases the overall number of binaries from channel \RomanNumeralCaps 1. For these two models, a heavy BH is favored because it reserves more orbital energy in the same orbit. The $\lambda_{\rm CE}^{X_{\rm H}=0.3}$ model results in wider post-CE orbits, which increases the probability of having merger delay times longer than the Hubble time. In addition, the $\lambda_{\rm CE}^{X_{\rm H}=0.3}$ model produces high-mass progenitor stars for the NSs, which are reflected by the relatively small number of light NSs. %In \texttt{POSYDON} simulations, after the CE phase, the BH -- He-star binary surviving hte CE is matched within the CO-HeMS grids. When using the $0.01\,X_{\rm H}$ core-envelope boundary definition, the He cores that survived CE evolution are more compact than the He stars evolved from He main-sequence stars at the same mass in the CO-HeMS grids. It is possible that the post-CE orbital separations are too small for the matched He stars and the binaries are classified to have initial RLOF. We consider them as mergers. 
Perhaps most notably, for the two model variations with less efficient CE evolution, models $\alpha_{\rm CE}=0.5$ and $\lambda_{\rm CE}^{X_{\rm H}=0.01}$, it is channel \RomanNumeralCaps 2  that becomes the dominant one, with the fraction of merging NSBH binaries coming from channel \RomanNumeralCaps 1 becoming $26.2\%$ and $47.6\%$, respectively. 
%In the case of $0.01\,X_{\rm H}$ model and $\alpha_{\rm CE}=0.5$, the number of NSBHs from channel \RomanNumeralCaps 2 exceeds it from channel \RomanNumeralCaps 1.

%We need to mention that with $0.01\,X_{\rm H}$ core-envelope boundary definition, about one third of binaries surviving CE are matched to initial Roche-lobe overflow at the next step in the CO-HeMS grids.

\begin{figure*}\center
\includegraphics[width=\textwidth]{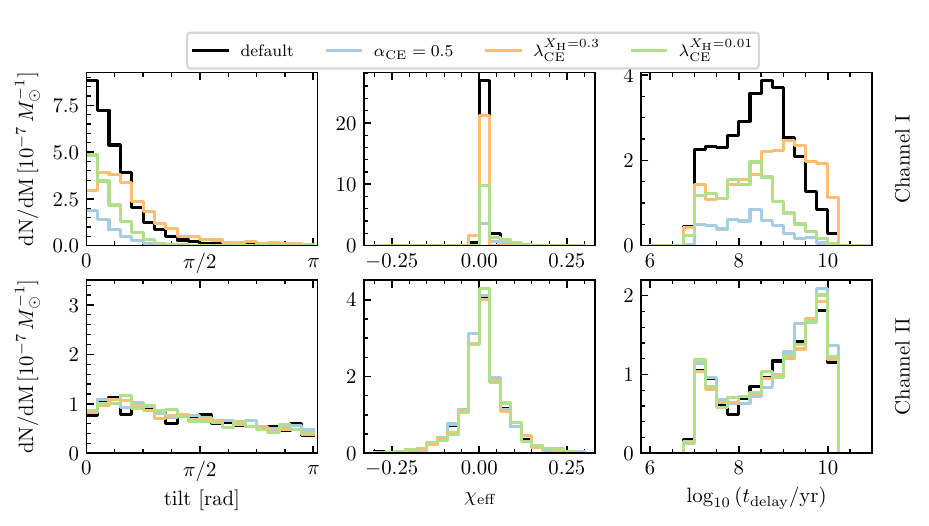}\\
\caption{The orbital tilt, effective spin and delay time distributions for variations in CE evolution in channel \RomanNumeralCaps 1 (top) and channel \RomanNumeralCaps 2 (bottom). 
}
\label{fig:CE_orbit}
\end{figure*}

%\begin{deluxetable*}{cccccccc}
%\tablecaption{Fractions of electromagnetic counterparts for different assumed NS radii\label{tab:1}}

%\tablenum{2}

%\tablehead{\colhead{Formoation Channel}&\colhead{$R_{\rm} = 13\,\rm{km}$}& \colhead{$R_{\rm} = 12\,\rm{km}$} & \colhead{$R_{\rm} = 11\,\rm{km}$} \\} 

%% All data must appear between the \startdata and \enddata commands
%\startdata
%Channel \RomanNumeralCaps 1 & $16.43\%$ & $10.24\%$ & $2.72\%$  \\
%Channel \RomanNumeralCaps 2 & $10.98\%$ & $4.32\%$ & $0.43\%$   \\
%Channel \RomanNumeralCaps 3 & 7.99  & 6.87  & 1.26   & 7.83  & 11.00 & 11.17 & 8.61 \\
%Total & $14.71\%$ & $8.37\%$ & $2.04\%$ \\
%\enddata
%\end{deluxetable*}

\begin{table*}[]
\centering
\fontsize{8pt}{8pt}\selectfont
\renewcommand{\arraystretch}{1.5}
\caption{EMC fractions for different model assumptions of NS radii and model variations\label{tab:emc}.}
\begin{tabular}{l|ccc|ccc|ccc}
\hline\hline
\multirow{2}{4em}{Models}  & \multicolumn{3}{|c|}{Channel \RomanNumeralCaps 1}  &  \multicolumn{3}{|c|}{Channel \RomanNumeralCaps 2}  &  \multicolumn{3}{|c}{Total} \\

  & $R_{\rm NS}$=13\,km&$R_{\rm NS}$=12\,km&$R_{\rm NS}$=11\,km& $R_{\rm NS}$=13\,km&$R_{\rm NS}$=12\,km&$R_{\rm NS}$=11\,km& $R_{\rm NS}$=13\,km&$R_{\rm NS}$=12\,km&$R_{\rm NS}$=11\,km\\
\hline
default & $15.80\%$ & $8.71\%$ & $1.69\%$ & $9.83\%$ & $3.31\%$& $0.62\%$  & $13.92\%$ & $7.01\%$& $1.35\%$\\
$\alpha_{\rm CE}=0.5$ & $3.71\%$ & $0.60\%$ & $0$ & $10.84\%$ & $3.95\%$& $0.55\%$  & $8.97\%$ & $3.07\%$& $0.41\%$ \\
$\lambda_{\rm CE}^{X_{\rm H}=0.3}$ & $8.43\%$ & $5.17\%$ & $1.94\%$ & $11.15\%$ & $3.72\%$& $0.22\%$  & $9.46\%$ & $4.62\%$& $1.29\%$ \\
$\lambda_{\rm CE}^{X_{\rm H}=0.01}$ & $9.18\%$ & $1.24\%$ & $0.19\%$ & $10.66\%$ & $3.31\%$& $0.39\%$  & $9.96\%$ & $2.32\%$& $0.30\%$ \\
$\sigma_{\rm CCSN} = 150\,\rm{km\,s^{-1}}$ & $19.52\%$ & $10.92\%$ & $3.08\%$ & $15.36\%$ & $7.64\%$& $1.21\%$  & $18.64\%$ & $10.23\%$& $2.68\%$ \\
$\sigma_{\rm CCSN} = 450\,\rm{km\,s^{-1}}$ & $14.50\%$ & $8.37\%$ & $1.84\%$ & $5.98\%$ & $1.75\%$& $0$  & $11.79\%$ & $6.26\%$& $1.25\%$ \\
Fryer-rapid & $0.015\%$ & $0$ & $0$ & $5.08\%$ & $0$& $0$  & $1.47\%$ & $0$& $0$\\
Patton\&Sukhbold & $0.025\%$ & $0.013\%$ & $0$ & $0.29\%$ & $0.084\%$& $0.042\%$  & $0.087\%$ & $0.029\%$& $0.0097\%$\\
\hline
\end{tabular}

\end{table*}

In Figure \ref{fig:CE_orbit}, we show the distributions of the final orbit tilt angle, effective spin, and delay time for different parameters associated with CE evolution. The variations in CE evolution have limited impact on the distribution of tilt angles and effective spins, other than the overall change in the normalization of the distribution. The merger delay time distribution, on the other hand, is sensitive to the $\alpha_{\rm CE}$ parameter and core-envelope definition. A smaller $\alpha_{\rm CE}$ or a smaller $\lambda_{\rm CE}$ means more orbital energy is needed to eject the envelope, suggesting a closer post-CE orbit, thus a shorter delay time. Under these two models, most binaries have delay time below $\sim 10^{8.5}\,\rm{yr}$. On the contrary, we can see that the $\lambda_{\rm CE}^{X_{\rm H}=0.3}$ model predicts more merging NSBHs with long delay time $\gtrsim 10^{9.5}\,\rm{yr}$ because the core-envelope boundary moves outward, the envelope's binding energy decreases and CE evolution leads to less dramatic orbital shrinkage. 
%We can see that the majority has delay time between $10^{7}-10^{8.5}\,\rm{yr}$ for $\alpha_{\rm CE} = 0.5$ and $10^{8.5}-10^{10}\,\rm{yr}$ for $\alpha = 2.0$. The delay time peak at about $10^{7}\,\rm{yr}$ and $10^{10}\,\rm{yr}$. The first peak corresponds to the binaries with eccentricities near one and the overall large final orbital separations lead to long delay time. 

We notice that with the $\lambda_{\rm CE}^{X_{\rm H}=0.3}$ model, merging NSBHs can also form through double CE evolution. These binaries have initial mass ratios close to unity and orbital periods of several tens of days. They will enter a contact phase after both stars have evolved off the main sequence, which we assume leads to a double CE phase. When the CE is ejected successfully, two stripped He cores will find themselves to be left in an orbit of a few days. Afterward, the two stars will remain detached till the formation of NSBHs. However, the number of the merging NSBHs originating from this channel is extremely limited, which only account for $0.1\%$ of the total. We thus do not discuss this channel any further.

\subsection{Variations in core-collapse prescriptions}

The properties of the compact object formed from a star naturally depend on adopted prescription for core-collapse process. These prescriptions map the properties of the structure of the star right before the core collapse to the properties of the formed compact object, with different prescriptions predicting not only different compact object properties, but even different compact object type, for the same progenitor star structure. In this study, beside the default \citet{2012ApJ...749...91F} delayed prescription, we consider two varying core-collapse prescriptions: the \citet{2012ApJ...749...91F} rapid prescription \citep{2012ApJ...749...91F} and the \citet{2020MNRAS.499.2803P} prescription. 

\begin{figure*}\center
\includegraphics[width=\textwidth]{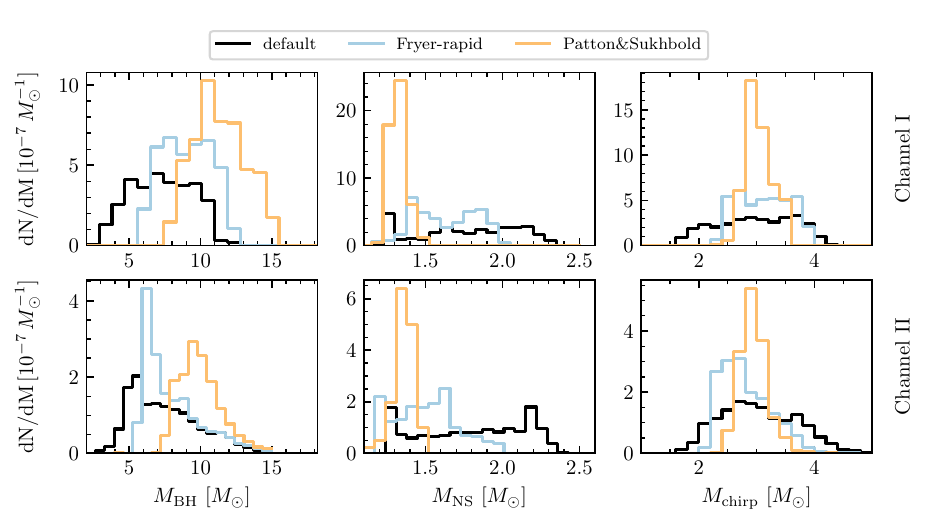}\\
\caption{The same as Figure \ref{fig:CE_mass} but with different core-collapse mechanisms. The default model corresponds to Fryer-delayed prescription. 
}
\label{fig:SN_mass}
\end{figure*}

Similar to Figure \ref{fig:CE_mass}, we show the mass distributions for the two variation models compared with the default model in Figure \ref{fig:SN_mass}. The \citet{2012ApJ...749...91F} rapid prescription produces a mass gap of $2.5-5\,M_{\sun}$ between NSs and BHs. It generates BHs within a mass range of $\sim 6-12\,M_{\sun}$ in channel \RomanNumeralCaps 1 and an abundance of BHs with masses of $\sim 6-7\,M_{\sun}$ in channel \RomanNumeralCaps 2. The \citet{2020MNRAS.499.2803P} prescription predicts a BH mass range of $\sim 8-15\,M_{\sun}$ in both channels. Few NSs from the \citet{2012ApJ...749...91F} rapid model exceed $2\,M_{\sun}$. For the \citet{2020MNRAS.499.2803P} prescription, the NS masses accumulate within $1.1-1.5\,M_{\sun}$. As for the chirp mass, the \citet{2012ApJ...749...91F} delayed prescription predicts a distribution between $1.5-4\,M_{\sun}$. The \citet{2012ApJ...749...91F} rapid prescription provides a narrower range between $2.5-4\,M_{\sun}$ because of the deficiency of the BHs in the mass gap and heavy NSs above $2\,M_{\sun}$. The \citet{2020MNRAS.499.2803P} prescription predicts a chirp mass range of $2.5-3.5\,M_{\sun}$ and a peak at $\sim 3\,M_{\sun}$ . 

\begin{figure*}\center
\includegraphics[width=\textwidth]{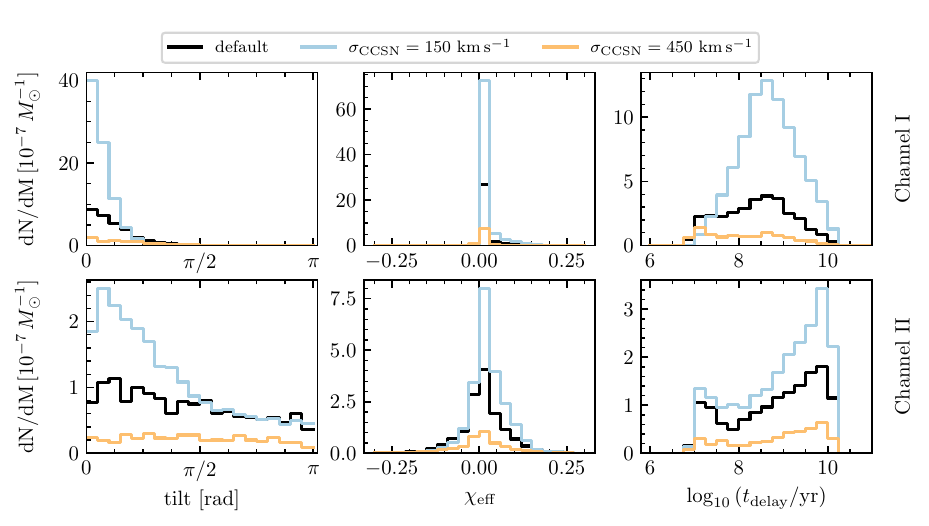}\\
\caption{The same as Figure \ref{fig:CE_orbit} but with different natal kick velocity distributions for CCSN. The default model corresponds to $\sigma_{\rm CCSN} = 265\,\rm{km\,s^{-1}}$.
}
\label{fig:SN_orbit}
\end{figure*}

Apart from the core-collapse prescription, a potential natal kick imparted onto the newly-formed compact object also plays an important role in the formation of double compact objects. We consider two different values for the dispersion of the assumed Maxwellian kick velocity distribution for CCSN, $\sigma_{\rm CCSN} = 150\,\rm{km\,s^{-1}}$ and $\sigma_{\rm CCSN} = 450\,\rm{km\,s^{-1}}$ (the default value is $\sigma_{\rm CCSN} = 265\,\rm{km\,s^{-1}}$).   
Figure \ref{fig:SN_orbit} shows the tilt, effective spin, and delay time for the three different dispersion values of the CCSN natal kick velocity distribution. Natal kicks significantly affect both the orbital properties of the NSBH systems and the overall merger rate. Smaller kicks decrease the possibility of binary disruptions or excessive delay times caused by SN, thus enhancing the formation rate of merging NSBHs. Furthermore, small kicks produce a large number of merging NSBHs with small tilt angles. In channel \RomanNumeralCaps 2, compared with the default model, the model of $\sigma_{\rm CCSN} = 150\,\rm{km\,s^{-1}}$ predicts more systems with aligned BHs but almost the same number of anti-aligned systems. Larger kicks generate a flatter distribution of tilt angles but the small total number limits the number of anti-aligned systems. The peak below $10^{8}\,\rm{yr}$ in the delay time distribution corresponds to highly eccentric binaries. In contrast, we can see that in channel \RomanNumeralCaps 1, there is no significant secondary peak below short delay times ($10^{8}\,\rm{yr}$). The binaries in these channel tend to have short orbits and high orbital velocities right before the formation of the second compact object. Hence, is it is very unlikely to form NSBH binaries with extreme eccentricities, which are required to achieve merger delay times of the order of $10^7\,\rm yr$. Kicks in excess of $\gtrsim 1000\,\rm km/s$ would be necessary for this scenario, which is why we only see a mild excess of very short merger delay times for the highest dispersion value of our assumed kick distribution.  

\subsection{Fraction of associated electromagnetic counterparts}
Table \ref{tab:emc} lists the predicted fractions of NSBH mergers that could potentially produce EMCs under different models and for three assumed NS radii. We estimate the fractions in the range of $0-18.64\%$. The $\alpha_{\rm CE}=0.5$ model and the $\lambda_{\rm CE}^{X_{\rm H}=0.01}$ model yield lower fractions compared with the default model because in channel \RomanNumeralCaps 1, these two models predict a limited population of low-mass BHs ($\lesssim 6\,M_{\sun}$). The $\lambda_{\rm CE}^{X_{\rm H}=0.01}$ model also gives a lower fraction, which can be attributed to a reduced proportion of low-mass NSs ($\lesssim 1.5\,M_{\sun}$) in the model. The model of $\sigma_{\rm CCSN} = 150\,\rm{km\,s^{-1}}$ predicts the highest fraction of NSBH mergers that are accompanied by an EMC. The variations of core-collapse prescriptions exert the most substantial impact, as they yield different mass distributions for the compact objects. The \citet{2012ApJ...749...91F} rapid model does not produce BHs within the mass gap of $2.5-5\,M_{\sun}$. In the absence of low-mass BHs, the EMC fraction experiences a significant decline. Only when assuming the largest NS radius does the fraction in channel \RomanNumeralCaps 2 become not marginal ($5.08\%$). The BH masses are below $7.07\,M_{\sun}$ and their spins are in the range of $0.02-0.13$. The inclination toward channel \RomanNumeralCaps 2 is attributed to the low-mass BHs within this channel possessing slightly higher spins. The \citet{2020MNRAS.499.2803P} prescription predicts BH masses above $\sim 8\,M_{\sun}$, leading to negligible EMC fractions.

\section{Discussion \label{sec:discussion}}
\subsection{Uncertainties in stellar and binary physics \label{sec:discussion:uncertainties}}
Apart from the model variations in how we treat the core collapse and the CE phases we discussed in the last section, other not fully-understood stellar and binary evolution physics can also affect the BPS results in a significant way. The accretion efficiency onto a non-degenerate star is such an example for the formation of merging NSBH binaries. In our pre-computed grids of detailed binary evolution models, we have adopted a rotation-dependent accretion model \citep{2013ApJ...764..166D}, which assumes the accretion would be suppressed if the accretor reaches critical rotation. While this is a physically self-consistent way to treat the accretion of matter and angular momentum onto a star, some authors \citep{1991ApJ...370..597P,1991ApJ...370..604P} suggest that a critically rotating star can still accrete mass because the viscous coupling between the star and the accretion disk can regulate the angular momentum transport without affecting the mass accretion. In this case, the mass transfer efficiency can be much higher than our model. However, it is not clear whether the accretion would still be limited by other factors, such as the inflation of the accreting star due to high mass-transfer rate. Detailed binary evolution models that connect the accretion efficiency onto a star with such accretion disk  models are not yet available. 

The treatment of accretion onto a BH may also affect the properties of the NSBH population, especially the spin of the BH, which in turn determines whether the merger of the NSBH binary will be accompanied by an EMC. Radiation, general-relativistic, magneto-hydrodynamics simulation of supercritical accretion disk around BHs \citep[e.g.,][]{2016MNRAS.456.3929S} suggest that the accretion rate may significantly exceed the Eddington limit. This possibility has been explored in the context of BH high-mass X-ray binaries \citep[][]{2008ApJ...689L...9M,2022RAA....22c5023Q} and coalescing binary BH populations \citep[e.g.,][]{2021A&A...647A.153B,2023MNRAS.520.5724B}. However, the effect of super-Eddington accretion on the formation and the properties of merging NSBH, and especially the probability of being accompanied by EMCs, is less well studied. We leave the exploration of different assumption regarding the accretion physics onto both non-degenerate stars and compact objects for a future study.

%The stellar and binary grids in \texttt{POSYDON} are built with stellar evolution code \texttt{MESA}, which enables us to explore the impact of stellar physics on BPS studies by adopting different wind prescriptions and overshooting parameters in a self-consistent way. But this kind of exploration is limited by the capacity of the current version of \texttt{POSYDON}. Thorough investigations on the stellar physics can be done in a future work.

\subsection{An additional formation channel}

%Under our default settings, we only identify two channels for the merging NSBH formation. 

We found that in some binaries with initial binary mass ratio close to unity, after a stable mass transfer from the primary to the secondary, the secondary evolves off the main sequence before the primary explodes and starts a reverse mass-transfer phase\footnote{In the revision of \texttt{MESA} used for the calculation of {\tt POSYDON}'s grids of binary evolution models (\texttt{MESA} revision 11701), as well as the current revision at the time of writing this paper (r23.05.1), the modeling of reverse mass transfer in the \texttt{Kolb} scheme is not supported. To implement the proper modeling of reverse mass transfer within the \texttt{Kolb} scheme, we had to make small modifications in the \texttt{MESA} code base. These modifications are documented here: \href{https://github.com/MESAHub/mesa/issues/545}{https://github.com/MESAHub/mesa/issues/545}, and will be included in future revisions of {\tt MESA}}. At the beginning of the reverse mass-transfer phase, the accretor has been almost or totally stripped. It accretes hydrogen from the secondary, quickly creating a thin layer of hydrogen at the surface. The accretion tends to make the star spin up and expand. However, the accretor will reach critical rotation and cease further accretion before it has a chance to expand significantly. The remaining of the mass-transfer phase will proceed effectively as fully non-conservative mass-transfer.

%The accretion would make the star spin up and expand, while nearly at the mean time, to avoid exceeding critical rotation, the star would lose some mass to contract again. And the oscillation of star radius may lead to numerical issues. 

%To deal with it, our first attempt is to simplify the mass transfer process. Because the accretor would reach critical rotation quickly and accrete little afterward, we assume no accretion and no mass loss for the accretor, making it in an equilibrium. If the timescale of mass transfer is short enough, the accreted mass would be negligible and the stage of the accretor would not change dramatically. Under these assumptions, most of the reverse mass transfer would be stable and this additional formation channel produces $<1\%$ NSBHs.

The behavior we described above depends sensitively on our assumption of mass-transfer efficiency on a non-degenerate star (see discussion in Section~\ref{sec:discussion:uncertainties}). If we consider instead the possibility that the accretion continues onto critically rotating stars, we would expect that the stripped He stars, after accreting a small amount of hydrogen, will swell up substantially and fill the Roche lobe. This would result into a CE evolution quickly after the onset of reverse mass transfer. Under the assumption that the reverse mass transfer would become unstable and initiate CE evolution immediately, we evolve these binaries further with \texttt{POSYDON}. It turned out that this evolutionary path can be important for the formation of merging NSBHs. The interesting part is that after the CE phase, two He stars will be left in a close orbit and the two stars can be effectively spun up by tidal interactions. The BHs formed through this channel can have high spins at birth. Therefore, a thorough study on the accretion physics is needed to have the whole picture of binary evolution and the formation of double compact objects. 

%\textbf{The latest Initial-final population runs show that the number of NSBH from this channel reduces a lot compared with nearest neighbor population (from ~80 to ~9). One exception is IF runs with PattonSukhbold, the number of this channel is not changed that much. }

\subsection{Local merger rate density}
In this study, we focus on models at solar metallicity. Thus, we are not calculating the merger rate density that can be directly compared with the estimate from GW detection. However, multiple works have shown that in contrast to BBH, the NSBH formation rate has limited connection with metallicity \citep{2019MNRAS.490.3740N, 2021ApJ...912L..23R, 2022MNRAS.tmp.1776B}. In order to get a general idea on the local merger rate density from our simulation, we make the estimation by only integrating the star formation rate (SFR) around solar metallicity. To do that, we follow the procedure described in \citet{2023NatAs.tmp..142B} integrating SFR in a metallicity range of $[0.5Z_{\sun},2Z_{\sun}]$. In this way, we underestimate the contribution of NSBH mergers at high redshifts where star-formation at low metallicity is important. Under these assumption, our default model predicts a local merger rate density $60.11\, \rm{Gpc}^{-3}\,\rm{yr}^{-1}$, while the model variations we considered generate local merger rate densities in the range of $17.00\,\rm{Gpc}^{-3}\,\rm{yr}^{-1}$ to $155.64\,\rm{Gpc}^{-3}\,\rm{yr}^{-1}$.

\subsection{Observation of NSBH populations with electromagnetic waves}

Before merging, the NS may appear as a radio pulsar, making the system a pulsar-BH binary. Detection of pulsar-BH binaries can give us valuable insights on the formation and evolution of NSBH binaries, extending our objects to different population of NSBH binaries. BPS studies have estimated several to thousands of pulsar-BH binaries in the Milky Way \citep{2021MNRAS.504.3682C,2021ApJ...920...81S,2021arXiv210904512P}. They could be observed by the radio telescopes like the Square Kilometre Array \citep[SKA,][]{2004NewAR..48..993K}, MeerKAT \citep{2009arXiv0910.2935B}, the Five-Hundred Metre Aperture Spherical Radio Telescope \citep[FAST,][]{2011IJMPD..20..989N} and Arecibo radio telescope \citep{2006ApJ...637..446C}. Pulsar-BH binaries with orbital periods less than a day may be observed by both radio telescopes and GW detector LISA \citep{2021MNRAS.504.3682C}. These joint observations could be further used to constrain model uncertainties in the formation of NSBH systems. 

\section{Summary} \label{sec:sum}

In this study, we investigated the formation channels and population properties of merging NSBH systems formed from isolated binary evolution at solar metallicity using the newly developed BPS code \texttt{POSYDON}. \texttt{POSYDON} incorporates detailed stellar and binary models calculated based on the binary-star evolution code \texttt{MESA}. The main findings of our work are:
\begin{itemize}
    
\item We identify two main formation channels of NSBH mergers. In channel \RomanNumeralCaps 1 ($\sim70\%$), the binaries undergo a stable mass transfer prior to the formation of the BH, a CE phase after the formation of the BH, and a subsequent case-BB stable mass-transfer phase (channel \RomanNumeralCaps 1a, $\sim94\%$) or no case-BB mass transfer (channel \RomanNumeralCaps 1b, $\sim6\%$). In channel \RomanNumeralCaps 2 ($\sim30\%$), the binaries go through only stable mass transfer. Most binaries within channel \RomanNumeralCaps 2 experience two stable mass-transfer phases, one prior to and another after the BH formation (channel \RomanNumeralCaps 2a, $\sim 95.4\%$). In addition, a small portion of the binaries, which contain massive primary stars above $\sim 60\,M_{\sun}$, get stripped because of strong stellar winds and avoid mass transfer before the BH formation (channel \RomanNumeralCaps 2b, $\sim 4.6\%$).

\item  Independently from the formation channels and the explored model uncertainties, our models suggest that the BH always form first in merging NSBH systems. The primary reason for this difference compared with other NSBH BPS studies in the literature is our adopted model of rotation-dependent accretion efficiency onto non-degenerate stars, which generally results in highly non-conservative mass transfer.

\item Most binaries in channel \RomanNumeralCaps 1 have initial orbital periods $\gtrsim 50\,\rm d$ and mass ratios $\gtrsim 0.5$. 
%Due to the CE phase which efficiently shrinks the binaries, this channel could produce close NSBH systems. 
Given the initially large orbital separation of these binaries, the majority of the BHs form with near-zero spins, assuming efficient angular momentum transport within the stars. After their formation, BHs can still be spun up due to mass accretion during the case-BB mass-transfer phase. However, since we assume that the accretion onto a compact object is Eddington limited, this effect is not significant. 

\item Compared with channel \RomanNumeralCaps 1, most of the binaries formed through channel \RomanNumeralCaps 2 have initial orbital periods $\lesssim 50\,\rm d$ and a wider spread of initial mass ratios. Those binaries end up in relatively wide orbits before the second core collapse, as the second stable mass-transfer phase does not shrink the orbit as efficiently as the CE. Natal kicks during the second SN can alter significantly those orbits and in some cases result in highly eccentric NSBH binaries. It is the high eccentricity ($>0.5$) of the orbits after the NS formation that causes the binary to merge within a Hubble time. Furthermore, binaries in channel \RomanNumeralCaps 2, due to tides in their close initial orbits and the further possible shrinking of the orbit during the first mass-transfer episode, have primary stars that can retain some angular momentum till the end of their lives.
%Nevertheless, the primary stars would lose most of their outer layer by mass transfer and stellar winds, along with most of the stellar angular momentum. 
The angular momentum preserved in the core contributes to non-zero, small spins ($\lesssim 0.2$) for the BHs. The BHs can be further spun up moderately because of the mass-transfer phase following the BH formation. 
 
\item Binaries from channel \RomanNumeralCaps 1 mostly have small orbital tilt angles while for channel \RomanNumeralCaps 2, binaries with large tilt angles $(>\pi/2\,\rm{rad})$ account for a significant fraction. The reason for the latter is that the NSBH binaries that experienced a sufficiently high kick to become highly eccentric manage to merge within a Hubble time in channel \RomanNumeralCaps 2 and the kick can easily tilt the orbit in such wide systems. This feature also contributes to the distinctive distributions of effective spin for the two channels. For channel \RomanNumeralCaps 1, the negligible BH spins and the small tilt angles make the distribution of effective spin peak at zero, with an excess at positive values. In contrast, for channel \RomanNumeralCaps 2, the effective spins extend to $\sim -0.2$ and $\sim 0.2$. 

\item With the information of compact object masses, BH spins, and orbital tilt angles, we estimate the fraction of merging NSBHs that we expect to be accompanied by an EMC. We find that assuming a stiff equation of state for the NSs ($R_{\rm NS} = 13\,\rm{km}$), the EMC fraction is about $13.92\%$, while for a soft equation of state ($R_{\rm NS} = 11\,\rm{km} $) it is about $1.35\%$. The binaries that have potential EMC favor light BHs and NSs. Approximately $85\%$ of BHs are within the mass gap of $2.5-5\, M_{\sun}$ when assuming $R_{\rm NS} = 13\,\rm{km}$, and in the other two cases, all BHs are expected to fall within the mass gap. Incorporating uncertainties in both the population modeling and the neutron star equation of state, we estimate that between $0$ and $18.6\%$ of NSBH mergers could potentially exhibit an EMC.

\item In addition to our default BPS model, we investigated a range of model uncertainties, including different prescriptions and model parameters for the CE and core collapse phase. 
%to explore the impact of the variations on the population of merging NSBH. 
Different parameters for CE evolution affect channel \RomanNumeralCaps 1 in terms of the absolute number of merging NSBHs and the compact object masses. The $X_{\rm H}=0.01$ core-envelope boundary definition results in a smaller core and thus higher binding energy of the envelope compared with the default $0.1\,X_{\rm H}$, while a smaller $\alpha_{\rm CE}$ means more orbital energy is needed to overcome the binding energy. These two variations predict less merging NSBHs and preference for massive BHs $\sim 10\,M_{\sun}$. The $\lambda_{\rm CE}^{X_{\rm H}=0.3}$ model leads to a larger post-CE orbital separation. The formation rate of merging NSBHs is lower because of excessive delay time for binaries going through CE evolution.
%, and a larger fraction of the merging NSBHs have delay time $\gtrsim 10^{9.5}\,\rm{yr}$. 
Variations in the core-collapse prescriptions affect both channels concerning the formation efficiency of merging NSBHs, as well as their mass distributions and orbital properties. With our default \citet{2012ApJ...749...91F} delayed model, most merging NSBHs have BHs less massive than $\sim 11\,M_{\sun}$, chirp masses in a range of $1.5-4\,M_{\sun}$, and NS masses within $1.26-2.4\,M_{\sun}$. The \citet{2012ApJ...749...91F} rapid core collapse prescription produces the mass gap between NSs and BHs and NSs less massive than $\sim 2\,M_{\sun}$. The \citet{2020MNRAS.499.2803P} prescription produces a narrow NS mass range of $1.1-1.5\,M_{\sun}$. Assuming a Maxwellian distribution of natal kick velocities with a smaller dispersion ($150\,\rm{km\,s^{-1}}$ versus the $265\,\rm{km\,s^{-1}}$ of our default model) predicts more merging NSBHs with a larger proportion of them exhibiting small tilt angles. Besides, a smaller kick velocity dispersion produces more merging NSBHs with long delay times, as it is less efficient at forming very eccentric NSBH binaries. 

\end{itemize}

The utilization of detailed stellar and binary models in BPS studies ensures enhanced modelling accuracy and self-consist treatment of stellar evolution and binary interactions, providing us with more accurate and informative results. As we expect a growing number of GW events to be detected in the future, it becomes imperative to employ increasingly accurate BPS models\textemdash a crucial component to unravel the origin and formation mechanisms of observed double compact objects and refine our understanding of underlying stellar and binary physics.

\begin{acknowledgements}
We thank Christopher Berry for helpful comments. The POSYDON project is supported primarily by two sources: the Swiss National Science Foundation (PI Fragos, project numbers PP00P2\_211006 and CRSII5\_213497) and the Gordon and Betty Moore Foundation (PI Kalogera, grant award GBMF8477). ZX acknowledges support from the Chinese Scholarship Council (CSC). 
S.S.B., T.F., M.K., and Z.X. were supported by the project number PP00P2\_211006. 
S.S.B. was also supported by the project number CRSII5\_213497. 
A.D., K.A.R., P.M.S. and M.S. were supported by the project number GBMF8477. 
KK acknowledges support from the Spanish State Research Agency, through the María de Maeztu Program for Centers and Units of Excellence in R\&D, No. CEX2020-001058-M. 
E.Z. acknowledges funding support from the European Research Council (ERC) under the European Union’s Horizon 2020 research and innovation programme (Grant agreement No. 772086). 
The computations were performed at Northwestern University on the Trident computer cluster (funded by the GBMF8477 award) and at the University of Geneva on the Yggdrasil computer cluster. This research was partly supported by the computational resources and staff contributions provided5for the Quest high-performance computing facility at Northwestern University, jointly supported by the Office of the Provost, the Office for Research and Northwestern University Information Technology
\end{acknowledgements}

\bibliographystyle{aa}
\bibliography{references.bib}

\begin{thebibliography}{138}
\expandafter\ifx\csname natexlab\endcsname\relax\def\natexlab#1{#1}\fi

\bibitem[{{Abbott} {et~al.}(2018){Abbott}, {Abbott}, {Abbott}, {Abernathy}, {Acernese}, {Ackley}, {Adams}, {Adams}, {Addesso}, {Adhikari}, {Adya}, {Affeldt}, {Agathos}, {Agatsuma}, {Aggarwal}, {Aguiar}, {Aiello}, {Ain}, {Ajith}, {Akutsu}, {Allen}, {Allocca}, {Altin}, {Ananyeva}, {Anderson}, {Anderson}, {Ando}, {Appert}, {Arai}, {Araya}, {Araya}, {Areeda}, {Arnaud}, {Arun}, {Asada}, {Ascenzi}, {Ashton}, {Aso}, {Ast}, {Aston}, {Astone}, {Atsuta}, {Aufmuth}, {Aulbert}, {Avila-Alvarez}, {Awai}, {Babak}, {Bacon}, {Bader}, {Baiotti}, {Baker}, {Baldaccini}, {Ballardin}, {Ballmer}, {Barayoga}, {Barclay}, {Barish}, {Barker}, {Barone}, {Barr}, {Barsotti}, {Barsuglia}, {Barta}, {Bartlett}, {Barton}, {Bartos}, {Bassiri}, {Basti}, {Batch}, {Baune}, {Bavigadda}, {Bazzan}, {B{\'e}csy}, {Beer}, {Bejger}, {Belahcene}, {Belgin}, {Bell}, {Berger}, {Bergmann}, {Berry}, {Bersanetti}, {Bertolini}, {Betzwieser}, {Bhagwat}, {Bhandare}, {Bilenko}, {Billingsley}, {Billman}, {Birch}, {Birney}, {Birnholtz}, {Biscans}, {Bisht},
  {Bitossi}, {Biwer}, {Bizouard}, {Blackburn}, {Blackman}, {Blair}, {Blair}, {Blair}, {Bloemen}, {Bock}, {Boer}, {Bogaert}, {Bohe}, {Bondu}, {Bonnand}, {Boom}, {Bork}, {Boschi}, {Bose}, {Bouffanais}, {Bozzi}, {Bradaschia}, {Brady}, {Braginsky}, {Branchesi}, {Brau}, {Briant}, {Brillet}, {Brinkmann}, {Brisson}, {Brockill}, {Broida}, {Brooks}, {Brown}, {Brown}, {Brown}, {Brunett}, {Buchanan}, {Buikema}, {Bulik}, {Bulten}, {Buonanno}, {Buskulic}, {Buy}, {Byer}, {Cabero}, {Cadonati}, {Cagnoli}, {Cahillane}, {Calder{\'o}n Bustillo}, {Callister}, {Calloni}, {Camp}, {Cannon}, {Cao}, {Cao}, {Capano}, {Capocasa}, {Carbognani}, {Caride}, {Casanueva Diaz}, {Casentini}, {Caudill}, {Cavagli{\`a}}, {Cavalier}, {Cavalieri}, {Cella}, {Cepeda}, {Cerboni Baiardi}, {Cerretani}, {Cesarini}, {Chamberlin}, {Chan}, {Chao}, {Charlton}, {Chassande-Mottin}, {Cheeseboro}, {Chen}, {Chen}, {Cheng}, {Chincarini}, {Chiummo}, {Chmiel}, {Cho}, {Cho}, {Chow}, {Christensen}, {Chu}, {Chua}, {Chua}, {Chung}, {Ciani}, {Clara}, {Clark}, {Cleva},
  {Cocchieri}, {Coccia}, {Cohadon}, {Colla}, {Collette}, {Cominsky}, {Constancio}, {Conti}, {Cooper}, {Corbitt}, {Cornish}, {Corsi}, {Cortese}, {Costa}, {Coughlin}, {Coughlin}, {Coulon}, {Countryman}, {Couvares}, {Covas}, {Cowan}, {Coward}, {Cowart}, {Coyne}, {Coyne}, {Creighton}, {Creighton}, {Cripe}, {Crowder}, {Cullen}, {Cumming}, {Cunningham}, {Cuoco}, {Dal Canton}, {Danilishin}, {D'Antonio}, {Danzmann}, {Dasgupta}, {da Silva Costa}, {Dattilo}, {Dave}, {Davier}, {Davies}, {Davis}, {Daw}, {Day}, {Day}, {de}, {Debra}, {Debreczeni}, {Degallaix}, {de Laurentis}, {Del{\'e}glise}, {Del Pozzo}, {Denker}, {Dent}, {Dergachev}, {De Rosa}, {Derosa}, {Desalvo}, {Devine}, {Dhurandhar}, {D{\'\i}az}, {di Fiore}, {di Giovanni}, {di Girolamo}, {di Lieto}, {di Pace}, {di Palma}, {di Virgilio}, {Doctor}, {Doi}, {Dolique}, {Donovan}, {Dooley}, {Doravari}, {Dorrington}, {Douglas}, {Dovale {\'A}lvarez}, {Downes}, {Drago}, {Drever}, {Driggers}, {Du}, {Ducrot}, {Dwyer}, {Eda}, {Edo}, {Edwards}, {Effler}, {Eggenstein}, {Ehrens},
  {Eichholz}, {Eikenberry}, {Eisenstein}, {Essick}, {Etienne}, {Etzel}, {Evans}, {Evans}, {Everett}, {Factourovich}, {Fafone}, {Fair}, {Fairhurst}, {Fan}, {Farinon}, {Farr}, {Farr}, {Fauchon-Jones}, {Favata}, {Fays}, {Fehrmann}, {Fejer}, {Fern{\'a}ndez Galiana}, {Ferrante}, {Ferreira}, {Ferrini}, {Fidecaro}, {Fiori}, {Fiorucci}, {Fisher}, {Flaminio}, {Fletcher}, {Fong}, {Forsyth}, {Fournier}, {Frasca}, {Frasconi}, {Frei}, {Freise}, {Frey}, {Frey}, {Fries}, {Fritschel}, {Frolov}, {Fujii}, {Fujimoto}, {Fulda}, {Fyffe}, {Gabbard}, {Gadre}, {Gaebel}, {Gair}, {Gammaitoni}, {Gaonkar}, {Garufi}, {Gaur}, {Gayathri}, {Gehrels}, {Gemme}, {Genin}, {Gennai}, {George}, {Gergely}, {Germain}, {Ghonge}, {Ghosh}, {Ghosh}, {Ghosh}, {Giaime}, {Giardina}, {Giazotto}, {Gill}, {Glaefke}, {Goetz}, {Goetz}, {Gondan}, {Gonz{\'a}lez}, {Gonzalez Castro}, {Gopakumar}, {Gorodetsky}, {Gossan}, {Gosselin}, {Gouaty}, {Grado}, {Graef}, {Granata}, {Grant}, {Gras}, {Gray}, {Greco}, {Green}, {Groot}, {Grote}, {Grunewald}, {Guidi}, {Guo},
  {Gupta}, {Gupta}, {Gushwa}, {Gustafson}, {Gustafson}, {Hacker}, {Hagiwara}, {Hall}, {Hall}, {Hammond}, {Haney}, {Hanke}, {Hanks}, {Hanna}, {Hannam}, {Hanson}, {Hardwick}, {Harms}, {Harry}, {Harry}, {Hart}, {Hartman}, {Haster}, {Haughian}, {Hayama}, {Healy}, {Heidmann}, {Heintze}, {Heitmann}, {Hello}, {Hemming}, {Hendry}, {Heng}, {Hennig}, {Henry}, {Heptonstall}, {Heurs}, {Hild}, {Hirose}, {Hoak}, {Hofman}, {Holt}, {Holz}, {Hopkins}, {Hough}, {Houston}, {Howell}, {Hu}, {Huerta}, {Huet}, {Hughey}, {Husa}, {Huttner}, {Huynh-Dinh}, {Indik}, {Ingram}, {Inta}, {Ioka}, {Isa}, {Isac}, {Isi}, {Isogai}, {Itoh}, {Iyer}, {Izumi}, {Jacqmin}, {Jani}, {Jaranowski}, {Jawahar}, {Jim{\'e}nez-Forteza}, {Johnson}, {Jones}, {Jones}, {Jonker}, {Ju}, {Junker}, {Kagawa}, {Kajita}, {Kakizaki}, {Kalaghatgi}, {Kalogera}, {Kamiizumi}, {Kanda}, {Kandhasamy}, {Kanemura}, {Kaneyama}, {Kang}, {Kanner}, {Karki}, {Karvinen}, {Kasprzack}, {Kataoka}, {Katsavounidis}, {Katzman}, {Kaufer}, {Kaur}, {Kawabe}, {Kawai}, {Kawamura},
  {K{\'e}f{\'e}lian}, {Keitel}, {Kelley}, {Kennedy}, {Key}, {Khalili}, {Khan}, {Khan}, {Khan}, {Khazanov}, {Kijbunchoo}, {Kim}, {Kim}, {Kim}, {Kim}, {Kim}, {Kim}, {Kimbrell}, {Kimura}, {King}, {King}, {Kirchhoff}, {Kissel}, {Klein}, {Kleybolte}, {Klimenko}, {Koch}, {Koehlenbeck}, {Kojima}, {Kokeyama}, {Koley}, {Komori}, {Kondrashov}, {Kontos}, {Korobko}, {Korth}, {Kotake}, {Kowalska}, {Kozak}, {Kr{\"a}mer}, {Kringel}, {Krishnan}, {Kr{\'o}lak}, {Kuehn}, {Kumar}, {Kumar}, {Kumar}, {Kuo}, {Kuroda}, {Kutynia}, {Kuwahara}, {Lackey}, {Landry}, {Lang}, {Lange}, {Lantz}, {Lanza}, {Lartaux-Vollard}, {Lasky}, {Laxen}, {Lazzarini}, {Lazzaro}, {Leaci}, {Leavey}, {Lebigot}, {Lee}, {Lee}, {Lee}, {Lee}, {Lee}, {Lehmann}, {Lenon}, {Leonardi}, {Leong}, {Leroy}, {Letendre}, {Levin}, {Li}, {Libson}, {Littenberg}, {Liu}, {Lockerbie}, {Lombardi}, {London}, {Lord}, {Lorenzini}, {Loriette}, {Lormand}, {Losurdo}, {Lough}, {Lousto}, {Lovelace}, {L{\"u}ck}, {Lundgren}, {Lynch}, {Ma}, {Macfoy}, {Machenschalk}, {Macinnis}, {MacLeod},
  {Maga{\~n}a-Sandoval}, {Majorana}, {Maksimovic}, {Malvezzi}, {Man}, {Mandic}, {Mangano}, {Mano}, {Mansell}, {Manske}, {Mantovani}, {Marchesoni}, {Marchio}, {Marion}, {M{\'a}rka}, {M{\'a}rka}, {Markosyan}, {Maros}, {Martelli}, {Martellini}, {Martin}, {Martynov}, {Mason}, {Masserot}, {Massinger}, {Masso-Reid}, {Mastrogiovanni}, {Matichard}, {Matone}, {Matsumoto}, {Matsushima}, {Mavalvala}, {Mazumder}, {McCarthy}, {McClelland}, {McCormick}, {McGrath}, {McGuire}, {McIntyre}, {McIver}, {McManus}, {McRae}, {McWilliams}, {Meacher}, {Meadors}, {Meidam}, {Melatos}, {Mendell}, {Mendoza-Gandara}, {Mercer}, {Merilh}, {Merzougui}, {Meshkov}, {Messenger}, {Messick}, {Metzdorff}, {Meyers}, {Mezzani}, {Miao}, {Michel}, {Michimura}, {Middleton}, {Mikhailov}, {Milano}, {Miller}, {Miller}, {Miller}, {Miller}, {Millhouse}, {Minenkov}, {Ming}, {Mirshekari}, {Mishra}, {Mitrofanov}, {Mitselmakher}, {Mittleman}, {Miyakawa}, {Miyamoto}, {Miyamoto}, {Miyoki}, {Moggi}, {Mohan}, {Mohapatra}, {Montani}, {Moore}, {Moore}, {Moraru},
  {Moreno}, {Morii}, {Morisaki}, {Moriwaki}, {Morriss}, {Mours}, {Mow-Lowry}, {Mueller}, {Muir}, {Mukherjee}, {Mukherjee}, {Mukherjee}, {Mukund}, {Mullavey}, {Munch}, {Muniz}, {Murray}, {Mytidis}, {Nagano}, {Nakamura}, {Nakamura}, {Nakano}, {Nakano}, {Nakano}, {Nakao}, {Napier}, {Nardecchia}, {Narikawa}, {Naticchioni}, {Nelemans}, {Nelson}, {Neri}, {Nery}, {Neunzert}, {Newport}, {Newton}, {Nguyen}, {Ni}, {Nielsen}, {Nissanke}, {Nitz}, {Noack}, {Nocera}, {Nolting}, {Normandin}, {Nuttall}, {Oberling}, {Ochsner}, {Oelker}, {Ogin}, {Oh}, {Oh}, {Ohashi}, {Ohishi}, {Ohkawa}, {Ohme}, {Okutomi}, {Oliver}, {Ono}, {Ono}, {Oohara}, {Oppermann}, {Oram}, {O'Reilly}, {O'Shaughnessy}, {Ottaway}, {Overmier}, {Owen}, {Pace}, {Page}, {Pai}, {Pai}, {Palamos}, {Palashov}, {Palomba}, {Pal-Singh}, {Pan}, {Pankow}, {Pannarale}, {Pant}, {Paoletti}, {Paoli}, {Papa}, {Paris}, {Parker}, {Pascucci}, {Pasqualetti}, {Passaquieti}, {Passuello}, {Patricelli}, {Pearlstone}, {Pedraza}, {Pedurand}, {Pekowsky}, {Pele}, {Pe{\~n}a Arellano},
  {Penn}, {Perez}, {Perreca}, {Perri}, {Pfeiffer}, {Phelps}, {Piccinni}, {Pichot}, {Piergiovanni}, {Pierro}, {Pillant}, {Pinard}, {Pinto}, {Pitkin}, {Poe}, {Poggiani}, {Popolizio}, {Post}, {Powell}, {Prasad}, {Pratt}, {Predoi}, {Prestegard}, {Prijatelj}, {Principe}, {Privitera}, {Prodi}, {Prokhorov}, {Puncken}, {Punturo}, {Puppo}, {P{\"u}rrer}, {Qi}, {Qin}, {Qiu}, {Quetschke}, {Quintero}, {Quitzow-James}, {Raab}, {Rabeling}, {Radkins}, {Raffai}, {Raja}, {Rajan}, {Rakhmanov}, {Rapagnani}, {Raymond}, {Razzano}, {Re}, {Read}, {Regimbau}, {Rei}, {Reid}, {Reitze}, {Rew}, {Reyes}, {Rhoades}, {Ricci}, {Riles}, {Rizzo}, {Robertson}, {Robie}, {Robinet}, {Rocchi}, {Rolland}, {Rollins}, {Roma}, {Romano}, {Romie}, {Rosi{\'n}ska}, {Rowan}, {R{\"u}diger}, {Ruggi}, {Ryan}, {Sachdev}, {Sadecki}, {Sadeghian}, {Sago}, {Saijo}, {Saito}, {Sakai}, {Sakellariadou}, {Salconi}, {Saleem}, {Salemi}, {Samajdar}, {Sammut}, {Sampson}, {Sanchez}, {Sandberg}, {Sanders}, {Sasaki}, {Sassolas}, {Sathyaprakash}, {Sato}, {Sato}, {Saulson},
  {Sauter}, {Savage}, {Sawadsky}, {Schale}, {Scheuer}, {Schmidt}, {Schmidt}, {Schmidt}, {Schnabel}, {Schofield}, {Sch{\"o}nbeck}, {Schreiber}, {Schuette}, {Schutz}, {Schwalbe}, {Scott}, {Scott}, {Sekiguchi}, {Sekiguchi}, {Sellers}, {Sengupta}, {Sentenac}, {Sequino}, {Sergeev}, {Setyawati}, {Shaddock}, {Shaffer}, {Shahriar}, {Shapiro}, {Shawhan}, {Sheperd}, {Shibata}, {Shikano}, {Shimoda}, {Shoda}, {Shoemaker}, {Shoemaker}, {Siellez}, {Siemens}, {Sieniawska}, {Sigg}, {Silva}, {Singer}, {Singer}, {Singh}, {Singh}, {Singhal}, {Sintes}, {Slagmolen}, {Smith}, {Smith}, {Smith}, {Somiya}, {Son}, {Sorazu}, {Sorrentino}, {Souradeep}, {Spencer}, {Srivastava}, {Staley}, {Steinke}, {Steinlechner}, {Steinlechner}, {Steinmeyer}, {Stephens}, {Stevenson}, {Stone}, {Strain}, {Straniero}, {Stratta}, {Strigin}, {Sturani}, {Stuver}, {Sugimoto}, {Summerscales}, {Sun}, {Sunil}, {Sutton}, {Suzuki}, {Swinkels}, {Szczepa{\'n}czyk}, {Tacca}, {Tagoshi}, {Takada}, {Takahashi}, {Takahashi}, {Takamori}, {Talukder}, {Tanaka}, {Tanaka},
  {Tanaka}, {Tanner}, {T{\'a}pai}, {Taracchini}, {Tatsumi}, {Taylor}, {Telada}, {Theeg}, {Thomas}, {Thomas}, {Thomas}, {Thorne}, {Thrane}, {Tippens}, {Tiwari}, {Tiwari}, {Tokmakov}, {Toland}, {Tomaru}, {Tomlinson}, {Tonelli}, {Tornasi}, {Torrie}, {T{\"o}yr{\"a}}, {Travasso}, {Traylor}, {Trifir{\`o}}, {Trinastic}, {Tringali}, {Trozzo}, {Tse}, {Tso}, {Tsubono}, {Tsuzuki}, {Turconi}, {Tuyenbayev}, {Uchiyama}, {Uehara}, {Ueki}, {Ueno}, {Ugolini}, {Unnikrishnan}, {Urban}, {Ushiba}, {Usman}, {Vahlbruch}, {Vajente}, {Valdes}, {van Bakel}, {van Beuzekom}, {van den Brand}, {van den Broeck}, {Vander-Hyde}, {van der Schaaf}, {van Heijningen}, {van Putten}, {van Veggel}, {Vardaro}, {Varma}, {Vass}, {Vas{\'u}th}, {Vecchio}, {Vedovato}, {Veitch}, {Veitch}, {Venkateswara}, {Venugopalan}, {Verkindt}, {Vetrano}, {Vicer{\'e}}, {Viets}, {Vinciguerra}, {Vine}, {Vinet}, {Vitale}, {Vo}, {Vocca}, {Vorvick}, {Voss}, {Vousden}, {Vyatchanin}, {Wade}, {Wade}, {Wade}, {Wakamatsu}, {Walker}, {Wallace}, {Walsh}, {Wang}, {Wang}, {Wang},
  {Wang}, {Ward}, {Warner}, {Was}, {Watchi}, {Weaver}, {Wei}, {Weinert}, {Weinstein}, {Weiss}, {Wen}, {We{\ss}els}, {Westphal}, {Wette}, {Whelan}, {Whiting}, {Whittle}, {Williams}, {Williams}, {Williamson}, {Willis}, {Willke}, {Wimmer}, {Winkler}, {Wipf}, {Wittel}, {Woan}, {Woehler}, {Worden}, {Wright}, {Wu}, {Wu}, {Yam}, {Yamamoto}, {Yamamoto}, {Yamamoto}, {Yancey}, {Yano}, {Yap}, {Yokoyama}, {Yokozawa}, {Yoon}, {Yu}, {Yu}, {Yuzurihara}, {Yvert}, {Zadro{\.z}ny}, {Zangrando}, {Zanolin}, {Zeidler}, {Zendri}, {Zevin}, {Zhang}, {Zhang}, {Zhang}, {Zhang}, {Zhao}, {Zhou}, {Zhou}, {Zhu}, {Zhu}, {Zucker}, {Zweizig}, {Kagra Collaboration}, \& {VIRGO Collaboration}}]{2018LRR....21....3A}
{Abbott}, B.~P., {Abbott}, R., {Abbott}, T.~D., {et~al.} 2018, Living Reviews in Relativity, 21, 3

\bibitem[{{Abbott} {et~al.}(2016){Abbott}, {Abbott}, {Abbott}, {Abernathy}, {Acernese}, {Ackley}, {Adams}, {Adams}, {Addesso}, {Adhikari}, {Adya}, {Affeldt}, {Agathos}, {Agatsuma}, {Aggarwal}, {Aguiar}, {Aiello}, {Ain}, {Ajith}, {Allen}, {Allocca}, {Altin}, {Anderson}, {Anderson}, {Arai}, {Arain}, {Araya}, {Arceneaux}, {Areeda}, {Arnaud}, {Arun}, {Ascenzi}, {Ashton}, {Ast}, {Aston}, {Astone}, {Aufmuth}, {Aulbert}, {Babak}, {Bacon}, {Bader}, {Baker}, {Baldaccini}, {Ballardin}, {Ballmer}, {Barayoga}, {Barclay}, {Barish}, {Barker}, {Barone}, {Barr}, {Barsotti}, {Barsuglia}, {Barta}, {Bartlett}, {Barton}, {Bartos}, {Bassiri}, {Basti}, {Batch}, {Baune}, {Bavigadda}, {Bazzan}, {Behnke}, {Bejger}, {Belczynski}, {Bell}, {Bell}, {Berger}, {Bergman}, {Bergmann}, {Berry}, {Bersanetti}, {Bertolini}, {Betzwieser}, {Bhagwat}, {Bhandare}, {Bilenko}, {Billingsley}, {Birch}, {Birney}, {Birnholtz}, {Biscans}, {Bisht}, {Bitossi}, {Biwer}, {Bizouard}, {Blackburn}, {Blair}, {Blair}, {Blair}, {Bloemen}, {Bock}, {Bodiya}, {Boer},
  {Bogaert}, {Bogan}, {Bohe}, {Bojtos}, {Bond}, {Bondu}, {Bonnand}, {Boom}, {Bork}, {Boschi}, {Bose}, {Bouffanais}, {Bozzi}, {Bradaschia}, {Brady}, {Braginsky}, {Branchesi}, {Brau}, {Briant}, {Brillet}, {Brinkmann}, {Brisson}, {Brockill}, {Brooks}, {Brown}, {Brown}, {Brown}, {Buchanan}, {Buikema}, {Bulik}, {Bulten}, {Buonanno}, {Buskulic}, {Buy}, {Byer}, {Cabero}, {Cadonati}, {Cagnoli}, {Cahillane}, {Bustillo}, {Callister}, {Calloni}, {Camp}, {Cannon}, {Cao}, {Capano}, {Capocasa}, {Carbognani}, {Caride}, {Casanueva Diaz}, {Casentini}, {Caudill}, {Cavagli{\`a}}, {Cavalier}, {Cavalieri}, {Cella}, {Cepeda}, {Baiardi}, {Cerretani}, {Cesarini}, {Chakraborty}, {Chalermsongsak}, {Chamberlin}, {Chan}, {Chao}, {Charlton}, {Chassande-Mottin}, {Chen}, {Chen}, {Cheng}, {Chincarini}, {Chiummo}, {Cho}, {Cho}, {Chow}, {Christensen}, {Chu}, {Chua}, {Chung}, {Ciani}, {Clara}, {Clark}, {Cleva}, {Coccia}, {Cohadon}, {Colla}, {Collette}, {Cominsky}, {Constancio}, {Conte}, {Conti}, {Cook}, {Corbitt}, {Cornish}, {Corsi},
  {Cortese}, {Costa}, {Coughlin}, {Coughlin}, {Coulon}, {Countryman}, {Couvares}, {Cowan}, {Coward}, {Cowart}, {Coyne}, {Coyne}, {Craig}, {Creighton}, {Creighton}, {Cripe}, {Crowder}, {Cruise}, {Cumming}, {Cunningham}, {Cuoco}, {Dal Canton}, {Danilishin}, {D'Antonio}, {Danzmann}, {Darman}, {Da Silva Costa}, {Dattilo}, {Dave}, {Daveloza}, {Davier}, {Davies}, {Daw}, {Day}, {De}, {DeBra}, {Debreczeni}, {Degallaix}, {De Laurentis}, {Del{\'e}glise}, {Del Pozzo}, {Denker}, {Dent}, {Dereli}, {Dergachev}, {DeRosa}, {De Rosa}, {DeSalvo}, {Dhurandhar}, {D{\'\i}az}, {Di Fiore}, {Di Giovanni}, {Di Lieto}, {Di Pace}, {Di Palma}, {Di Virgilio}, {Dojcinoski}, {Dolique}, {Donovan}, {Dooley}, {Doravari}, {Douglas}, {Downes}, {Drago}, {Drever}, {Driggers}, {Du}, {Ducrot}, {Dwyer}, {Edo}, {Edwards}, {Effler}, {Eggenstein}, {Ehrens}, {Eichholz}, {Eikenberry}, {Engels}, {Essick}, {Etzel}, {Evans}, {Evans}, {Everett}, {Factourovich}, {Fafone}, {Fair}, {Fairhurst}, {Fan}, {Fang}, {Farinon}, {Farr}, {Farr}, {Favata}, {Fays},
  {Fehrmann}, {Fejer}, {Feldbaum}, {Ferrante}, {Ferreira}, {Ferrini}, {Fidecaro}, {Finn}, {Fiori}, {Fiorucci}, {Fisher}, {Flaminio}, {Fletcher}, {Fong}, {Fournier}, {Franco}, {Frasca}, {Frasconi}, {Frede}, {Frei}, {Freise}, {Frey}, {Frey}, {Fricke}, {Fritschel}, {Frolov}, {Fulda}, {Fyffe}, {Gabbard}, {Gair}, {Gammaitoni}, {Gaonkar}, {Garufi}, {Gatto}, {Gaur}, {Gehrels}, {Gemme}, {Gendre}, {Genin}, {Gennai}, {George}, {Gergely}, {Germain}, {Ghosh}, {Ghosh}, {Ghosh}, {Giaime}, {Giardina}, {Giazotto}, {Gill}, {Glaefke}, {Gleason}, {Goetz}, {Goetz}, {Gondan}, {Gonz{\'a}lez}, {Castro}, {Gopakumar}, {Gordon}, {Gorodetsky}, {Gossan}, {Gosselin}, {Gouaty}, {Graef}, {Graff}, {Granata}, {Grant}, {Gras}, {Gray}, {Greco}, {Green}, {Greenhalgh}, {Groot}, {Grote}, {Grunewald}, {Guidi}, {Guo}, {Gupta}, {Gupta}, {Gushwa}, {Gustafson}, {Gustafson}, {Hacker}, {Hall}, {Hall}, {Hammond}, {Haney}, {Hanke}, {Hanks}, {Hanna}, {Hannam}, {Hanson}, {Hardwick}, {Harms}, {Harry}, {Harry}, {Hart}, {Hartman}, {Haster}, {Haughian},
  {Healy}, {Heefner}, {Heidmann}, {Heintze}, {Heinzel}, {Heitmann}, {Hello}, {Hemming}, {Hendry}, {Heng}, {Hennig}, {Heptonstall}, {Heurs}, {Hild}, {Hoak}, {Hodge}, {Hofman}, {Hollitt}, {Holt}, {Holz}, {Hopkins}, {Hosken}, {Hough}, {Houston}, {Howell}, {Hu}, {Huang}, {Huerta}, {Huet}, {Hughey}, {Husa}, {Huttner}, {Huynh-Dinh}, {Idrisy}, {Indik}, {Ingram}, {Inta}, {Isa}, {Isac}, {Isi}, {Islas}, {Isogai}, {Iyer}, {Izumi}, {Jacobson}, {Jacqmin}, {Jang}, {Jani}, {Jaranowski}, {Jawahar}, {Jim{\'e}nez-Forteza}, {Johnson}, {Johnson-McDaniel}, {Jones}, {Jones}, {Jonker}, {Ju}, {Haris}, {Kalaghatgi}, {Kalogera}, {Kandhasamy}, {Kang}, {Kanner}, {Karki}, {Kasprzack}, {Katsavounidis}, {Katzman}, {Kaufer}, {Kaur}, {Kawabe}, {Kawazoe}, {K{\'e}f{\'e}lian}, {Kehl}, {Keitel}, {Kelley}, {Kells}, {Kennedy}, {Keppel}, {Key}, {Khalaidovski}, {Khalili}, {Khan}, {Khan}, {Khan}, {Khazanov}, {Kijbunchoo}, {Kim}, {Kim}, {Kim}, {Kim}, {Kim}, {Kim}, {King}, {King}, {Kinzel}, {Kissel}, {Kleybolte}, {Klimenko}, {Koehlenbeck}, {Kokeyama},
  {Koley}, {Kondrashov}, {Kontos}, {Koranda}, {Korobko}, {Korth}, {Kowalska}, {Kozak}, {Kringel}, {Krishnan}, {Kr{\'o}lak}, {Krueger}, {Kuehn}, {Kumar}, {Kumar}, {Kuo}, {Kutynia}, {Kwee}, {Lackey}, {Landry}, {Lange}, {Lantz}, {Lasky}, {Lazzarini}, {Lazzaro}, {Leaci}, {Leavey}, {Lebigot}, {Lee}, {Lee}, {Lee}, {Lee}, {Lenon}, {Leonardi}, {Leong}, {Leroy}, {Letendre}, {Levin}, {Levine}, {Li}, {Libson}, {Littenberg}, {Lockerbie}, {Logue}, {Lombardi}, {London}, {Lord}, {Lorenzini}, {Loriette}, {Lormand}, {Losurdo}, {Lough}, {Lousto}, {Lovelace}, {L{\"u}ck}, {Lundgren}, {Luo}, {Lynch}, {Ma}, {MacDonald}, {Machenschalk}, {MacInnis}, {Macleod}, {Maga{\~n}a-Sandoval}, {Magee}, {Mageswaran}, {Majorana}, {Maksimovic}, {Malvezzi}, {Man}, {Mandel}, {Mandic}, {Mangano}, {Mansell}, {Manske}, {Mantovani}, {Marchesoni}, {Marion}, {M{\'a}rka}, {M{\'a}rka}, {Markosyan}, {Maros}, {Martelli}, {Martellini}, {Martin}, {Martin}, {Martynov}, {Marx}, {Mason}, {Masserot}, {Massinger}, {Masso-Reid}, {Matichard}, {Matone}, {Mavalvala},
  {Mazumder}, {Mazzolo}, {McCarthy}, {McClelland}, {McCormick}, {McGuire}, {McIntyre}, {McIver}, {McManus}, {McWilliams}, {Meacher}, {Meadors}, {Meidam}, {Melatos}, {Mendell}, {Mendoza-Gandara}, {Mercer}, {Merilh}, {Merzougui}, {Meshkov}, {Messenger}, {Messick}, {Meyers}, {Mezzani}, {Miao}, {Michel}, {Middleton}, {Mikhailov}, {Milano}, {Miller}, {Millhouse}, {Minenkov}, {Ming}, {Mirshekari}, {Mishra}, {Mitra}, {Mitrofanov}, {Mitselmakher}, {Mittleman}, {Moggi}, {Mohan}, {Mohapatra}, {Montani}, {Moore}, {Moore}, {Moraru}, {Moreno}, {Morriss}, {Mossavi}, {Mours}, {Mow-Lowry}, {Mueller}, {Mueller}, {Muir}, {Mukherjee}, {Mukherjee}, {Mukherjee}, {Mukund}, {Mullavey}, {Munch}, {Murphy}, {Murray}, {Mytidis}, {Nardecchia}, {Naticchioni}, {Nayak}, {Necula}, {Nedkova}, {Nelemans}, {Neri}, {Neunzert}, {Newton}, {Nguyen}, {Nielsen}, {Nissanke}, {Nitz}, {Nocera}, {Nolting}, {Normandin}, {Nuttall}, {Oberling}, {Ochsner}, {O'Dell}, {Oelker}, {Ogin}, {Oh}, {Oh}, {Ohme}, {Oliver}, {Oppermann}, {Oram}, {O'Reilly},
  {O'Shaughnessy}, {Ott}, {Ottaway}, {Ottens}, {Overmier}, {Owen}, {Pai}, {Pai}, {Palamos}, {Palashov}, {Palomba}, {Pal-Singh}, {Pan}, {Pan}, {Pankow}, {Pannarale}, {Pant}, {Paoletti}, {Paoli}, {Papa}, {Paris}, {Parker}, {Pascucci}, {Pasqualetti}, {Passaquieti}, {Passuello}, {Patricelli}, {Patrick}, {Pearlstone}, {Pedraza}, {Pedurand}, {Pekowsky}, {Pele}, {Penn}, {Perreca}, {Pfeiffer}, {Phelps}, {Piccinni}, {Pichot}, {Pickenpack}, {Piergiovanni}, {Pierro}, {Pillant}, {Pinard}, {Pinto}, {Pitkin}, {Poeld}, {Poggiani}, {Popolizio}, {Post}, {Powell}, {Prasad}, {Predoi}, {Premachandra}, {Prestegard}, {Price}, {Prijatelj}, {Principe}, {Privitera}, {Prix}, {Prodi}, {Prokhorov}, {Puncken}, {Punturo}, {Puppo}, {P{\"u}rrer}, {Qi}, {Qin}, {Quetschke}, {Quintero}, {Quitzow-James}, {Raab}, {Rabeling}, {Radkins}, {Raffai}, {Raja}, {Rakhmanov}, {Ramet}, {Rapagnani}, {Raymond}, {Razzano}, {Re}, {Read}, {Reed}, {Regimbau}, {Rei}, {Reid}, {Reitze}, {Rew}, {Reyes}, {Ricci}, {Riles}, {Robertson}, {Robie}, {Robinet}, {Rocchi},
  {Rolland}, {Rollins}, {Roma}, {Romano}, {Romano}, {Romanov}, {Romie}, {Rosi{\'n}ska}, {Rowan}, {R{\"u}diger}, {Ruggi}, {Ryan}, {Sachdev}, {Sadecki}, {Sadeghian}, {Salconi}, {Saleem}, {Salemi}, {Samajdar}, {Sammut}, {Sampson}, {Sanchez}, {Sandberg}, {Sandeen}, {Sanders}, {Sanders}, {Sassolas}, {Sathyaprakash}, {Saulson}, {Sauter}, {Savage}, {Sawadsky}, {Schale}, {Schilling}, {Schmidt}, {Schmidt}, {Schnabel}, {Schofield}, {Sch{\"o}nbeck}, {Schreiber}, {Schuette}, {Schutz}, {Scott}, {Scott}, {Sellers}, {Sengupta}, {Sentenac}, {Sequino}, {Sergeev}, {Serna}, {Setyawati}, {Sevigny}, {Shaddock}, {Shaffer}, {Shah}, {Shahriar}, {Shaltev}, {Shao}, {Shapiro}, {Shawhan}, {Sheperd}, {Shoemaker}, {Shoemaker}, {Siellez}, {Siemens}, {Sigg}, {Silva}, {Simakov}, {Singer}, {Singer}, {Singh}, {Singh}, {Singhal}, {Sintes}, {Slagmolen}, {Smith}, {Smith}, {Smith}, {Smith}, {Son}, {Sorazu}, {Sorrentino}, {Souradeep}, {Srivastava}, {Staley}, {Steinke}, {Steinlechner}, {Steinlechner}, {Steinmeyer}, {Stephens}, {Stevenson}, {Stone},
  {Strain}, {Straniero}, {Stratta}, {Strauss}, {Strigin}, {Sturani}, {Stuver}, {Summerscales}, {Sun}, {Sutton}, {Swinkels}, {Szczepa{\'n}czyk}, {Tacca}, {Talukder}, {Tanner}, {T{\'a}pai}, {Tarabrin}, {Taracchini}, {Taylor}, {Theeg}, {Thirugnanasambandam}, {Thomas}, {Thomas}, {Thomas}, {Thorne}, {Thorne}, {Thrane}, {Tiwari}, {Tiwari}, {Tokmakov}, {Tomlinson}, {Tonelli}, {Torres}, {Torrie}, {T{\"o}yr{\"a}}, {Travasso}, {Traylor}, {Trifir{\`o}}, {Tringali}, {Trozzo}, {Tse}, {Turconi}, {Tuyenbayev}, {Ugolini}, {Unnikrishnan}, {Urban}, {Usman}, {Vahlbruch}, {Vajente}, {Valdes}, {Vallisneri}, {van Bakel}, {van Beuzekom}, {van den Brand}, {Van Den Broeck}, {Vander-Hyde}, {van der Schaaf}, {van Heijningen}, {van Veggel}, {Vardaro}, {Vass}, {Vas{\'u}th}, {Vaulin}, {Vecchio}, {Vedovato}, {Veitch}, {Veitch}, {Venkateswara}, {Verkindt}, {Vetrano}, {Vicer{\'e}}, {Vinciguerra}, {Vine}, {Vinet}, {Vitale}, {Vo}, {Vocca}, {Vorvick}, {Voss}, {Vousden}, {Vyatchanin}, {Wade}, {Wade}, {Wade}, {Waldman}, {Walker}, {Wallace},
  {Walsh}, {Wang}, {Wang}, {Wang}, {Wang}, {Wang}, {Ward}, {Ward}, {Warner}, {Was}, {Weaver}, {Wei}, {Weinert}, {Weinstein}, {Weiss}, {Welborn}, {Wen}, {We{\ss}els}, {Westphal}, {Wette}, {Whelan}, {Whitcomb}, {White}, {Whiting}, {Wiesner}, {Wilkinson}, {Willems}, {Williams}, {Williams}, {Williamson}, {Willis}, {Willke}, {Wimmer}, {Winkelmann}, {Winkler}, {Wipf}, {Wiseman}, {Wittel}, {Woan}, {Worden}, {Wright}, {Wu}, {Yablon}, {Yakushin}, {Yam}, {Yamamoto}, {Yancey}, {Yap}, {Yu}, {Yvert}, {Zadro{\.Z}ny}, {Zangrando}, {Zanolin}, {Zendri}, {Zevin}, {Zhang}, {Zhang}, {Zhang}, {Zhang}, {Zhao}, {Zhou}, {Zhou}, {Zhu}, {Zucker}, {Zuraw}, {Zweizig}, {LIGO Scientific Collaboration}, \& {Virgo Collaboration}}]{2016PhRvL.116f1102A}
{Abbott}, B.~P., {Abbott}, R., {Abbott}, T.~D., {et~al.} 2016, \prl, 116, 061102

\bibitem[{{Abbott} {et~al.}(2017{\natexlab{a}}){Abbott}, {Abbott}, {Abbott}, {Abernathy}, {Ackley}, {Adams}, {Addesso}, {Adhikari}, {Adya}, {Affeldt}, {Aggarwal}, {Aguiar}, {Ain}, {Ajith}, {Allen}, {Altin}, {Anderson}, {Anderson}, {Arai}, {Araya}, {Arceneaux}, {Areeda}, {Arun}, {Ashton}, {Ast}, {Aston}, {Aufmuth}, {Aulbert}, {Babak}, {Baker}, {Ballmer}, {Barayoga}, {Barclay}, {Barish}, {Barker}, {Barr}, {Barsotti}, {Bartlett}, {Bartos}, {Bassiri}, {Batch}, {Baune}, {Bell}, {Berger}, {Bergmann}, {Berry}, {Betzwieser}, {Bhagwat}, {Bhandare}, {Bilenko}, {Billingsley}, {Birch}, {Birney}, {Biscans}, {Bisht}, {Biwer}, {Blackburn}, {Blair}, {Blair}, {Blair}, {Bock}, {Bogan}, {Bohe}, {Bond}, {Bork}, {Bose}, {Brady}, {Braginsky}, {Brau}, {Brinkmann}, {Brockill}, {Broida}, {Brooks}, {Brown}, {Brown}, {Brown}, {Brunett}, {Buchanan}, {Buikema}, {Buonanno}, {Byer}, {Cabero}, {Cadonati}, {Cahillane}, {Calder{\'o}n Bustillo}, {Callister}, {Camp}, {Cannon}, {Cao}, {Capano}, {Caride}, {Caudill}, {Cavagli{\`a}}, {Cepeda},
  {Chamberlin}, {Chan}, {Chao}, {Charlton}, {Cheeseboro}, {Chen}, {Chen}, {Cheng}, {Cho}, {Cho}, {Chow}, {Christensen}, {Chu}, {Chung}, {Ciani}, {Clara}, {Clark}, {Collette}, {Cominsky}, {Constancio}, {Cook}, {Corbitt}, {Cornish}, {Corsi}, {Costa}, {Coughlin}, {Coughlin}, {Countryman}, {Couvares}, {Cowan}, {Coward}, {Cowart}, {Coyne}, {Coyne}, {Craig}, {Creighton}, {Cripe}, {Crowder}, {Cumming}, {Cunningham}, {Dal Canton}, {Danilishin}, {Danzmann}, {Darman}, {Dasgupta}, {Da Silva Costa}, {Dave}, {Davies}, {Daw}, {De}, {DeBra}, {Del Pozzo}, {Denker}, {Dent}, {Dergachev}, {DeRosa}, {DeSalvo}, {Devine}, {Dhurandhar}, {D{\'\i}az}, {Di Palma}, {Donovan}, {Dooley}, {Doravari}, {Douglas}, {Downes}, {Drago}, {Drever}, {Driggers}, {Dwyer}, {Edo}, {Edwards}, {Effler}, {Eggenstein}, {Ehrens}, {Eichholz}, {Eikenberry}, {Engels}, {Essick}, {Etzel}, {Evans}, {Evans}, {Everett}, {Factourovich}, {Fair}, {Fairhurst}, {Fan}, {Fang}, {Farr}, {Farr}, {Favata}, {Fays}, {Fehrmann}, {Fejer}, {Fenyvesi}, {Ferreira}, {Fisher},
  {Fletcher}, {Frei}, {Freise}, {Frey}, {Fritschel}, {Frolov}, {Fulda}, {Fyffe}, {Gabbard}, {Gair}, {Gaonkar}, {Gaur}, {Gehrels}, {Geng}, {George}, {Gergely}, {Ghosh}, {Ghosh}, {Giaime}, {Giardina}, {Gill}, {Glaefke}, {Goetz}, {Goetz}, {Gondan}, {Gonz{\'a}lez}, {Gopakumar}, {Gordon}, {Gorodetsky}, {Gossan}, {Graef}, {Graff}, {Grant}, {Gras}, {Gray}, {Green}, {Grote}, {Grunewald}, {Guo}, {Gupta}, {Gupta}, {Gushwa}, {Gustafson}, {Gustafson}, {Hacker}, {Hall}, {Hall}, {Hammond}, {Haney}, {Hanke}, {Hanks}, {Hanna}, {Hannam}, {Hanson}, {Hardwick}, {Harry}, {Harry}, {Hart}, {Hartman}, {Haster}, {Haughian}, {Heintze}, {Hendry}, {Heng}, {Hennig}, {Henry}, {Heptonstall}, {Heurs}, {Hild}, {Hoak}, {Holt}, {Holz}, {Hopkins}, {Hough}, {Houston}, {Howell}, {Hu}, {Huang}, {Huerta}, {Hughey}, {Husa}, {Huttner}, {Huynh-Dinh}, {Indik}, {Ingram}, {Inta}, {Isa}, {Isi}, {Isogai}, {Iyer}, {Izumi}, {Jang}, {Jani}, {Jawahar}, {Jian}, {Jim{\'e}nez-Forteza}, {Johnson}, {Jones}, {Jones}, {Ju}, {Haris}, {Kalaghatgi}, {Kalogera},
  {Kandhasamy}, {Kang}, {Kanner}, {Kapadia}, {Karki}, {Karvinen}, {Kasprzack}, {Katsavounidis}, {Katzman}, {Kaufer}, {Kaur}, {Kawabe}, {Kehl}, {Keitel}, {Kelley}, {Kells}, {Kennedy}, {Key}, {Khalili}, {Khan}, {Khan}, {Khazanov}, {Kijbunchoo}, {Kim}, {Kim}, {Kim}, {Kim}, {Kim}, {Kim}, {Kim}, {Kimbrell}, {King}, {King}, {Kissel}, {Klein}, {Kleybolte}, {Klimenko}, {Koehlenbeck}, {Kondrashov}, {Kontos}, {Korobko}, {Korth}, {Kozak}, {Kringel}, {Krueger}, {Kuehn}, {Kumar}, {Kumar}, {Kuo}, {Lackey}, {Landry}, {Lange}, {Lantz}, {Lasky}, {Laxen}, {Lazzarini}, {Leavey}, {Lebigot}, {Lee}, {Lee}, {Lee}, {Lee}, {Lenon}, {Leong}, {Levin}, {Lewis}, {Li}, {Libson}, {Littenberg}, {Lockerbie}, {Lombardi}, {London}, {Lord}, {Lormand}, {Lough}, {L{\"u}ck}, {Lundgren}, {Lynch}, {Ma}, {Machenschalk}, {MacInnis}, {Macleod}, {Maga{\~n}a-Sandoval}, {Maga{\~n}a Zertuche}, {Magee}, {Mandic}, {Mangano}, {Mansell}, {Manske}, {M{\'a}rka}, {M{\'a}rka}, {Markosyan}, {Maros}, {Martin}, {Martynov}, {Mason}, {Massinger}, {Masso-Reid},
  {Matichard}, {Matone}, {Mavalvala}, {Mazumder}, {McCarthy}, {McClelland}, {McCormick}, {McGuire}, {McIntyre}, {McIver}, {McManus}, {McRae}, {McWilliams}, {Meacher}, {Meadors}, {Melatos}, {Mendell}, {Mercer}, {Merilh}, {Meshkov}, {Messenger}, {Messick}, {Meyers}, {Miao}, {Middleton}, {Mikhailov}, {Miller}, {Miller}, {Miller}, {Miller}, {Millhouse}, {Ming}, {Mirshekari}, {Mishra}, {Mitra}, {Mitrofanov}, {Mitselmakher}, {Mittleman}, {Mohapatra}, {Moore}, {Moore}, {Moraru}, {Moreno}, {Morriss}, {Mossavi}, {Mow-Lowry}, {Mueller}, {Muir}, {Mukherjee}, {Mukherjee}, {Mukherjee}, {Mukund}, {Mullavey}, {Munch}, {Murphy}, {Murray}, {Mytidis}, {Nayak}, {Nedkova}, {Nelson}, {Neunzert}, {Newton}, {Nguyen}, {Nielsen}, {Nitz}, {Nolting}, {Normandin}, {Nuttall}, {Oberling}, {Ochsner}, {O'Dell}, {Oelker}, {Ogin}, {Oh}, {Oh}, {Ohme}, {Oliver}, {Oppermann}, {Oram}, {O'Reilly}, {O'Shaughnessy}, {Ottaway}, {Overmier}, {Owen}, {Pai}, {Pai}, {Palamos}, {Palashov}, {Pal-Singh}, {Pan}, {Pankow}, {Pannarale}, {Pant}, {Papa}, {Paris},
  {Parker}, {Pascucci}, {Patrick}, {Pearlstone}, {Pedraza}, {Pekowsky}, {Pele}, {Penn}, {Perreca}, {Perri}, {Phelps}, {Pierro}, {Pinto}, {Pitkin}, {Poe}, {Post}, {Powell}, {Prasad}, {Predoi}, {Prestegard}, {Price}, {Prijatelj}, {Principe}, {Privitera}, {Prokhorov}, {Puncken}, {P{\"u}rrer}, {Qi}, {Qin}, {Qiu}, {Quetschke}, {Quintero}, {Quitzow-James}, {Raab}, {Rabeling}, {Radkins}, {Raffai}, {Raja}, {Rajan}, {Rakhmanov}, {Raymond}, {Read}, {Reed}, {Reid}, {Reitze}, {Rew}, {Reyes}, {Riles}, {Rizzo}, {Robertson}, {Robie}, {Rollins}, {Roma}, {Romanov}, {Romie}, {Rowan}, {R{\"u}diger}, {Ryan}, {Sachdev}, {Sadecki}, {Sadeghian}, {Sakellariadou}, {Saleem}, {Salemi}, {Samajdar}, {Sammut}, {Sanchez}, {Sandberg}, {Sandeen}, {Sanders}, {Sathyaprakash}, {Saulson}, {Sauter}, {Savage}, {Sawadsky}, {Schale}, {Schilling}, {Schmidt}, {Schmidt}, {Schnabel}, {Schofield}, {Sch{\"o}nbeck}, {Schreiber}, {Schuette}, {Schutz}, {Scott}, {Scott}, {Sellers}, {Sengupta}, {Sergeev}, {Shaddock}, {Shaffer}, {Shahriar}, {Shaltev},
  {Shapiro}, {Shawhan}, {Sheperd}, {Shoemaker}, {Shoemaker}, {Siellez}, {Siemens}, {Sigg}, {Silva}, {Singer}, {Singer}, {Singh}, {Singh}, {Sintes}, {Slagmolen}, {Smith}, {Smith}, {Smith}, {Son}, {Sorazu}, {Souradeep}, {Srivastava}, {Staley}, {Steinke}, {Steinlechner}, {Steinlechner}, {Steinmeyer}, {Stephens}, {Stone}, {Strain}, {Strauss}, {Strigin}, {Sturani}, {Stuver}, {Summerscales}, {Sun}, {Sunil}, {Sutton}, {Szczepa{\'n}czyk}, {Talukder}, {Tanner}, {T{\'a}pai}, {Tarabrin}, {Taracchini}, {Taylor}, {Theeg}, {Thirugnanasambandam}, {Thomas}, {Thomas}, {Thomas}, {Thorne}, {Thrane}, {Tiwari}, {Tokmakov}, {Toland}, {Tomlinson}, {Tornasi}, {Torres}, {Torrie}, {T{\"o}yr{\"a}}, {Traylor}, {Trifir{\`o}}, {Tse}, {Tuyenbayev}, {Ugolini}, {Unnikrishnan}, {Urban}, {Usman}, {Vahlbruch}, {Vajente}, {Valdes}, {Vander-Hyde}, {van Veggel}, {Vass}, {Vaulin}, {Vecchio}, {Veitch}, {Veitch}, {Venkateswara}, {Vinciguerra}, {Vine}, {Vitale}, {Vo}, {Vorvick}, {Voss}, {Vousden}, {Vyatchanin}, {Wade}, {Wade}, {Wade}, {Walker},
  {Wallace}, {Walsh}, {Wang}, {Wang}, {Wang}, {Wang}, {Ward}, {Warner}, {Weaver}, {Weinert}, {Weinstein}, {Weiss}, {Wen}, {We{\ss}els}, {Westphal}, {Wette}, {Whelan}, {Whiting}, {Williams}, {Williamson}, {Willis}, {Willke}, {Wimmer}, {Winkler}, {Wipf}, {Wittel}, {Woan}, {Woehler}, {Worden}, {Wright}, {Wu}, {Wu}, {Yablon}, {Yam}, {Yamamoto}, {Yancey}, {Yu}, {Zanolin}, {Zevin}, {Zhang}, {Zhang}, {Zhang}, {Zhao}, {Zhou}, {Zhou}, {Zhu}, {Zucker}, {Zuraw}, {Zweizig}, {(LIGO Scientific Collaboration}, \& {Harms}}]{2017CQGra..34d4001A}
{Abbott}, B.~P., {Abbott}, R., {Abbott}, T.~D., {et~al.} 2017{\natexlab{a}}, Classical and Quantum Gravity, 34, 044001

\bibitem[{{Abbott} {et~al.}(2017{\natexlab{b}}){Abbott}, {Abbott}, {Abbott}, {Acernese}, {Ackley}, {Adams}, {Adams}, {Addesso}, {Adhikari}, {Adya}, {Affeldt}, {Afrough}, {Agarwal}, {Agathos}, {Agatsuma}, {Aggarwal}, {Aguiar}, {Aiello}, {Ain}, {Ajith}, {Allen}, {Allen}, {Allocca}, {Altin}, {Amato}, {Ananyeva}, {Anderson}, {Anderson}, {Angelova}, {Antier}, {Appert}, {Arai}, {Araya}, {Areeda}, {Arnaud}, {Arun}, {Ascenzi}, {Ashton}, {Ast}, {Aston}, {Astone}, {Atallah}, {Aufmuth}, {Aulbert}, {AultONeal}, {Austin}, {Avila-Alvarez}, {Babak}, {Bacon}, {Bader}, {Bae}, {Bailes}, {Baker}, {Baldaccini}, {Ballardin}, {Ballmer}, {Banagiri}, {Barayoga}, {Barclay}, {Barish}, {Barker}, {Barkett}, {Barone}, {Barr}, {Barsotti}, {Barsuglia}, {Barta}, {Barthelmy}, {Bartlett}, {Bartos}, {Bassiri}, {Basti}, {Batch}, {Bawaj}, {Bayley}, {Bazzan}, {B{\'e}csy}, {Beer}, {Bejger}, {Belahcene}, {Bell}, {Berger}, {Bergmann}, {Bernuzzi}, {Bero}, {Berry}, {Bersanetti}, {Bertolini}, {Betzwieser}, {Bhagwat}, {Bhandare}, {Bilenko},
  {Billingsley}, {Billman}, {Birch}, {Birney}, {Birnholtz}, {Biscans}, {Biscoveanu}, {Bisht}, {Bitossi}, {Biwer}, {Bizouard}, {Blackburn}, {Blackman}, {Blair}, {Blair}, {Blair}, {Bloemen}, {Bock}, {Bode}, {Boer}, {Bogaert}, {Bohe}, {Bondu}, {Bonilla}, {Bonnand}, {Boom}, {Bork}, {Boschi}, {Bose}, {Bossie}, {Bouffanais}, {Bozzi}, {Bradaschia}, {Brady}, {Branchesi}, {Brau}, {Briant}, {Brillet}, {Brinkmann}, {Brisson}, {Brockill}, {Broida}, {Brooks}, {Brown}, {Brown}, {Brunett}, {Buchanan}, {Buikema}, {Bulik}, {Bulten}, {Buonanno}, {Buskulic}, {Buy}, {Byer}, {Cabero}, {Cadonati}, {Cagnoli}, {Cahillane}, {Calder{\'o}n Bustillo}, {Callister}, {Calloni}, {Camp}, {Canepa}, {Canizares}, {Cannon}, {Cao}, {Cao}, {Capano}, {Capocasa}, {Carbognani}, {Caride}, {Carney}, {Carullo}, {Casanueva Diaz}, {Casentini}, {Caudill}, {Cavagli{\`a}}, {Cavalier}, {Cavalieri}, {Cella}, {Cepeda}, {Cerd{\'a}-Dur{\'a}n}, {Cerretani}, {Cesarini}, {Chamberlin}, {Chan}, {Chao}, {Charlton}, {Chase}, {Chassande-Mottin}, {Chatterjee},
  {Chatziioannou}, {Cheeseboro}, {Chen}, {Chen}, {Chen}, {Cheng}, {Chia}, {Chincarini}, {Chiummo}, {Chmiel}, {Cho}, {Cho}, {Chow}, {Christensen}, {Chu}, {Chua}, {Chua}, {Chung}, {Chung}, {Ciani}, {Ciolfi}, {Cirelli}, {Cirone}, {Clara}, {Clark}, {Clearwater}, {Cleva}, {Cocchieri}, {Coccia}, {Cohadon}, {Cohen}, {Colla}, {Collette}, {Cominsky}, {Constancio}, {Conti}, {Cooper}, {Corban}, {Corbitt}, {Cordero-Carri{\'o}n}, {Corley}, {Cornish}, {Corsi}, {Cortese}, {Costa}, {Coughlin}, {Coughlin}, {Coulon}, {Countryman}, {Couvares}, {Covas}, {Cowan}, {Coward}, {Cowart}, {Coyne}, {Coyne}, {Creighton}, {Creighton}, {Cripe}, {Crowder}, {Cullen}, {Cumming}, {Cunningham}, {Cuoco}, {Dal Canton}, {D{\'a}lya}, {Danilishin}, {D'Antonio}, {Danzmann}, {Dasgupta}, {Da Silva Costa}, {Dattilo}, {Dave}, {Davier}, {Davis}, {Daw}, {Day}, {De}, {DeBra}, {Degallaix}, {De Laurentis}, {Del{\'e}glise}, {Del Pozzo}, {Demos}, {Denker}, {Dent}, {De Pietri}, {Dergachev}, {De Rosa}, {DeRosa}, {De Rossi}, {DeSalvo}, {de Varona}, {Devenson},
  {Dhurandhar}, {D{\'\i}az}, {Dietrich}, {Di Fiore}, {Di Giovanni}, {Di Girolamo}, {Di Lieto}, {Di Pace}, {Di Palma}, {Di Renzo}, {Doctor}, {Dolique}, {Donovan}, {Dooley}, {Doravari}, {Dorrington}, {Douglas}, {Dovale {\'A}lvarez}, {Downes}, {Drago}, {Dreissigacker}, {Driggers}, {Du}, {Ducrot}, {Dudi}, {Dupej}, {Dwyer}, {Edo}, {Edwards}, {Effler}, {Eggenstein}, {Ehrens}, {Eichholz}, {Eikenberry}, {Eisenstein}, {Essick}, {Estevez}, {Etienne}, {Etzel}, {Evans}, {Evans}, {Factourovich}, {Fafone}, {Fair}, {Fairhurst}, {Fan}, {Farinon}, {Farr}, {Farr}, {Fauchon-Jones}, {Favata}, {Fays}, {Fee}, {Fehrmann}, {Feicht}, {Fejer}, {Fernandez-Galiana}, {Ferrante}, {Ferreira}, {Ferrini}, {Fidecaro}, {Finstad}, {Fiori}, {Fiorucci}, {Fishbach}, {Fisher}, {Fitz-Axen}, {Flaminio}, {Fletcher}, {Fong}, {Font}, {Forsyth}, {Forsyth}, {Fournier}, {Frasca}, {Frasconi}, {Frei}, {Freise}, {Frey}, {Frey}, {Fries}, {Fritschel}, {Frolov}, {Fulda}, {Fyffe}, {Gabbard}, {Gadre}, {Gaebel}, {Gair}, {Gammaitoni}, {Ganija}, {Gaonkar},
  {Garcia-Quiros}, {Garufi}, {Gateley}, {Gaudio}, {Gaur}, {Gayathri}, {Gehrels}, {Gemme}, {Genin}, {Gennai}, {George}, {George}, {Gergely}, {Germain}, {Ghonge}, {Ghosh}, {Ghosh}, {Ghosh}, {Giaime}, {Giardina}, {Giazotto}, {Gill}, {Glover}, {Goetz}, {Goetz}, {Gomes}, {Goncharov}, {Gonz{\'a}lez}, {Gonzalez Castro}, {Gopakumar}, {Gorodetsky}, {Gossan}, {Gosselin}, {Gouaty}, {Grado}, {Graef}, {Granata}, {Grant}, {Gras}, {Gray}, {Greco}, {Green}, {Gretarsson}, {Groot}, {Grote}, {Grunewald}, {Gruning}, {Guidi}, {Guo}, {Gupta}, {Gupta}, {Gushwa}, {Gustafson}, {Gustafson}, {Halim}, {Hall}, {Hall}, {Hamilton}, {Hammond}, {Haney}, {Hanke}, {Hanks}, {Hanna}, {Hannam}, {Hannuksela}, {Hanson}, {Hardwick}, {Harms}, {Harry}, {Harry}, {Hart}, {Haster}, {Haughian}, {Healy}, {Heidmann}, {Heintze}, {Heitmann}, {Hello}, {Hemming}, {Hendry}, {Heng}, {Hennig}, {Heptonstall}, {Heurs}, {Hild}, {Hinderer}, {Ho}, {Hoak}, {Hofman}, {Holt}, {Holz}, {Hopkins}, {Horst}, {Hough}, {Houston}, {Howell}, {Hreibi}, {Hu}, {Huerta}, {Huet},
  {Hughey}, {Husa}, {Huttner}, {Huynh-Dinh}, {Indik}, {Inta}, {Intini}, {Isa}, {Isac}, {Isi}, {Iyer}, {Izumi}, {Jacqmin}, {Jani}, {Jaranowski}, {Jawahar}, {Jim{\'e}nez-Forteza}, {Johnson}, {Johnson-McDaniel}, {Jones}, {Jones}, {Jonker}, {Ju}, {Junker}, {Kalaghatgi}, {Kalogera}, {Kamai}, {Kandhasamy}, {Kang}, {Kanner}, {Kapadia}, {Karki}, {Karvinen}, {Kasprzack}, {Kastaun}, {Katolik}, {Katsavounidis}, {Katzman}, {Kaufer}, {Kawabe}, {K{\'e}f{\'e}lian}, {Keitel}, {Kemball}, {Kennedy}, {Kent}, {Key}, {Khalili}, {Khan}, {Khan}, {Khan}, {Khazanov}, {Kijbunchoo}, {Kim}, {Kim}, {Kim}, {Kim}, {Kim}, {Kim}, {Kimbrell}, {King}, {King}, {Kinley-Hanlon}, {Kirchhoff}, {Kissel}, {Kleybolte}, {Klimenko}, {Knowles}, {Koch}, {Koehlenbeck}, {Koley}, {Kondrashov}, {Kontos}, {Korobko}, {Korth}, {Kowalska}, {Kozak}, {Kr{\"a}mer}, {Kringel}, {Krishnan}, {Kr{\'o}lak}, {Kuehn}, {Kumar}, {Kumar}, {Kumar}, {Kuo}, {Kutynia}, {Kwang}, {Lackey}, {Lai}, {Landry}, {Lang}, {Lange}, {Lantz}, {Lanza}, {Larson}, {Lartaux-Vollard}, {Lasky},
  {Laxen}, {Lazzarini}, {Lazzaro}, {Leaci}, {Leavey}, {Lee}, {Lee}, {Lee}, {Lee}, {Lee}, {Lehmann}, {Lenon}, {Leon}, {Leonardi}, {Leroy}, {Letendre}, {Levin}, {Li}, {Linker}, {Littenberg}, {Liu}, {Liu}, {Lo}, {Lockerbie}, {London}, {Lord}, {Lorenzini}, {Loriette}, {Lormand}, {Losurdo}, {Lough}, {Lousto}, {Lovelace}, {L{\"u}ck}, {Lumaca}, {Lundgren}, {Lynch}, {Ma}, {Macas}, {Macfoy}, {Machenschalk}, {MacInnis}, {Macleod}, {Maga{\~n}a Hernandez}, {Maga{\~n}a-Sandoval}, {Maga{\~n}a Zertuche}, {Magee}, {Majorana}, {Maksimovic}, {Man}, {Mandic}, {Mangano}, {Mansell}, {Manske}, {Mantovani}, {Marchesoni}, {Marion}, {M{\'a}rka}, {M{\'a}rka}, {Markakis}, {Markosyan}, {Markowitz}, {Maros}, {Marquina}, {Marsh}, {Martelli}, {Martellini}, {Martin}, {Martin}, {Martynov}, {Marx}, {Mason}, {Massera}, {Masserot}, {Massinger}, {Masso-Reid}, {Mastrogiovanni}, {Matas}, {Matichard}, {Matone}, {Mavalvala}, {Mazumder}, {McCarthy}, {McClelland}, {McCormick}, {McCuller}, {McGuire}, {McIntyre}, {McIver}, {McManus}, {McNeill}, {McRae},
  {McWilliams}, {Meacher}, {Meadors}, {Mehmet}, {Meidam}, {Mejuto-Villa}, {Melatos}, {Mendell}, {Mercer}, {Merilh}, {Merzougui}, {Meshkov}, {Messenger}, {Messick}, {Metzdorff}, {Meyers}, {Miao}, {Michel}, {Middleton}, {Mikhailov}, {Milano}, {Miller}, {Miller}, {Miller}, {Millhouse}, {Milovich-Goff}, {Minazzoli}, {Minenkov}, {Ming}, {Mishra}, {Mitra}, {Mitrofanov}, {Mitselmakher}, {Mittleman}, {Moffa}, {Moggi}, {Mogushi}, {Mohan}, {Mohapatra}, {Molina}, {Montani}, {Moore}, {Moraru}, {Moreno}, {Morisaki}, {Morriss}, {Mours}, {Mow-Lowry}, {Mueller}, {Muir}, {Mukherjee}, {Mukherjee}, {Mukherjee}, {Mukund}, {Mullavey}, {Munch}, {Mu{\~n}iz}, {Muratore}, {Murray}, {Nagar}, {Napier}, {Nardecchia}, {Naticchioni}, {Nayak}, {Neilson}, {Nelemans}, {Nelson}, {Nery}, {Neunzert}, {Nevin}, {Newport}, {Newton}, {Ng}, {Nguyen}, {Nguyen}, {Nichols}, {Nielsen}, {Nissanke}, {Nitz}, {Noack}, {Nocera}, {Nolting}, {North}, {Nuttall}, {Oberling}, {O'Dea}, {Ogin}, {Oh}, {Oh}, {Ohme}, {Okada}, {Oliver}, {Oppermann}, {Oram}, {O'Reilly},
  {Ormiston}, {Ortega}, {O'Shaughnessy}, {Ossokine}, {Ottaway}, {Overmier}, {Owen}, {Pace}, {Page}, {Page}, {Pai}, {Pai}, {Palamos}, {Palashov}, {Palomba}, {Pal-Singh}, {Pan}, {Pan}, {Pang}, {Pang}, {Pankow}, {Pannarale}, {Pant}, {Paoletti}, {Paoli}, {Papa}, {Parida}, {Parker}, {Pascucci}, {Pasqualetti}, {Passaquieti}, {Passuello}, {Patil}, {Patricelli}, {Pearlstone}, {Pedraza}, {Pedurand}, {Pekowsky}, {Pele}, {Penn}, {Perez}, {Perreca}, {Perri}, {Pfeiffer}, {Phelps}, {Piccinni}, {Pichot}, {Piergiovanni}, {Pierro}, {Pillant}, {Pinard}, {Pinto}, {Pirello}, {Pitkin}, {Poe}, {Poggiani}, {Popolizio}, {Porter}, {Post}, {Powell}, {Prasad}, {Pratt}, {Pratten}, {Predoi}, {Prestegard}, {Prijatelj}, {Principe}, {Privitera}, {Prix}, {Prodi}, {Prokhorov}, {Puncken}, {Punturo}, {Puppo}, {P{\"u}rrer}, {Qi}, {Quetschke}, {Quintero}, {Quitzow-James}, {Raab}, {Rabeling}, {Radkins}, {Raffai}, {Raja}, {Rajan}, {Rajbhandari}, {Rakhmanov}, {Ramirez}, {Ramos-Buades}, {Rapagnani}, {Raymond}, {Razzano}, {Read}, {Regimbau}, {Rei},
  {Reid}, {Reitze}, {Ren}, {Reyes}, {Ricci}, {Ricker}, {Rieger}, {Riles}, {Rizzo}, {Robertson}, {Robie}, {Robinet}, {Rocchi}, {Rolland}, {Rollins}, {Roma}, {Romano}, {Romano}, {Romel}, {Romie}, {Rosi{\'n}ska}, {Ross}, {Rowan}, {R{\"u}diger}, {Ruggi}, {Rutins}, {Ryan}, {Sachdev}, {Sadecki}, {Sadeghian}, {Sakellariadou}, {Salconi}, {Saleem}, {Salemi}, {Samajdar}, {Sammut}, {Sampson}, {Sanchez}, {Sanchez}, {Sanchis-Gual}, {Sandberg}, {Sanders}, {Sassolas}, {Sathyaprakash}, {Saulson}, {Sauter}, {Savage}, {Sawadsky}, {Schale}, {Scheel}, {Scheuer}, {Schmidt}, {Schmidt}, {Schnabel}, {Schofield}, {Sch{\"o}nbeck}, {Schreiber}, {Schuette}, {Schulte}, {Schutz}, {Schwalbe}, {Scott}, {Scott}, {Seidel}, {Sellers}, {Sengupta}, {Sentenac}, {Sequino}, {Sergeev}, {Shaddock}, {Shaffer}, {Shah}, {Shahriar}, {Shaner}, {Shao}, {Shapiro}, {Shawhan}, {Sheperd}, {Shoemaker}, {Shoemaker}, {Siellez}, {Siemens}, {Sieniawska}, {Sigg}, {Silva}, {Singer}, {Singh}, {Singhal}, {Sintes}, {Slagmolen}, {Smith}, {Smith}, {Smith}, {Somala},
  {Son}, {Sonnenberg}, {Sorazu}, {Sorrentino}, {Souradeep}, {Spencer}, {Srivastava}, {Staats}, {Staley}, {Steinke}, {Steinlechner}, {Steinlechner}, {Steinmeyer}, {Stevenson}, {Stone}, {Stops}, {Strain}, {Stratta}, {Strigin}, {Strunk}, {Sturani}, {Stuver}, {Summerscales}, {Sun}, {Sunil}, {Suresh}, {Sutton}, {Swinkels}, {Szczepa{\'n}czyk}, {Tacca}, {Tait}, {Talbot}, {Talukder}, {Tanner}, {T{\'a}pai}, {Taracchini}, {Tasson}, {Taylor}, {Taylor}, {Tewari}, {Theeg}, {Thies}, {Thomas}, {Thomas}, {Thomas}, {Thorne}, {Thorne}, {Thrane}, {Tiwari}, {Tiwari}, {Tokmakov}, {Toland}, {Tonelli}, {Tornasi}, {Torres-Forn{\'e}}, {Torrie}, {T{\"o}yr{\"a}}, {Travasso}, {Traylor}, {Trinastic}, {Tringali}, {Trozzo}, {Tsang}, {Tse}, {Tso}, {Tsukada}, {Tsuna}, {Tuyenbayev}, {Ueno}, {Ugolini}, {Unnikrishnan}, {Urban}, {Usman}, {Vahlbruch}, {Vajente}, {Valdes}, {Vallisneri}, {van Bakel}, {van Beuzekom}, {van den Brand}, {Van Den Broeck}, {Vander-Hyde}, {van der Schaaf}, {van Heijningen}, {van Veggel}, {Vardaro}, {Varma}, {Vass},
  {Vas{\'u}th}, {Vecchio}, {Vedovato}, {Veitch}, {Veitch}, {Venkateswara}, {Venugopalan}, {Verkindt}, {Vetrano}, {Vicer{\'e}}, {Viets}, {Vinciguerra}, {Vine}, {Vinet}, {Vitale}, {Vo}, {Vocca}, {Vorvick}, {Vyatchanin}, {Wade}, {Wade}, {Wade}, {Walet}, {Walker}, {Wallace}, {Walsh}, {Wang}, {Wang}, {Wang}, {Wang}, {Wang}, {Ward}, {Warner}, {Was}, {Watchi}, {Weaver}, {Wei}, {Weinert}, {Weinstein}, {Weiss}, {Wen}, {Wessel}, {We{\ss}els}, {Westerweck}, {Westphal}, {Wette}, {Whelan}, {Whitcomb}, {Whiting}, {Whittle}, {Wilken}, {Williams}, {Williams}, {Williamson}, {Willis}, {Willke}, {Wimmer}, {Winkler}, {Wipf}, {Wittel}, {Woan}, {Woehler}, {Wofford}, {Wong}, {Worden}, {Wright}, {Wu}, {Wysocki}, {Xiao}, {Yamamoto}, {Yancey}, {Yang}, {Yap}, {Yazback}, {Yu}, {Yu}, {Yvert}, {Zadro{\.Z}ny}, {Zanolin}, {Zelenova}, {Zendri}, {Zevin}, {Zhang}, {Zhang}, {Zhang}, {Zhang}, {Zhao}, {Zhou}, {Zhou}, {Zhu}, {Zhu}, {Zimmerman}, {Zucker}, {Zweizig}, {LIGO Scientific Collaboration}, \& {Virgo Collaboration}}]{2017PhRvL.119p1101A}
{Abbott}, B.~P., {Abbott}, R., {Abbott}, T.~D., {et~al.} 2017{\natexlab{b}}, \prl, 119, 161101

\bibitem[{{Abbott} {et~al.}(2021{\natexlab{a}}){Abbott}, {Abbott}, {Abraham}, {Acernese}, {Ackley}, {Adams}, {Adams}, {Adhikari}, {Adya}, {Affeldt}, {Agarwal}, {Agathos}, {Agatsuma}, {Aggarwal}, {Aguiar}, {Aiello}, {Ain}, {Ajith}, {Akutsu}, {Aleman}, {Allen}, {Allocca}, {Altin}, {Amato}, {Anand}, {Ananyeva}, {Anderson}, {Anderson}, {Ando}, {Angelova}, {Ansoldi}, {Antelis}, {Antier}, {Appert}, {Arai}, {Arai}, {Arai}, {Araki}, {Araya}, {Araya}, {Areeda}, {Ar{\`e}ne}, {Aritomi}, {Arnaud}, {Aronson}, {Arun}, {Asada}, {Asali}, {Ashton}, {Aso}, {Aston}, {Astone}, {Aubin}, {Aufmuth}, {Aultoneal}, {Austin}, {Babak}, {Badaracco}, {Bader}, {Bae}, {Bae}, {Baer}, {Bagnasco}, {Bai}, {Baiotti}, {Baird}, {Bajpai}, {Ball}, {Ballardin}, {Ballmer}, {Bals}, {Balsamo}, {Baltus}, {Banagiri}, {Bankar}, {Bankar}, {Barayoga}, {Barbieri}, {Barish}, {Barker}, {Barneo}, {Barone}, {Barr}, {Barsotti}, {Barsuglia}, {Barta}, {Bartlett}, {Barton}, {Bartos}, {Bassiri}, {Basti}, {Bawaj}, {Bayley}, {Baylor}, {Bazzan}, {B{\'e}csy},
  {Bedakihale}, {Bejger}, {Belahcene}, {Benedetto}, {Beniwal}, {Benjamin}, {Benkel}, {Bennett}, {Bentley}, {Benyaala}, {Bergamin}, {Berger}, {Bernuzzi}, {Berry}, {Bersanetti}, {Bertolini}, {Betzwieser}, {Bhandare}, {Bhandari}, {Bhattacharjee}, {Bhaumik}, {Bidler}, {Bilenko}, {Billingsley}, {Birney}, {Birnholtz}, {Biscans}, {Bischi}, {Biscoveanu}, {Bisht}, {Biswas}, {Bitossi}, {Bizouard}, {Blackburn}, {Blackman}, {Blair}, {Blair}, {Blair}, {Bobba}, {Bode}, {Boer}, {Bogaert}, {Boldrini}, {Bondu}, {Bonilla}, {Bonnand}, {Booker}, {Boom}, {Bork}, {Boschi}, {Bose}, {Bose}, {Bossilkov}, {Boudart}, {Bouffanais}, {Bozzi}, {Bradaschia}, {Brady}, {Bramley}, {Branch}, {Branchesi}, {Brau}, {Breschi}, {Briant}, {Briggs}, {Brillet}, {Brinkmann}, {Brockill}, {Brooks}, {Brooks}, {Brown}, {Brunett}, {Bruno}, {Bruntz}, {Bryant}, {Buikema}, {Bulik}, {Bulten}, {Buonanno}, {Buscicchio}, {Buskulic}, {Byer}, {Cadonati}, {Caesar}, {Cagnoli}, {Cahillane}, {Cain}, {Calder{\'o}n Bustillo}, {Callaghan}, {Callister}, {Calloni}, {Camp},
  {Canepa}, {Cannavacciuolo}, {Cannon}, {Cao}, {Cao}, {Cao}, {Capocasa}, {Capote}, {Carapella}, {Carbognani}, {Carlin}, {Carney}, {Carpinelli}, {Carullo}, {Carver}, {Casanueva Diaz}, {Casentini}, {Castaldi}, {Caudill}, {Cavagli{\`a}}, {Cavalier}, {Cavalieri}, {Cella}, {Cerd{\'a}-Dur{\'a}n}, {Cesarini}, {Chaibi}, {Chakravarti}, {Champion}, {Chan}, {Chan}, {Chan}, {Chan}, {Chandra}, {Chanial}, {Chao}, {Charlton}, {Chase}, {Chassande-Mottin}, {Chatterjee}, {Chaturvedi}, {Chatziioannou}, {Chen}, {Chen}, {Chen}, {Chen}, {Chen}, {Chen}, {Chen}, {Chen}, {Chen}, {Cheng}, {Cheong}, {Cheung}, {Chia}, {Chiadini}, {Chiang}, {Chierici}, {Chincarini}, {Chiofalo}, {Chiummo}, {Cho}, {Cho}, {Choate}, {Choudhary}, {Choudhary}, {Christensen}, {Chu}, {Chu}, {Chu}, {Chua}, {Chung}, {Ciani}, {Ciecielag}, {Cie{\'s}lar}, {Cifaldi}, {Ciobanu}, {Ciolfi}, {Cipriano}, {Cirone}, {Clara}, {Clark}, {Clark}, {Clarke}, {Clearwater}, {Clesse}, {Cleva}, {Coccia}, {Cohadon}, {Cohen}, {Cohen}, {Colleoni}, {Collette}, {Colpi}, {Compton},
  {Constancio}, {Conti}, {Cooper}, {Corban}, {Corbitt}, {Cordero-Carri{\'o}n}, {Corezzi}, {Corley}, {Cornish}, {Corre}, {Corsi}, {Cortese}, {Costa}, {Cotesta}, {Coughlin}, {Coughlin}, {Coulon}, {Countryman}, {Cousins}, {Couvares}, {Covas}, {Coward}, {Cowart}, {Coyne}, {Coyne}, {Creighton}, {Creighton}, {Criswell}, {Croquette}, {Crowder}, {Cudell}, {Cullen}, {Cumming}, {Cummings}, {Cuoco}, {Cury{\l}o}, {Dal Canton}, {D{\'a}lya}, {Dana}, {Daneshgaranbajastani}, {D'Angelo}, {Danilishin}, {D'Antonio}, {Danzmann}, {Darsow-Fromm}, {Dasgupta}, {Datrier}, {Dattilo}, {Dave}, {Davier}, {Davies}, {Davis}, {Daw}, {Dean}, {Debra}, {Deenadayalan}, {Degallaix}, {de Laurentis}, {Del{\'e}glise}, {Del Favero}, {de Lillo}, {de Lillo}, {Del Pozzo}, {Demarchi}, {de Matteis}, {D'Emilio}, {Demos}, {Dent}, {Depasse}, {de Pietri}, {De Rosa}, {de Rossi}, {Desalvo}, {de Simone}, {Dhurandhar}, {D{\'\i}az}, {Diaz-Ortiz}, {Didio}, {Dietrich}, {di Fiore}, {di Fronzo}, {di Giorgio}, {di Giovanni}, {di Girolamo}, {di Lieto}, {Ding}, {di
  Pace}, {di Palma}, {di Renzo}, {Divakarla}, {Dmitriev}, {Doctor}, {D'Onofrio}, {Donovan}, {Dooley}, {Doravari}, {Dorrington}, {Drago}, {Driggers}, {Drori}, {Du}, {Ducoin}, {Dupej}, {Durante}, {D'Urso}, {Duverne}, {Dwyer}, {Easter}, {Ebersold}, {Eddolls}, {Edelman}, {Edo}, {Edy}, {Effler}, {Eguchi}, {Eichholz}, {Eikenberry}, {Eisenmann}, {Eisenstein}, {Ejlli}, {Enomoto}, {Errico}, {Essick}, {Estell{\'e}s}, {Estevez}, {Etienne}, {Etzel}, {Evans}, {Evans}, {Ewing}, {Fafone}, {Fair}, {Fairhurst}, {Fan}, {Farah}, {Farinon}, {Farr}, {Farr}, {Farrow}, {Fauchon-Jones}, {Favata}, {Fays}, {Fazio}, {Feicht}, {Fejer}, {Feng}, {Fenyvesi}, {Ferguson}, {Fernandez-Galiana}, {Ferrante}, {Ferreira}, {Fidecaro}, {Figura}, {Fiori}, {Fishbach}, {Fisher}, {Fittipaldi}, {Fiumara}, {Flaminio}, {Floden}, {Flynn}, {Fong}, {Font}, {Fornal}, {Forsyth}, {Franke}, {Frasca}, {Frasconi}, {Frederick}, {Frei}, {Freise}, {Frey}, {Fritschel}, {Frolov}, {Fronz{\'e}}, {Fujii}, {Fujikawa}, {Fukunaga}, {Fukushima}, {Fulda}, {Fyffe}, {Gabbard},
  {Gadre}, {Gaebel}, {Gair}, {Gais}, {Galaudage}, {Gamba}, {Ganapathy}, {Ganguly}, {Gao}, {Gaonkar}, {Garaventa}, {Garc{\'\i}a-N{\'u}{\~n}ez}, {Garc{\'\i}a-Quir{\'o}s}, {Garufi}, {Gateley}, {Gaudio}, {Gayathri}, {Ge}, {Gemme}, {Gennai}, {George}, {Gergely}, {Gewecke}, {Ghonge}, {Ghosh}, {Ghosh}, {Ghosh}, {Ghosh}, {Ghosh}, {Giacomazzo}, {Giacoppo}, {Giaime}, {Giardina}, {Gibson}, {Gier}, {Giesler}, {Giri}, {Gissi}, {Glanzer}, {Gleckl}, {Godwin}, {Goetz}, {Goetz}, {Gohlke}, {Goncharov}, {Gonz{\'a}lez}, {Gopakumar}, {Gosselin}, {Gouaty}, {Grace}, {Grado}, {Granata}, {Granata}, {Grant}, {Gras}, {Grassia}, {Gray}, {Gray}, {Greco}, {Green}, {Green}, {Gretarsson}, {Gretarsson}, {Griffith}, {Griffiths}, {Griggs}, {Grignani}, {Grimaldi}, {Grimes}, {Grimm}, {Grote}, {Grunewald}, {Gruning}, {Guerrero}, {Guidi}, {Guimaraes}, {Guix{\'e}}, {Gulati}, {Guo}, {Guo}, {Gupta}, {Gupta}, {Gupta}, {Gustafson}, {Gustafson}, {Guzman}, {Ha}, {Haegel}, {Hagiwara}, {Haino}, {Halim}, {Hall}, {Hamilton}, {Hammond}, {Han}, {Haney},
  {Hanks}, {Hanna}, {Hannam}, {Hannuksela}, {Hansen}, {Hansen}, {Hanson}, {Harder}, {Hardwick}, {Haris}, {Harms}, {Harry}, {Harry}, {Hartwig}, {Hasegawa}, {Haskell}, {Hasskew}, {Haster}, {Hattori}, {Haughian}, {Hayakawa}, {Hayama}, {Hayes}, {Healy}, {Heidmann}, {Heintze}, {Heinze}, {Heinzel}, {Heitmann}, {Hellman}, {Hello}, {Helmling-Cornell}, {Hemming}, {Hendry}, {Heng}, {Hennes}, {Hennig}, {Hennig}, {Hernandez Vivanco}, {Heurs}, {Hild}, {Hill}, {Himemoto}, {Hinderer}, {Hines}, {Hiranuma}, {Hirata}, {Hirose}, {Ho}, {Hochheim}, {Hofman}, {Hohmann}, {Holgado}, {Holland}, {Hollows}, {Holmes}, {Holt}, {Holz}, {Hong}, {Hopkins}, {Hough}, {Howell}, {Hoy}, {Hoyland}, {Hreibi}, {Hsieh}, {Hsu}, {Huang}, {Huang}, {Huang}, {Huang}, {Huang}, {Huang}, {H{\"u}bner}, {Huddart}, {Huerta}, {Hughey}, {Hui}, {Hui}, {Husa}, {Huttner}, {Huxford}, {Huynh-Dinh}, {Ide}, {Idzkowski}, {Iess}, {Ikenoue}, {Imam}, {Inayoshi}, {Inchauspe}, {Ingram}, {Inoue}, {Intini}, {Ioka}, {Isi}, {Isleif}, {Ito}, {Itoh}, {Iyer}, {Izumi},
  {Jaberianhamedan}, {Jacqmin}, {Jadhav}, {Jadhav}, {James}, {Jan}, {Jani}, {Janssens}, {Janthalur}, {Jaranowski}, {Jariwala}, {Jaume}, {Jenkins}, {Jeon}, {Jeunon}, {Jia}, {Jiang}, {Jin}, {Johns}, {Jones}, {Jones}, {Jones}, {Jones}, {Jones}, {Jonker}, {Ju}, {Jung}, {Jung}, {Junker}, {Kaihotsu}, {Kajita}, {Kakizaki}, {Kalaghatgi}, {Kalogera}, {Kamai}, {Kamiizumi}, {Kanda}, {Kandhasamy}, {Kang}, {Kanner}, {Kao}, {Kapadia}, {Kapasi}, {Karat}, {Karathanasis}, {Karki}, {Kashyap}, {Kasprzack}, {Kastaun}, {Katsanevas}, {Katsavounidis}, {Katzman}, {Kaur}, {Kawabe}, {Kawaguchi}, {Kawai}, {Kawasaki}, {K{\'e}f{\'e}lian}, {Keitel}, {Key}, {Khadka}, {Khalili}, {Khan}, {Khan}, {Khazanov}, {Khetan}, {Khursheed}, {Kijbunchoo}, {Kim}, {Kim}, {Kim}, {Kim}, {Kim}, {Kim}, {Kimball}, {Kimura}, {King}, {Kinley-Hanlon}, {Kirchhoff}, {Kissel}, {Kita}, {Kitazawa}, {Kleybolte}, {Klimenko}, {Knee}, {Knowles}, {Knyazev}, {Koch}, {Koekoek}, {Kojima}, {Kokeyama}, {Koley}, {Kolitsidou}, {Kolstein}, {Komori}, {Kondrashov}, {Kong}, {Kontos},
  {Koper}, {Korobko}, {Kotake}, {Kovalam}, {Kozak}, {Kozakai}, {Kozu}, {Kringel}, {Krishnendu}, {Kr{\'o}lak}, {Kuehn}, {Kuei}, {Kumar}, {Kumar}, {Kumar}, {Kumar}, {Kume}, {Kuns}, {Kuo}, {Kuo}, {Kuromiya}, {Kuroyanagi}, {Kusayanagi}, {Kwak}, {Kwang}, {Laghi}, {Lalande}, {Lam}, {Lamberts}, {Landry}, {Landry}, {Lane}, {Lang}, {Lange}, {Lantz}, {La Rosa}, {Lartaux-Vollard}, {Lasky}, {Laxen}, {Lazzarini}, {Lazzaro}, {Leaci}, {Leavey}, {Lecoeuche}, {Lee}, {Lee}, {Lee}, {Lee}, {Lee}, {Lee}, {Lehmann}, {Lema{\^\i}tre}, {Leon}, {Leonardi}, {Leroy}, {Letendre}, {Levin}, {Leviton}, {Li}, {Li}, {Li}, {Li}, {Li}, {Li}, {Lin}, {Lin}, {Lin}, {Lin}, {Lin}, {Linde}, {Linker}, {Linley}, {Littenberg}, {Liu}, {Liu}, {Liu}, {Liu}, {Llorens-Monteagudo}, {Lo}, {Lockwood}, {Lollie}, {London}, {Longo}, {Lopez}, {Lorenzini}, {Loriette}, {Lormand}, {Losurdo}, {Lough}, {Lousto}, {Lovelace}, {L{\"u}ck}, {Lumaca}, {Lundgren}, {Luo}, {Macas}, {Macinnis}, {MacLeod}, {MacMillan}, {Macquet}, {Maga{\~n}a Hernandez}, {Maga{\~n}a-Sandoval},
  {Magazz{\`u}}, {Magee}, {Maggiore}, {Majorana}, {Makarem}, {Maksimovic}, {Maliakal}, {Malik}, {Man}, {Mandic}, {Mangano}, {Mango}, {Mansell}, {Manske}, {Mantovani}, {Mapelli}, {Marchesoni}, {Marchio}, {Marion}, {Mark}, {M{\'a}rka}, {M{\'a}rka}, {Markakis}, {Markosyan}, {Markowitz}, {Maros}, {Marquina}, {Marsat}, {Martelli}, {Martin}, {Martin}, {Martinez}, {Martinez}, {Martinovic}, {Martynov}, {Marx}, {Masalehdan}, {Mason}, {Massera}, {Masserot}, {Massinger}, {Masso-Reid}, {Mastrogiovanni}, {Matas}, {Mateu-Lucena}, {Matichard}, {Matiushechkina}, {Mavalvala}, {McCann}, {McCarthy}, {McClelland}, {McClincy}, {McCormick}, {McCuller}, {McGhee}, {McGuire}, {McIsaac}, {McIver}, {McManus}, {McRae}, {McWilliams}, {Meacher}, {Mehmet}, {Mehta}, {Melatos}, {Melchor}, {Mendell}, {Menendez-Vazquez}, {Menoni}, {Mercer}, {Mereni}, {Merfeld}, {Merilh}, {Merritt}, {Merzougui}, {Meshkov}, {Messenger}, {Messick}, {Meyers}, {Meylahn}, {Mhaske}, {Miani}, {Miao}, {Michaloliakos}, {Michel}, {Michimura}, {Middleton}, {Milano},
  {Miller}, {Millhouse}, {Mills}, {Milotti}, {Milovich-Goff}, {Minazzoli}, {Minenkov}, {Mio}, {Mir}, {Mishkin}, {Mishra}, {Mishra}, {Mistry}, {Mitra}, {Mitrofanov}, {Mitselmakher}, {Mittleman}, {Miyakawa}, {Miyamoto}, {Miyazaki}, {Miyo}, {Miyoki}, {Mo}, {Mogushi}, {Mohapatra}, {Mohite}, {Molina}, {Molina-Ruiz}, {Mondin}, {Montani}, {Moore}, {Moraru}, {Morawski}, {More}, {Moreno}, {Moreno}, {Mori}, {Morisaki}, {Moriwaki}, {Mours}, {Mow-Lowry}, {Mozzon}, {Muciaccia}, {Mukherjee}, {Mukherjee}, {Mukherjee}, {Mukherjee}, {Mukund}, {Mullavey}, {Munch}, {Mu{\~n}iz}, {Murray}, {Musenich}, {Nadji}, {Nagano}, {Nagano}, {Nagar}, {Nakamura}, {Nakano}, {Nakano}, {Nakashima}, {Nakayama}, {Nardecchia}, {Narikawa}, {Naticchioni}, {Nayak}, {Nayak}, {Negishi}, {Neil}, {Neilson}, {Nelemans}, {Nelson}, {Nery}, {Neunzert}, {Ng}, {Ng}, {Nguyen}, {Nguyen}, {Nguyen}, {Nguyen Quynh}, {Ni}, {Nichols}, {Nishizawa}, {Nissanke}, {Nocera}, {Noh}, {Norman}, {North}, {Nozaki}, {Nuttall}, {Oberling}, {O'Brien}, {Obuchi}, {O'Dell}, {Ogaki},
  {Oganesyan}, {Oh}, {Oh}, {Oh}, {Ohashi}, {Ohishi}, {Ohkawa}, {Ohme}, {Ohta}, {Okada}, {Okutani}, {Okutomi}, {Olivetto}, {Oohara}, {Ooi}, {Oram}, {O'Reilly}, {Ormiston}, {Ormsby}, {Ortega}, {O'Shaughnessy}, {O'Shea}, {Oshino}, {Ossokine}, {Osthelder}, {Otabe}, {Ottaway}, {Overmier}, {Pace}, {Pagano}, {Page}, {Pagliaroli}, {Pai}, {Pai}, {Palamos}, {Palashov}, {Palomba}, {Pan}, {Panda}, {Pang}, {Pang}, {Pankow}, {Pannarale}, {Pant}, {Paoletti}, {Paoli}, {Paolone}, {Parisi}, {Park}, {Parker}, {Pascucci}, {Pasqualetti}, {Passaquieti}, {Passuello}, {Patel}, {Patricelli}, {Payne}, {Pechsiri}, {Pedraza}, {Pegoraro}, {Pele}, {Pe{\~n}a Arellano}, {Penn}, {Perego}, {Pereira}, {Pereira}, {Perez}, {P{\'e}rigois}, {Perreca}, {Perri{\`e}s}, {Petermann}, {Petterson}, {Pfeiffer}, {Pham}, {Phukon}, {Piccinni}, {Pichot}, {Piendibene}, {Piergiovanni}, {Pierini}, {Pierro}, {Pillant}, {Pilo}, {Pinard}, {Pinto}, {Piotrzkowski}, {Piotrzkowski}, {Pirello}, {Pitkin}, {Placidi}, {Plastino}, {Pluchar}, {Poggiani}, {Polini}, {Pong},
  {Ponrathnam}, {Popolizio}, {Porter}, {Powell}, {Pracchia}, {Pradier}, {Prajapati}, {Prasai}, {Prasanna}, {Pratten}, {Prestegard}, {Principe}, {Prodi}, {Prokhorov}, {Prosposito}, {Prudenzi}, {Puecher}, {Punturo}, {Puosi}, {Puppo}, {P{\"u}rrer}, {Qi}, {Quetschke}, {Quinonez}, {Quitzow-James}, {Raab}, {Raaijmakers}, {Radkins}, {Radulesco}, {Raffai}, {Rail}, {Raja}, {Rajan}, {Ramirez}, {Ramirez}, {Ramos-Buades}, {Rana}, {Rapagnani}, {Rapol}, {Ratto}, {Ray}, {Raymond}, {Raza}, {Razzano}, {Read}, {Rees}, {Regimbau}, {Rei}, {Reid}, {Reitze}, {Relton}, {Rettegno}, {Ricci}, {Richardson}, {Richardson}, {Richardson}, {Ricker}, {Riemenschneider}, {Riles}, {Rizzo}, {Robertson}, {Robie}, {Robinet}, {Rocchi}, {Rocha}, {Rodriguez}, {Rodriguez-Soto}, {Rolland}, {Rollins}, {Roma}, {Romanelli}, {Romano}, {Romel}, {Romero}, {Romero-Shaw}, {Romie}, {Rose}, {Rosi{\'n}ska}, {Rosofsky}, {Ross}, {Rowan}, {Rowlinson}, {Roy}, {Roy}, {Rozza}, {Ruggi}, {Ryan}, {Sachdev}, {Sadecki}, {Sadiq}, {Sago}, {Saito}, {Saito}, {Sakai}, {Sakai},
  {Sakellariadou}, {Sakuno}, {Salafia}, {Salconi}, {Saleem}, {Salemi}, {Samajdar}, {Sanchez}, {Sanchez}, {Sanchez}, {Sanchis-Gual}, {Sanders}, {Sanuy}, {Saravanan}, {Sarin}, {Sassolas}, {Satari}, {Sathyaprakash}, {Sato}, {Sato}, {Sauter}, {Savage}, {Savant}, {Sawada}, {Sawant}, {Sawant}, {Sayah}, {Schaetzl}, {Scheel}, {Scheuer}, {Schindler-Tyka}, {Schmidt}, {Schnabel}, {Schneewind}, {Schofield}, {Sch{\"o}nbeck}, {Schulte}, {Schutz}, {Schwartz}, {Scott}, {Scott}, {Seglar-Arroyo}, {Seidel}, {Sekiguchi}, {Sekiguchi}, {Sellers}, {Sengupta}, {Sennett}, {Sentenac}, {Seo}, {Sequino}, {Sergeev}, {Setyawati}, {Shaffer}, {Shahriar}, {Shams}, {Shao}, {Sharifi}, {Sharma}, {Sharma}, {Shawhan}, {Shcheblanov}, {Shen}, {Shibagaki}, {Shikauchi}, {Shimizu}, {Shimoda}, {Shimode}, {Shink}, {Shinkai}, {Shishido}, {Shoda}, {Shoemaker}, {Shoemaker}, {Shukla}, {Shyamsundar}, {Sieniawska}, {Sigg}, {Singer}, {Singh}, {Singh}, {Singha}, {Sintes}, {Sipala}, {Skliris}, {Slagmolen}, {Slaven-Blair}, {Smetana}, {Smith}, {Smith}, {Somala},
  {Somiya}, {Son}, {Soni}, {Soni}, {Sorazu}, {Sordini}, {Sorrentino}, {Sorrentino}, {Sotani}, {Soulard}, {Souradeep}, {Sowell}, {Spagnuolo}, {Spencer}, {Spera}, {Srivastava}, {Srivastava}, {Staats}, {Stachie}, {Steer}, {Steinlechner}, {Steinlechner}, {Stops}, {Stevenson}, {Stover}, {Strain}, {Strang}, {Stratta}, {Strunk}, {Sturani}, {Stuver}, {S{\"u}dbeck}, {Sudhagar}, {Sudhir}, {Sugimoto}, {Suh}, {Summerscales}, {Sun}, {Sun}, {Sunil}, {Sur}, {Suresh}, {Sutton}, {Suzuki}, {Suzuki}, {Swinkels}, {Szczepa{\'n}czyk}, {Szewczyk}, {Tacca}, {Tagoshi}, {Tait}, {Takahashi}, {Takahashi}, {Takamori}, {Takano}, {Takeda}, {Takeda}, {Talbot}, {Tanaka}, {Tanaka}, {Tanaka}, {Tanaka}, {Tanaka}, {Tanasijczuk}, {Tanioka}, {Tanner}, {Tao}, {Tapia}, {Tapia San Martin}, {Tasson}, {Telada}, {Tenorio}, {Terkowski}, {Test}, {Thirugnanasambandam}, {Thomas}, {Thomas}, {Thompson}, {Thondapu}, {Thorne}, {Thrane}, {Tiwari}, {Tiwari}, {Tiwari}, {Toland}, {Tolley}, {Tomaru}, {Tomigami}, {Tomura}, {Tonelli}, {Torres-Forn{\'e}}, {Torrie},
  {Tosta E Melo}, {T{\"o}yr{\"a}}, {Trapananti}, {Travasso}, {Traylor}, {Tringali}, {Tripathee}, {Troiano}, {Trovato}, {Trozzo}, {Trudeau}, {Tsai}, {Tsai}, {Tsang}, {Tsang}, {Tsao}, {Tse}, {Tso}, {Tsubono}, {Tsuchida}, {Tsukada}, {Tsuna}, {Tsutsui}, {Tsuzuki}, {Turconi}, {Tuyenbayev}, {Ubhi}, {Uchikata}, {Uchiyama}, {Udall}, {Ueda}, {Uehara}, {Ueno}, {Ueshima}, {Ugolini}, {Unnikrishnan}, {Uraguchi}, {Urban}, {Ushiba}, {Usman}, {Utina}, {Vahlbruch}, {Vajente}, {Vajpeyi}, {Valdes}, {Valentini}, {Valsan}, {van Bakel}, {van Beuzekom}, {van den Brand}, {van den Broeck}, {Vander-Hyde}, {van der Schaaf}, {van Heijningen}, {Vanosky}, {van Putten}, {Vardaro}, {Vargas}, {Varma}, {Vas{\'u}th}, {Vecchio}, {Vedovato}, {Veitch}, {Veitch}, {Venkateswara}, {Venneberg}, {Venugopalan}, {Verkindt}, {Verma}, {Veske}, {Vetrano}, {Vicer{\'e}}, {Viets}, {Villa-Ortega}, {Vinet}, {Vitale}, {Vo}, {Vocca}, {von Reis}, {von Wrangel}, {Vorvick}, {Vyatchanin}, {Wade}, {Wade}, {Wagner}, {Walet}, {Walker}, {Wallace}, {Wallace}, {Walsh},
  {Wang}, {Wang}, {Wang}, {Ward}, {Warner}, {Was}, {Washimi}, {Washington}, {Watchi}, {Weaver}, {Wei}, {Weinert}, {Weinstein}, {Weiss}, {Weller}, {Wellmann}, {Wen}, {We{\ss}els}, {Westhouse}, {Wette}, {Whelan}, {White}, {Whiting}, {Whittle}, {Wilken}, {Williams}, {Williams}, {Williamson}, {Willis}, {Willke}, {Wilson}, {Winkler}, {Wipf}, {Wlodarczyk}, {Woan}, {Woehler}, {Wofford}, {Wong}, {Wu}, {Wu}, {Wu}, {Wu}, {Wysocki}, {Xiao}, {Xu}, {Yamada}, {Yamamoto}, {Yamamoto}, {Yamamoto}, {Yamamoto}, {Yamashita}, {Yamazaki}, {Yang}, {Yang}, {Yang}, {Yang}, {Yang}, {Yap}, {Yeeles}, {Yelikar}, {Ying}, {Yokogawa}, {Yokoyama}, {Yokozawa}, {Yoon}, {Yoshioka}, {Yu}, {Yu}, {Yuzurihara}, {Zadro{\.z}ny}, {Zanolin}, {Zappa}, {Zeidler}, {Zelenova}, {Zendri}, {Zevin}, {Zhan}, {Zhang}, {Zhang}, {Zhang}, {Zhang}, {Zhang}, {Zhao}, {Zhao}, {Zhao}, {Zhao}, {Zhou}, {Zhu}, {Zhu}, {Zimmerman}, {Zlochower}, {Zucker}, {Zweizig}, {Ligo Scientific Collaboration}, {VIRGO Collaboration}, \& {KAGRA Collaboration}}]{2021ApJ...915L...5A}
{Abbott}, R., {Abbott}, T.~D., {Abraham}, S., {et~al.} 2021{\natexlab{a}}, \apjl, 915, L5

\bibitem[{{Abbott} {et~al.}(2021{\natexlab{b}}){Abbott}, {Abbott}, {Abraham}, {Acernese}, {Ackley}, {Adams}, {Adams}, {Adhikari}, {Adya}, {Affeldt}, {Agathos}, {Agatsuma}, {Aggarwal}, {Aguiar}, {Aiello}, {Ain}, {Ajith}, {Akcay}, {Allen}, {Allocca}, {Altin}, {Amato}, {Anand}, {Ananyeva}, {Anderson}, {Anderson}, {Angelova}, {Ansoldi}, {Antelis}, {Antier}, {Appert}, {Arai}, {Araya}, {Areeda}, {Ar{\`e}ne}, {Arnaud}, {Aronson}, {Arun}, {Asali}, {Ascenzi}, {Ashton}, {Aston}, {Astone}, {Aubin}, {Aufmuth}, {AultONeal}, {Austin}, {Avendano}, {Babak}, {Badaracco}, {Bader}, {Bae}, {Baer}, {Bagnasco}, {Baird}, {Ball}, {Ballardin}, {Ballmer}, {Bals}, {Balsamo}, {Baltus}, {Banagiri}, {Bankar}, {Bankar}, {Barayoga}, {Barbieri}, {Barish}, {Barker}, {Barneo}, {Barnum}, {Barone}, {Barr}, {Barsotti}, {Barsuglia}, {Barta}, {Bartlett}, {Bartos}, {Bassiri}, {Basti}, {Bawaj}, {Bayley}, {Bazzan}, {Becher}, {B{\'e}csy}, {Bedakihale}, {Bejger}, {Belahcene}, {Beniwal}, {Benjamin}, {Bennett}, {Bentley}, {Bergamin}, {Berger}, {Bergmann},
  {Bernuzzi}, {Berry}, {Bersanetti}, {Bertolini}, {Betzwieser}, {Bhandare}, {Bhandari}, {Bhattacharjee}, {Bidler}, {Bilenko}, {Billingsley}, {Birney}, {Birnholtz}, {Biscans}, {Bischi}, {Biscoveanu}, {Bisht}, {Bitossi}, {Bizouard}, {Blackburn}, {Blackman}, {Blair}, {Blair}, {Blair}, {Blanch}, {Bobba}, {Bode}, {Boer}, {Boetzel}, {Bogaert}, {Boldrini}, {Bondu}, {Bonilla}, {Bonnand}, {Booker}, {Boom}, {Bork}, {Boschi}, {Bose}, {Bossilkov}, {Boudart}, {Bouffanais}, {Bozzi}, {Bradaschia}, {Brady}, {Bramley}, {Branchesi}, {Brau}, {Breschi}, {Briant}, {Briggs}, {Brighenti}, {Brillet}, {Brinkmann}, {Brockill}, {Brooks}, {Brooks}, {Brown}, {Brunett}, {Bruno}, {Bruntz}, {Buikema}, {Bulik}, {Bulten}, {Buonanno}, {Buscicchio}, {Buskulic}, {Byer}, {Cabero}, {Cadonati}, {Caesar}, {Cagnoli}, {Cahillane}, {Calder{\'o}n Bustillo}, {Callaghan}, {Callister}, {Calloni}, {Camp}, {Canepa}, {Cannon}, {Cao}, {Cao}, {Carapella}, {Carbognani}, {Carney}, {Carpinelli}, {Carullo}, {Carver}, {Casanueva Diaz}, {Casentini}, {Caudill},
  {Cavagli{\`a}}, {Cavalier}, {Cavalieri}, {Cella}, {Cerd{\'a}-Dur{\'a}n}, {Cesarini}, {Chaibi}, {Chakravarti}, {Chan}, {Chan}, {Chandra}, {Chanial}, {Chao}, {Charlton}, {Chase}, {Chassande-Mottin}, {Chatterjee}, {Chattopadhyay}, {Chaturvedi}, {Chatziioannou}, {Chen}, {Chen}, {Chen}, {Chen}, {Cheng}, {Cheong}, {Chia}, {Chiadini}, {Chierici}, {Chincarini}, {Chiummo}, {Cho}, {Cho}, {Cho}, {Choate}, {Christensen}, {Chu}, {Chua}, {Chung}, {Chung}, {Ciani}, {Ciecielag}, {Cie{\'s}lar}, {Cifaldi}, {Ciobanu}, {Ciolfi}, {Cipriano}, {Cirone}, {Clara}, {Clark}, {Clark}, {Clarke}, {Clearwater}, {Clesse}, {Cleva}, {Coccia}, {Cohadon}, {Cohen}, {Colleoni}, {Collette}, {Collins}, {Colpi}, {Constancio}, {Conti}, {Cooper}, {Corban}, {Corbitt}, {Cordero-Carri{\'o}n}, {Corezzi}, {Corley}, {Cornish}, {Corre}, {Corsi}, {Cortese}, {Costa}, {Cotesta}, {Coughlin}, {Coughlin}, {Coulon}, {Countryman}, {Cousins}, {Couvares}, {Covas}, {Coward}, {Cowart}, {Coyne}, {Coyne}, {Creighton}, {Creighton}, {Croquette}, {Crowder}, {Cudell},
  {Cullen}, {Cumming}, {Cummings}, {Cunningham}, {Cuoco}, {Cury{\l}o}, {Canton}, {D{\'a}lya}, {Dana}, {DaneshgaranBajastani}, {D'Angelo}, {Danila}, {Danilishin}, {D'Antonio}, {Danzmann}, {Darsow-Fromm}, {Dasgupta}, {Datrier}, {Dattilo}, {Dave}, {Davier}, {Davies}, {Davis}, {Daw}, {Dean}, {DeBra}, {Deenadayalan}, {Degallaix}, {De Laurentis}, {Del{\'e}glise}, {Del Favero}, {De Lillo}, {De Lillo}, {Del Pozzo}, {DeMarchi}, {De Matteis}, {D'Emilio}, {Demos}, {Denker}, {Dent}, {Depasse}, {De Pietri}, {De Rosa}, {De Rossi}, {DeSalvo}, {de Varona}, {Dhurandhar}, {D{\'\i}az}, {Diaz-Ortiz}, {Didio}, {Dietrich}, {Di Fiore}, {DiFronzo}, {Di Giorgio}, {Di Giovanni}, {Di Giovanni}, {Di Girolamo}, {Di Lieto}, {Ding}, {Di Pace}, {Di Palma}, {Di Renzo}, {Divakarla}, {Dmitriev}, {Doctor}, {D'Onofrio}, {Donovan}, {Dooley}, {Doravari}, {Dorrington}, {Downes}, {Drago}, {Driggers}, {Du}, {Ducoin}, {Dupej}, {Durante}, {D'Urso}, {Duverne}, {Dwyer}, {Easter}, {Eddolls}, {Edelman}, {Edo}, {Edy}, {Effler}, {Eichholz}, {Eikenberry},
  {Eisenmann}, {Eisenstein}, {Ejlli}, {Errico}, {Essick}, {Estell{\'e}s}, {Estevez}, {Etienne}, {Etzel}, {Evans}, {Evans}, {Ewing}, {Fafone}, {Fair}, {Fairhurst}, {Fan}, {Farah}, {Farinon}, {Farr}, {Farr}, {Fauchon-Jones}, {Favata}, {Fays}, {Fazio}, {Feicht}, {Fejer}, {Feng}, {Fenyvesi}, {Ferguson}, {Fernandez-Galiana}, {Ferrante}, {Ferreira}, {Fidecaro}, {Figura}, {Fiori}, {Fiorucci}, {Fishbach}, {Fisher}, {Fishner}, {Fittipaldi}, {Fitz-Axen}, {Fiumara}, {Flaminio}, {Floden}, {Flynn}, {Fong}, {Font}, {Forsyth}, {Fournier}, {Frasca}, {Frasconi}, {Frei}, {Freise}, {Frey}, {Frey}, {Fritschel}, {Frolov}, {Fronz{\'e}}, {Fulda}, {Fyffe}, {Gabbard}, {Gadre}, {Gaebel}, {Gair}, {Gais}, {Galaudage}, {Gamba}, {Ganapathy}, {Ganguly}, {Gaonkar}, {Garaventa}, {Garc{\'\i}a-Quir{\'o}s}, {Garufi}, {Gateley}, {Gaudio}, {Gayathri}, {Gemme}, {Gennai}, {George}, {George}, {George}, {Gergely}, {Ghonge}, {Ghosh}, {Ghosh}, {Ghosh}, {Giacomazzo}, {Giacoppo}, {Giaime}, {Giardina}, {Gibson}, {Gier}, {Gill}, {Giri}, {Glanzer},
  {Gleckl}, {Godwin}, {Goetz}, {Goetz}, {Gohlke}, {Goncharov}, {Gonz{\'a}lez}, {Gopakumar}, {Gossan}, {Gosselin}, {Gouaty}, {Grace}, {Grado}, {Granata}, {Granata}, {Grant}, {Gras}, {Grassia}, {Gray}, {Gray}, {Greco}, {Green}, {Green}, {Gretarsson}, {Griggs}, {Grignani}, {Grimaldi}, {Grimes}, {Grimm}, {Grote}, {Grunewald}, {Gruning}, {Guerrero}, {Guidi}, {Guimaraes}, {Guix{\'e}}, {Gulati}, {Guo}, {Gupta}, {Gupta}, {Gupta}, {Gustafson}, {Gustafson}, {Guzman}, {Haegel}, {Halim}, {Hall}, {Hamilton}, {Hammond}, {Haney}, {Hanke}, {Hanks}, {Hanna}, {Hannam}, {Hannuksela}, {Hannuksela}, {Hansen}, {Hansen}, {Hanson}, {Harder}, {Hardwick}, {Haris}, {Harms}, {Harry}, {Harry}, {Hartwig}, {Hasskew}, {Haster}, {Haughian}, {Hayes}, {Healy}, {Heidmann}, {Heintze}, {Heinze}, {Heinzel}, {Heitmann}, {Hellman}, {Hello}, {Helmling-Cornell}, {Hemming}, {Hendry}, {Heng}, {Hennes}, {Hennig}, {Hennig}, {Hernandez Vivanco}, {Heurs}, {Hild}, {Hill}, {Hines}, {Hochheim}, {Hofgard}, {Hofman}, {Hohmann}, {Holgado}, {Holland}, {Hollows},
  {Holmes}, {Holt}, {Holz}, {Hopkins}, {Horst}, {Hough}, {Howell}, {Hoy}, {Hoyland}, {Huang}, {H{\"u}bner}, {Huddart}, {Huerta}, {Hughey}, {Hui}, {Husa}, {Huttner}, {Hutzler}, {Huxford}, {Huynh-Dinh}, {Idzkowski}, {Iess}, {Imperato}, {Inchauspe}, {Ingram}, {Intini}, {Isi}, {Iyer}, {JaberianHamedan}, {Jacqmin}, {Jadhav}, {Jadhav}, {James}, {Jani}, {Janssens}, {Janthalur}, {Jaranowski}, {Jariwala}, {Jaume}, {Jenkins}, {Jeunon}, {Jiang}, {Johns}, {Johnson-McDaniel}, {Jones}, {Jones}, {Jones}, {Jones}, {Jones}, {Jonker}, {Ju}, {Junker}, {Kalaghatgi}, {Kalogera}, {Kamai}, {Kandhasamy}, {Kang}, {Kanner}, {Kapadia}, {Kapasi}, {Karathanasis}, {Karki}, {Kashyap}, {Kasprzack}, {Kastaun}, {Katsanevas}, {Katsavounidis}, {Katzman}, {Kawabe}, {K{\'e}f{\'e}lian}, {Keitel}, {Key}, {Khadka}, {Khalili}, {Khan}, {Khan}, {Khazanov}, {Khetan}, {Khursheed}, {Kijbunchoo}, {Kim}, {Kim}, {Kim}, {Kim}, {Kim}, {Kim}, {Kimball}, {King}, {Kinley-Hanlon}, {Kirchhoff}, {Kissel}, {Kleybolte}, {Klimenko}, {Knowles}, {Knyazev}, {Koch},
  {Koehlenbeck}, {Koekoek}, {Koley}, {Kolstein}, {Komori}, {Kondrashov}, {Kontos}, {Koper}, {Korobko}, {Korth}, {Kovalam}, {Kozak}, {Kr{\"a}mer}, {Kringel}, {Krishnendu}, {Kr{\'o}lak}, {Kuehn}, {Kumar}, {Kumar}, {Kumar}, {Kumar}, {Kuns}, {Kwang}, {Lackey}, {Laghi}, {Lalande}, {Lam}, {Lamberts}, {Landry}, {Lane}, {Lang}, {Lange}, {Lantz}, {Lanza}, {La Rosa}, {Lartaux-Vollard}, {Lasky}, {Laxen}, {Lazzarini}, {Lazzaro}, {Leaci}, {Leavey}, {Lecoeuche}, {Lee}, {Lee}, {Lee}, {Lee}, {Lehmann}, {Leon}, {Leroy}, {Letendre}, {Levin}, {Li}, {Li}, {Li}, {Li}, {Li}, {Linde}, {Linker}, {Linley}, {Littenberg}, {Liu}, {Liu}, {Llorens-Monteagudo}, {Lo}, {Lockwood}, {London}, {Longo}, {Lorenzini}, {Loriette}, {Lormand}, {Losurdo}, {Lough}, {Lousto}, {Lovelace}, {L{\"u}ck}, {Lumaca}, {Lundgren}, {Ma}, {Macas}, {MacInnis}, {Macleod}, {MacMillan}, {Macquet}, {Maga{\~n}a Hernandez}, {Maga{\~n}a-Sandoval}, {Magazz{\`u}}, {Magee}, {Majorana}, {Maksimovic}, {Maliakal}, {Malik}, {Man}, {Mandic}, {Mangano}, {Mansell}, {Manske},
  {Mantovani}, {Mapelli}, {Marchesoni}, {Marion}, {M{\'a}rka}, {M{\'a}rka}, {Markakis}, {Markosyan}, {Markowitz}, {Maros}, {Marquina}, {Marsat}, {Martelli}, {Martin}, {Martin}, {Martinez}, {Martinez}, {Martynov}, {Masalehdan}, {Mason}, {Massera}, {Masserot}, {Massinger}, {Masso-Reid}, {Mastrogiovanni}, {Matas}, {Mateu-Lucena}, {Matichard}, {Matiushechkina}, {Mavalvala}, {Maynard}, {McCann}, {McCarthy}, {McClelland}, {McCormick}, {McCuller}, {McGuire}, {McIsaac}, {McIver}, {McManus}, {McRae}, {McWilliams}, {Meacher}, {Meadors}, {Mehmet}, {Mehta}, {Melatos}, {Melchor}, {Mendell}, {Menendez-Vazquez}, {Mercer}, {Mereni}, {Merfeld}, {Merilh}, {Merritt}, {Merzougui}, {Meshkov}, {Messenger}, {Messick}, {Metzdorff}, {Meyers}, {Meylahn}, {Mhaske}, {Miani}, {Miao}, {Michaloliakos}, {Michel}, {Middleton}, {Milano}, {Miller}, {Millhouse}, {Mills}, {Milotti}, {Milovich-Goff}, {Minazzoli}, {Minenkov}, {Mir}, {Mishkin}, {Mishra}, {Mistry}, {Mitra}, {Mitrofanov}, {Mitselmakher}, {Mittleman}, {Mo}, {Mogushi}, {Mohapatra},
  {Mohite}, {Molina}, {Molina-Ruiz}, {Mondin}, {Montani}, {Moore}, {Moraru}, {Morawski}, {Moreno}, {Morisaki}, {Mours}, {Mow-Lowry}, {Mozzon}, {Muciaccia}, {Mukherjee}, {Mukherjee}, {Mukherjee}, {Mukherjee}, {Mukund}, {Mullavey}, {Munch}, {Mu{\~n}iz}, {Murray}, {Nadji}, {Nagar}, {Nardecchia}, {Naticchioni}, {Nayak}, {Neil}, {Neilson}, {Nelemans}, {Nelson}, {Nery}, {Neunzert}, {Nitz}, {Ng}, {Ng}, {Nguyen}, {Nguyen}, {Nguyen}, {Nichols}, {Nissanke}, {Nocera}, {Noh}, {North}, {Nothard}, {Nuttall}, {Oberling}, {O'Brien}, {O'Dell}, {Oganesyan}, {Ogin}, {Oh}, {Oh}, {Ohme}, {Ohta}, {Okada}, {Olivetto}, {Oppermann}, {Oram}, {O'Reilly}, {Ormiston}, {Ortega}, {O'Shaughnessy}, {Ossokine}, {Osthelder}, {Ottaway}, {Overmier}, {Owen}, {Pace}, {Pagano}, {Page}, {Pagliaroli}, {Pai}, {Pai}, {Palamos}, {Palashov}, {Palomba}, {Pan}, {Panda}, {Pang}, {Pankow}, {Pannarale}, {Pant}, {Paoletti}, {Paoli}, {Paolone}, {Parker}, {Pascucci}, {Pasqualetti}, {Passaquieti}, {Passuello}, {Patel}, {Patricelli}, {Payne}, {Pechsiri},
  {Pedraza}, {Pegoraro}, {Pele}, {Penn}, {Perego}, {Perez}, {P{\'e}rigois}, {Perreca}, {Perri{\`e}s}, {Petermann}, {Petterson}, {Pfeiffer}, {Pham}, {Phukon}, {Piccinni}, {Pichot}, {Piendibene}, {Piergiovanni}, {Pierini}, {Pierro}, {Pillant}, {Pilo}, {Pinard}, {Pinto}, {Piotrzkowski}, {Pirello}, {Pitkin}, {Placidi}, {Plastino}, {Pluchar}, {Poggiani}, {Polini}, {Pong}, {Ponrathnam}, {Popolizio}, {Porter}, {Poverman}, {Powell}, {Pracchia}, {Prajapati}, {Prasai}, {Prasanna}, {Pratten}, {Prestegard}, {Principe}, {Prodi}, {Prokhorov}, {Prosposito}, {Prudenzi}, {Puecher}, {Punturo}, {Puosi}, {Puppo}, {P{\"u}rrer}, {Qi}, {Quetschke}, {Quinonez}, {Quitzow-James}, {Raab}, {Raaijmakers}, {Radkins}, {Radulesco}, {Raffai}, {Rafferty}, {Rail}, {Raja}, {Rajan}, {Rajbhandari}, {Rakhmanov}, {Ramirez}, {Ramirez}, {Ramos-Buades}, {Rana}, {Rao}, {Rapagnani}, {Rapol}, {Ratto}, {Raymond}, {Razzano}, {Read}, {Regimbau}, {Rei}, {Reid}, {Reitze}, {Rettegno}, {Ricci}, {Richardson}, {Richardson}, {Richardson}, {Ricker},
  {Riemenschneider}, {Riles}, {Rizzo}, {Robertson}, {Robinet}, {Rocchi}, {Rocha}, {Rodriguez}, {Rodriguez-Soto}, {Rolland}, {Rollins}, {Roma}, {Romanelli}, {Romano}, {Romel}, {Romero}, {Romero-Shaw}, {Romie}, {Ronchini}, {Rose}, {Rose}, {Rose}, {Rosell}, {Rosi{\'n}ska}, {Rosofsky}, {Ross}, {Rowan}, {Rowlinson}, {Roy}, {Roy}, {Ruggi}, {Ryan}, {Sachdev}, {Sadecki}, {Sadiq}, {Sakellariadou}, {Salafia}, {Salconi}, {Saleem}, {Samajdar}, {Sanchez}, {Sanchez}, {Sanchez}, {Sanchis-Gual}, {Sanders}, {Sandles}, {Santiago}, {Santos}, {Saravanan}, {Sarin}, {Sassolas}, {Sathyaprakash}, {Sauter}, {Savage}, {Savant}, {Sawant}, {Sayah}, {Schaetzl}, {Schale}, {Scheel}, {Scheuer}, {Schindler-Tyka}, {Schmidt}, {Schnabel}, {Schofield}, {Sch{\"o}nbeck}, {Schreiber}, {Schulte}, {Schutz}, {Schwarm}, {Schwartz}, {Scott}, {Scott}, {Seglar-Arroyo}, {Seidel}, {Sellers}, {Sengupta}, {Sennett}, {Sentenac}, {Sequino}, {Sergeev}, {Setyawati}, {Shaffer}, {Shahriar}, {Sharifi}, {Sharma}, {Sharma}, {Shawhan}, {Shen}, {Shikauchi}, {Shink},
  {Shoemaker}, {Shoemaker}, {Shukla}, {ShyamSundar}, {Sieniawska}, {Sigg}, {Singer}, {Singh}, {Singh}, {Singha}, {Singhal}, {Sintes}, {Sipala}, {Skliris}, {Slagmolen}, {Slaven-Blair}, {Smetana}, {Smith}, {Smith}, {Somala}, {Son}, {Soni}, {Soni}, {Sorazu}, {Sordini}, {Sorrentino}, {Sorrentino}, {Soulard}, {Souradeep}, {Sowell}, {Spencer}, {Spera}, {Srivastava}, {Srivastava}, {Staats}, {Stachie}, {Steer}, {Steinhoff}, {Steinke}, {Steinlechner}, {Steinlechner}, {Steinmeyer}, {Stevenson}, {Stolle-McAllister}, {Stops}, {Stover}, {Strain}, {Stratta}, {Strunk}, {Sturani}, {Stuver}, {S{\"u}dbeck}, {Sudhagar}, {Sudhir}, {Suh}, {Summerscales}, {Sun}, {Sun}, {Sunil}, {Sur}, {Suresh}, {Sutton}, {Swinkels}, {Szczepa{\'n}czyk}, {Tacca}, {Tait}, {Talbot}, {Tanasijczuk}, {Tanner}, {Tao}, {Tapia}, {Tapia San Martin}, {Tasson}, {Taylor}, {Tenorio}, {Terkowski}, {Thirugnanasambandam}, {Thomas}, {Thomas}, {Thomas}, {Thompson}, {Thondapu}, {Thorne}, {Thrane}, {Tiwari}, {Tiwari}, {Tiwari}, {Toland}, {Tolley}, {Tonelli}, {Tornasi},
  {Torres-Forn{\'e}}, {Torrie}, {e Melo}, {T{\"o}yr{\"a}}, {Tran}, {Trapananti}, {Travasso}, {Traylor}, {Tringali}, {Tripathee}, {Trovato}, {Trudeau}, {Tsai}, {Tsang}, {Tse}, {Tso}, {Tsukada}, {Tsuna}, {Tsutsui}, {Turconi}, {Ubhi}, {Udall}, {Ueno}, {Ugolini}, {Unnikrishnan}, {Urban}, {Usman}, {Utina}, {Vahlbruch}, {Vajente}, {Vajpeyi}, {Valdes}, {Valentini}, {Valsan}, {van Bakel}, {van Beuzekom}, {van den Brand}, {Van Den Broeck}, {Vander-Hyde}, {van der Schaaf}, {van Heijningen}, {Vardaro}, {Vargas}, {Varma}, {Vass}, {Vas{\'u}th}, {Vecchio}, {Vedovato}, {Veitch}, {Veitch}, {Venkateswara}, {Venneberg}, {Venugopalan}, {Verkindt}, {Verma}, {Veske}, {Vetrano}, {Vicer{\'e}}, {Viets}, {Vijaykumar}, {Villa-Ortega}, {Vinet}, {Vitale}, {Vo}, {Vocca}, {Vorvick}, {Vyatchanin}, {Wade}, {Wade}, {Wade}, {Walet}, {Walker}, {Wallace}, {Wallace}, {Walsh}, {Wang}, {Wang}, {Wang}, {Wang}, {Ward}, {Warner}, {Was}, {Washington}, {Watchi}, {Weaver}, {Wei}, {Weinert}, {Weinstein}, {Weiss}, {Wellmann}, {Wen}, {We{\ss}els},
  {Westhouse}, {Wette}, {Whelan}, {White}, {White}, {Whiting}, {Whittle}, {Wilken}, {Williams}, {Williams}, {Williamson}, {Willis}, {Willke}, {Wilson}, {Wimmer}, {Winkler}, {Wipf}, {Woan}, {Woehler}, {Wofford}, {Wong}, {Wrangel}, {Wright}, {Wu}, {Wysocki}, {Xiao}, {Yamamoto}, {Yang}, {Yang}, {Yang}, {Yap}, {Yeeles}, {Yoon}, {Yu}, {Yu}, {Yuen}, {Zadro{\.Z}ny}, {Zanolin}, {Zelenova}, {Zendri}, {Zevin}, {Zhang}, {Zhang}, {Zhang}, {Zhang}, {Zhao}, {Zhao}, {Zheng}, {Zhou}, {Zhou}, {Zhu}, {Zimmerman}, {Zlochower}, {Zucker}, {Zweizig}, {LIGO Scientific Collaboration}, \& {Virgo Collaboration}}]{2021PhRvX..11b1053A}
{Abbott}, R., {Abbott}, T.~D., {Abraham}, S., {et~al.} 2021{\natexlab{b}}, Physical Review X, 11, 021053

\bibitem[{{Abbott} {et~al.}(2020){Abbott}, {Abbott}, {Abraham}, {Acernese}, {Ackley}, {Adams}, {Adhikari}, {Adya}, {Affeldt}, {Agathos}, {Agatsuma}, {Aggarwal}, {Aguiar}, {Aich}, {Aiello}, {Ain}, {Ajith}, {Akcay}, {Allen}, {Allocca}, {Altin}, {Amato}, {Anand}, {Ananyeva}, {Anderson}, {Anderson}, {Angelova}, {Ansoldi}, {Antier}, {Appert}, {Arai}, {Araya}, {Areeda}, {Ar{\`e}ne}, {Arnaud}, {Aronson}, {Arun}, {Asali}, {Ascenzi}, {Ashton}, {Aston}, {Astone}, {Aubin}, {Aufmuth}, {AultONeal}, {Austin}, {Avendano}, {Babak}, {Bacon}, {Badaracco}, {Bader}, {Bae}, {Baer}, {Baird}, {Baldaccini}, {Ballardin}, {Ballmer}, {Bals}, {Balsamo}, {Baltus}, {Banagiri}, {Bankar}, {Bankar}, {Barayoga}, {Barbieri}, {Barish}, {Barker}, {Barkett}, {Barneo}, {Barone}, {Barr}, {Barsotti}, {Barsuglia}, {Barta}, {Bartlett}, {Bartos}, {Bassiri}, {Basti}, {Bawaj}, {Bayley}, {Bazzan}, {B{\'e}csy}, {Bejger}, {Belahcene}, {Bell}, {Beniwal}, {Benjamin}, {Benkel}, {Bentley}, {Bergamin}, {Berger}, {Bergmann}, {Bernuzzi}, {Berry}, {Bersanetti},
  {Bertolini}, {Betzwieser}, {Bhandare}, {Bhandari}, {Bidler}, {Biggs}, {Bilenko}, {Billingsley}, {Birney}, {Birnholtz}, {Biscans}, {Bischi}, {Biscoveanu}, {Bisht}, {Bissenbayeva}, {Bitossi}, {Bizouard}, {Blackburn}, {Blackman}, {Blair}, {Blair}, {Blair}, {Bobba}, {Bode}, {Boer}, {Boetzel}, {Bogaert}, {Bondu}, {Bonilla}, {Bonnand}, {Booker}, {Boom}, {Bork}, {Boschi}, {Bose}, {Bossilkov}, {Bosveld}, {Bouffanais}, {Bozzi}, {Bradaschia}, {Brady}, {Bramley}, {Branchesi}, {Brau}, {Breschi}, {Briant}, {Briggs}, {Brighenti}, {Brillet}, {Brinkmann}, {Brito}, {Brockill}, {Brooks}, {Brooks}, {Brown}, {Brunett}, {Bruno}, {Bruntz}, {Buikema}, {Bulik}, {Bulten}, {Buonanno}, {Buskulic}, {Byer}, {Cabero}, {Cadonati}, {Cagnoli}, {Cahillane}, {Bustillo}, {Callaghan}, {Callister}, {Calloni}, {Camp}, {Canepa}, {Cannon}, {Cao}, {Cao}, {Carapella}, {Carbognani}, {Caride}, {Carney}, {Carullo}, {Diaz}, {Casentini}, {Casta{\~n}eda}, {Caudill}, {Cavagli{\`a}}, {Cavalier}, {Cavalieri}, {Cella}, {Cerd{\'a}-Dur{\'a}n}, {Cesarini},
  {Chaibi}, {Chakravarti}, {Chan}, {Chan}, {Chao}, {Charlton}, {Chase}, {Chassande-Mottin}, {Chatterjee}, {Chaturvedi}, {Chatziioannou}, {Chen}, {Chen}, {Chen}, {Cheng}, {Cheong}, {Chia}, {Chiadini}, {Chierici}, {Chincarini}, {Chiummo}, {Cho}, {Cho}, {Cho}, {Christensen}, {Chu}, {Chua}, {Chung}, {Chung}, {Ciani}, {Ciecielag}, {Cie{\'s}lar}, {Ciobanu}, {Ciolfi}, {Cipriano}, {Cirone}, {Clara}, {Clark}, {Clearwater}, {Clesse}, {Cleva}, {Coccia}, {Cohadon}, {Cohen}, {Colleoni}, {Collette}, {Collins}, {Colpi}, {Constancio}, {Conti}, {Cooper}, {Corban}, {Corbitt}, {Cordero-Carri{\'o}n}, {Corezzi}, {Corley}, {Cornish}, {Corre}, {Corsi}, {Cortese}, {Costa}, {Cotesta}, {Coughlin}, {Coughlin}, {Coulon}, {Countryman}, {Couvares}, {Covas}, {Coward}, {Cowart}, {Coyne}, {Coyne}, {Creighton}, {Creighton}, {Cripe}, {Croquette}, {Crowder}, {Cudell}, {Cullen}, {Cumming}, {Cummings}, {Cunningham}, {Cuoco}, {Curylo}, {Canton}, {D{\'a}lya}, {Dana}, {Daneshgaran-Bajastani}, {D'Angelo}, {Danilishin}, {D'Antonio}, {Danzmann},
  {Darsow-Fromm}, {Dasgupta}, {Datrier}, {Dattilo}, {Dave}, {Davier}, {Davies}, {Davis}, {Daw}, {DeBra}, {Deenadayalan}, {Degallaix}, {De Laurentis}, {Del{\'e}glise}, {Delfavero}, {De Lillo}, {Del Pozzo}, {DeMarchi}, {D'Emilio}, {Demos}, {Dent}, {De Pietri}, {De Rosa}, {De Rossi}, {DeSalvo}, {de Varona}, {Dhurandhar}, {D{\'\i}az}, {Diaz-Ortiz}, {Dietrich}, {Di Fiore}, {Di Fronzo}, {Di Giorgio}, {Di Giovanni}, {Di Giovanni}, {Di Girolamo}, {Di Lieto}, {Ding}, {Di Pace}, {Di Palma}, {Di Renzo}, {Divakarla}, {Dmitriev}, {Doctor}, {Donovan}, {Dooley}, {Doravari}, {Dorrington}, {Downes}, {Drago}, {Driggers}, {Du}, {Ducoin}, {Dupej}, {Durante}, {D'Urso}, {Dwyer}, {Easter}, {Eddolls}, {Edelman}, {Edo}, {Edy}, {Effler}, {Ehrens}, {Eichholz}, {Eikenberry}, {Eisenmann}, {Eisenstein}, {Ejlli}, {Errico}, {Essick}, {Estelles}, {Estevez}, {Etienne}, {Etzel}, {Evans}, {Evans}, {Ewing}, {Fafone}, {Fairhurst}, {Fan}, {Farinon}, {Farr}, {Farr}, {Fauchon-Jones}, {Favata}, {Fays}, {Fazio}, {Feicht}, {Fejer}, {Feng}, {Fenyvesi},
  {Ferguson}, {Fernandez-Galiana}, {Ferrante}, {Ferreira}, {Ferreira}, {Fidecaro}, {Fiori}, {Fiorucci}, {Fishbach}, {Fisher}, {Fittipaldi}, {Fitz-Axen}, {Fiumara}, {Flaminio}, {Floden}, {Flynn}, {Fong}, {Font}, {Forsyth}, {Fournier}, {Frasca}, {Frasconi}, {Frei}, {Freise}, {Frey}, {Frey}, {Fritschel}, {Frolov}, {Fronz{\`e}}, {Fulda}, {Fyffe}, {Gabbard}, {Gadre}, {Gaebel}, {Gair}, {Galaudage}, {Ganapathy}, {Ganguly}, {Gaonkar}, {Garc{\'\i}a-Quir{\'o}s}, {Garufi}, {Gateley}, {Gaudio}, {Gayathri}, {Gemme}, {Genin}, {Gennai}, {George}, {George}, {Gergely}, {Ghonge}, {Ghosh}, {Ghosh}, {Ghosh}, {Giacomazzo}, {Giaime}, {Giardina}, {Gibson}, {Gier}, {Gill}, {Glanzer}, {Gniesmer}, {Godwin}, {Goetz}, {Goetz}, {Gohlke}, {Goncharov}, {Gonz{\'a}lez}, {Gopakumar}, {Gossan}, {Gosselin}, {Gouaty}, {Grace}, {Grado}, {Granata}, {Grant}, {Gras}, {Grassia}, {Gray}, {Gray}, {Greco}, {Green}, {Green}, {Gretarsson}, {Griggs}, {Grignani}, {Grimaldi}, {Grimm}, {Grote}, {Grunewald}, {Gruning}, {Guidi}, {Guimaraes}, {Guix{\'e}},
  {Gulati}, {Guo}, {Gupta}, {Gupta}, {Gupta}, {Gustafson}, {Gustafson}, {Haegel}, {Halim}, {Hall}, {Hamilton}, {Hammond}, {Haney}, {Hanke}, {Hanks}, {Hanna}, {Hannam}, {Hannuksela}, {Hansen}, {Hanson}, {Harder}, {Hardwick}, {Haris}, {Harms}, {Harry}, {Harry}, {Hasskew}, {Haster}, {Haughian}, {Hayes}, {Healy}, {Heidmann}, {Heintze}, {Heinze}, {Heitmann}, {Hellman}, {Hello}, {Hemming}, {Hendry}, {Heng}, {Hennes}, {Hennig}, {Heurs}, {Hild}, {Hinderer}, {Hoback}, {Hochheim}, {Hofgard}, {Hofman}, {Holgado}, {Holland}, {Holt}, {Holz}, {Hopkins}, {Horst}, {Hough}, {Howell}, {Hoy}, {Huang}, {H{\"u}bner}, {Huerta}, {Huet}, {Hughey}, {Hui}, {Husa}, {Huttner}, {Huxford}, {Huynh-Dinh}, {Idzkowski}, {Iess}, {Inchauspe}, {Ingram}, {Intini}, {Isac}, {Isi}, {Iyer}, {Jacqmin}, {Jadhav}, {Jadhav}, {James}, {Jani}, {Janthalur}, {Jaranowski}, {Jariwala}, {Jaume}, {Jenkins}, {Jiang}, {Johns}, {Johnson-McDaniel}, {Jones}, {Jones}, {Jones}, {Jones}, {Jones}, {Jonker}, {Ju}, {Junker}, {Kalaghatgi}, {Kalogera}, {Kamai}, {Kandhasamy},
  {Kang}, {Kanner}, {Kapadia}, {Karki}, {Kashyap}, {Kasprzack}, {Kastaun}, {Katsanevas}, {Katsavounidis}, {Katzman}, {Kaufer}, {Kawabe}, {K{\'e}f{\'e}lian}, {Keitel}, {Keivani}, {Kennedy}, {Key}, {Khadka}, {Khalili}, {Khan}, {Khan}, {Khan}, {Khazanov}, {Khetan}, {Khursheed}, {Kijbunchoo}, {Kim}, {Kim}, {Kim}, {Kim}, {Kim}, {Kim}, {Kim}, {Kimball}, {King}, {Kinley-Hanlon}, {Kirchhoff}, {Kissel}, {Kleybolte}, {Klimenko}, {Knowles}, {Knyazev}, {Koch}, {Koehlenbeck}, {Koekoek}, {Koley}, {Kondrashov}, {Kontos}, {Koper}, {Korobko}, {Korth}, {Kovalam}, {Kozak}, {Kringel}, {Krishnendu}, {Kr{\'o}lak}, {Krupinski}, {Kuehn}, {Kumar}, {Kumar}, {Kumar}, {Kumar}, {Kumar}, {Kuo}, {Kutynia}, {Lackey}, {Laghi}, {Lalande}, {Lam}, {Lamberts}, {Landry}, {Landry}, {Lane}, {Lang}, {Lange}, {Lantz}, {Lanza}, {La Rosa}, {Lartaux-Vollard}, {Lasky}, {Laxen}, {Lazzarini}, {Lazzaro}, {Leaci}, {Leavey}, {Lecoeuche}, {Lee}, {Lee}, {Lee}, {Lee}, {Lee}, {Lehmann}, {Leroy}, {Letendre}, {Levin}, {Li}, {Li}, {li}, {Li}, {Li}, {Linde},
  {Linker}, {Linley}, {Littenberg}, {Liu}, {Liu}, {Llorens-Monteagudo}, {Lo}, {Lockwood}, {London}, {Longo}, {Lorenzini}, {Loriette}, {Lormand}, {Losurdo}, {Lough}, {Lousto}, {Lovelace}, {L{\"u}ck}, {Lumaca}, {Lundgren}, {Ma}, {Macas}, {Macfoy}, {MacInnis}, {Macleod}, {MacMillan}, {Macquet}, {Hernandez}, {Maga{\~n}a-Sandoval}, {Magee}, {Majorana}, {Maksimovic}, {Malik}, {Man}, {Mandic}, {Mangano}, {Mansell}, {Manske}, {Mantovani}, {Mapelli}, {Marchesoni}, {Marion}, {M{\'a}rka}, {M{\'a}rka}, {Markakis}, {Markosyan}, {Markowitz}, {Maros}, {Marquina}, {Marsat}, {Martelli}, {Martin}, {Martin}, {Martinez}, {Martynov}, {Masalehdan}, {Mason}, {Massera}, {Masserot}, {Massinger}, {Masso-Reid}, {Mastrogiovanni}, {Matas}, {Matichard}, {Mavalvala}, {Maynard}, {McCann}, {McCarthy}, {McClelland}, {McCormick}, {McCuller}, {McGuire}, {McIsaac}, {McIver}, {McManus}, {McRae}, {McWilliams}, {Meacher}, {Meadors}, {Mehmet}, {Mehta}, {Villa}, {Melatos}, {Mendell}, {Mercer}, {Mereni}, {Merfeld}, {Merilh}, {Merritt}, {Merzougui},
  {Meshkov}, {Messenger}, {Messick}, {Metzdorff}, {Meyers}, {Meylahn}, {Mhaske}, {Miani}, {Miao}, {Michaloliakos}, {Michel}, {Middleton}, {Milano}, {Miller}, {Millhouse}, {Mills}, {Milotti}, {Milovich-Goff}, {Minazzoli}, {Minenkov}, {Mishkin}, {Mishra}, {Mistry}, {Mitra}, {Mitrofanov}, {Mitselmakher}, {Mittleman}, {Mo}, {Mogushi}, {Mohapatra}, {Mohite}, {Molina-Ruiz}, {Mondin}, {Montani}, {Moore}, {Moraru}, {Morawski}, {Moreno}, {Morisaki}, {Mours}, {Mow-Lowry}, {Mozzon}, {Muciaccia}, {Mukherjee}, {Mukherjee}, {Mukherjee}, {Mukherjee}, {Mukund}, {Mullavey}, {Munch}, {Mu{\~n}iz}, {Murray}, {Nagar}, {Nardecchia}, {Naticchioni}, {Nayak}, {Neil}, {Neilson}, {Nelemans}, {Nelson}, {Nery}, {Neunzert}, {Ng}, {Ng}, {Nguyen}, {Nguyen}, {Nichols}, {Nichols}, {Nissanke}, {Nocera}, {Noh}, {North}, {Nothard}, {Nuttall}, {Oberling}, {O'Brien}, {Oganesyan}, {Ogin}, {Oh}, {Oh}, {Ohme}, {Ohta}, {Okada}, {Oliver}, {Olivetto}, {Oppermann}, {Oram}, {O'Reilly}, {Ormiston}, {Ortega}, {O'Shaughnessy}, {Ossokine}, {Osthelder},
  {Ottaway}, {Overmier}, {Owen}, {Pace}, {Pagano}, {Page}, {Pagliaroli}, {Pai}, {Pai}, {Palamos}, {Palashov}, {Palomba}, {Pan}, {Panda}, {Pang}, {Pankow}, {Pannarale}, {Pant}, {Paoletti}, {Paoli}, {Parida}, {Parker}, {Pascucci}, {Pasqualetti}, {Passaquieti}, {Passuello}, {Patricelli}, {Payne}, {Pearlstone}, {Pechsiri}, {Pedersen}, {Pedraza}, {Pele}, {Penn}, {Perego}, {Perez}, {P{\'e}rigois}, {Perreca}, {Perri{\`e}s}, {Petermann}, {Pfeiffer}, {Phelps}, {Phukon}, {Piccinni}, {Pichot}, {Piendibene}, {Piergiovanni}, {Pierro}, {Pillant}, {Pinard}, {Pinto}, {Piotrzkowski}, {Pirello}, {Pitkin}, {Plastino}, {Poggiani}, {Pong}, {Ponrathnam}, {Popolizio}, {Porter}, {Powell}, {Prajapati}, {Prasai}, {Prasanna}, {Pratten}, {Prestegard}, {Principe}, {Prodi}, {Prokhorov}, {Punturo}, {Puppo}, {P{\"u}rrer}, {Qi}, {Quetschke}, {Quinonez}, {Raab}, {Raaijmakers}, {Radkins}, {Radulesco}, {Raffai}, {Rafferty}, {Raja}, {Rajan}, {Rajbhandari}, {Rakhmanov}, {Ramirez}, {Ramos-Buades}, {Rana}, {Rao}, {Rapagnani}, {Raymond}, {Razzano},
  {Read}, {Regimbau}, {Rei}, {Reid}, {Reitze}, {Rettegno}, {Ricci}, {Richardson}, {Richardson}, {Ricker}, {Riemenschneider}, {Riles}, {Rizzo}, {Robertson}, {Robinet}, {Rocchi}, {Rodriguez-Soto}, {Rolland}, {Rollins}, {Roma}, {Romanelli}, {Romano}, {Romel}, {Romero-Shaw}, {Romie}, {Rose}, {Rose}, {Rose}, {Rosi{\'n}ska}, {Rosofsky}, {Ross}, {Rowan}, {Rowlinson}, {Roy}, {Roy}, {Roy}, {Ruggi}, {Rutins}, {Ryan}, {Sachdev}, {Sadecki}, {Sakellariadou}, {Salafia}, {Salconi}, {Saleem}, {Salemi}, {Samajdar}, {Sanchez}, {Sanchez}, {Sanchis-Gual}, {Sanders}, {Santiago}, {Santos}, {Sarin}, {Sassolas}, {Sathyaprakash}, {Sauter}, {Savage}, {Savant}, {Sawant}, {Sayah}, {Schaetzl}, {Schale}, {Scheel}, {Scheuer}, {Schmidt}, {Schnabel}, {Schofield}, {Sch{\"o}nbeck}, {Schreiber}, {Schulte}, {Schutz}, {Schwarm}, {Schwartz}, {Scott}, {Scott}, {Seidel}, {Sellers}, {Sengupta}, {Sennett}, {Sentenac}, {Sequino}, {Sergeev}, {Setyawati}, {Shaddock}, {Shaffer}, {Shahriar}, {Sharma}, {Sharma}, {Shawhan}, {Shen}, {Shikauchi}, {Shink},
  {Shoemaker}, {Shoemaker}, {Shukla}, {ShyamSundar}, {Siellez}, {Sieniawska}, {Sigg}, {Singer}, {Singh}, {Singh}, {Singha}, {Singhal}, {Sintes}, {Sipala}, {Skliris}, {Slagmolen}, {Slaven-Blair}, {Smetana}, {Smith}, {Smith}, {Somala}, {Son}, {Soni}, {Sorazu}, {Sordini}, {Sorrentino}, {Souradeep}, {Sowell}, {Spencer}, {Spera}, {Srivastava}, {Srivastava}, {Staats}, {Stachie}, {Standke}, {Steer}, {Steinhoff}, {Steinke}, {Steinlechner}, {Steinlechner}, {Steinmeyer}, {Stevenson}, {Stocks}, {Stops}, {Stover}, {Strain}, {Stratta}, {Strunk}, {Sturani}, {Stuver}, {Sudhagar}, {Sudhir}, {Summerscales}, {Sun}, {Sunil}, {Sur}, {Suresh}, {Sutton}, {Swinkels}, {Szczepa{\'n}czyk}, {Tacca}, {Tait}, {Talbot}, {Tanasijczuk}, {Tanner}, {Tao}, {T{\'a}pai}, {Tapia}, {San Martin}, {Tasson}, {Taylor}, {Tenorio}, {Terkowski}, {Thirugnanasambandam}, {Thomas}, {Thomas}, {Thompson}, {Thondapu}, {Thorne}, {Thrane}, {Tinsman}, {Saravanan}, {Tiwari}, {Tiwari}, {Tiwari}, {Toland}, {Tonelli}, {Tornasi}, {Torres-Forn{\'e}}, {Torrie}, {Tosta e
  Melo}, {T{\"o}yr{\"a}}, {Trail}, {Travasso}, {Traylor}, {Tringali}, {Tripathee}, {Trovato}, {Trudeau}, {Tsang}, {Tse}, {Tso}, {Tsukada}, {Tsuna}, {Tsutsui}, {Turconi}, {Ubhi}, {Ueno}, {Ugolini}, {Unnikrishnan}, {Urban}, {Usman}, {Utina}, {Vahlbruch}, {Vajente}, {Valdes}, {Valentini}, {van Bakel}, {van Beuzekom}, {van den Brand}, {Van Den Broeck}, {Vander-Hyde}, {van der Schaaf}, {Van Heijningen}, {van Veggel}, {Vardaro}, {Varma}, {Vass}, {Vas{\'u}th}, {Vecchio}, {Vedovato}, {Veitch}, {Veitch}, {Venkateswara}, {Venugopalan}, {Verkindt}, {Veske}, {Vetrano}, {Vicer{\'e}}, {Viets}, {Vinciguerra}, {Vine}, {Vinet}, {Vitale}, {Vivanco}, {Vo}, {Vocca}, {Vorvick}, {Vyatchanin}, {Wade}, {Wade}, {Wade}, {Walet}, {Walker}, {Wallace}, {Wallace}, {Walsh}, {Wang}, {Wang}, {Wang}, {Ward}, {Warden}, {Warner}, {Was}, {Watchi}, {Weaver}, {Wei}, {Weinert}, {Weinstein}, {Weiss}, {Wellmann}, {Wen}, {We{\ss}els}, {Westhouse}, {Wette}, {Whelan}, {Whiting}, {Whittle}, {Wilken}, {Williams}, {Willis}, {Willke}, {Winkler}, {Wipf},
  {Wittel}, {Woan}, {Woehler}, {Wofford}, {Wong}, {Wright}, {Wu}, {Wysocki}, {Xiao}, {Yamamoto}, {Yang}, {Yang}, {Yang}, {Yap}, {Yazback}, {Yeeles}, {Yu}, {Yu}, {Yuen}, {Zadro{\.z}ny}, {Zadro{\.z}ny}, {Zanolin}, {Zelenova}, {Zendri}, {Zevin}, {Zhang}, {Zhang}, {Zhang}, {Zhao}, {Zhao}, {Zhou}, {Zhou}, {Zhu}, {Zimmerman}, {Zucker}, {Zweizig}, {LIGO Scientific Collaboration}, \& {Virgo Collaboration}}]{2020ApJ...896L..44A}
{Abbott}, R., {Abbott}, T.~D., {Abraham}, S., {et~al.} 2020, \apjl, 896, L44

\bibitem[{{Abbott} {et~al.}(2021{\natexlab{c}}){Abbott}, {Abbott}, {Acernese}, {Ackley}, {Adams}, {Adhikari}, {Adhikari}, {Adya}, {Affeldt}, {Agarwal}, {Agathos}, {Agatsuma}, {Aggarwal}, {Aguiar}, {Aiello}, {Ain}, {Ajith}, {Akcay}, {Akutsu}, {Albanesi}, {Allocca}, {Altin}, {Amato}, {Anand}, {Anand}, {Ananyeva}, {Anderson}, {Anderson}, {Ando}, {Andrade}, {Andres}, {Andri{\'c}}, {Angelova}, {Ansoldi}, {Antelis}, {Antier}, {Appert}, {Arai}, {Arai}, {Arai}, {Araki}, {Araya}, {Araya}, {Areeda}, {Ar{\`e}ne}, {Aritomi}, {Arnaud}, {Arogeti}, {Aronson}, {Arun}, {Asada}, {Asali}, {Ashton}, {Aso}, {Assiduo}, {Aston}, {Astone}, {Aubin}, {Austin}, {Babak}, {Badaracco}, {Bader}, {Badger}, {Bae}, {Bae}, {Baer}, {Bagnasco}, {Bai}, {Baiotti}, {Baird}, {Bajpai}, {Ball}, {Ballardin}, {Ballmer}, {Balsamo}, {Baltus}, {Banagiri}, {Bankar}, {Barayoga}, {Barbieri}, {Barish}, {Barker}, {Barneo}, {Barone}, {Barr}, {Barsotti}, {Barsuglia}, {Barta}, {Bartlett}, {Barton}, {Bartos}, {Bassiri}, {Basti}, {Bawaj}, {Bayley}, {Baylor},
  {Bazzan}, {B{\'e}csy}, {Bedakihale}, {Bejger}, {Belahcene}, {Benedetto}, {Beniwal}, {Bennett}, {Bentley}, {BenYaala}, {Bergamin}, {Berger}, {Bernuzzi}, {Berry}, {Bersanetti}, {Bertolini}, {Betzwieser}, {Beveridge}, {Bhandare}, {Bhardwaj}, {Bhattacharjee}, {Bhaumik}, {Bilenko}, {Billingsley}, {Bini}, {Birney}, {Birnholtz}, {Biscans}, {Bischi}, {Biscoveanu}, {Bisht}, {Biswas}, {Bitossi}, {Bizouard}, {Blackburn}, {Blair}, {Blair}, {Blair}, {Bobba}, {Bode}, {Boer}, {Bogaert}, {Boldrini}, {Bonavena}, {Bondu}, {Bonilla}, {Bonnand}, {Booker}, {Boom}, {Bork}, {Boschi}, {Bose}, {Bose}, {Bossilkov}, {Boudart}, {Bouffanais}, {Bozzi}, {Bradaschia}, {Brady}, {Bramley}, {Branch}, {Branchesi}, {Brandt}, {Brau}, {Breschi}, {Briant}, {Briggs}, {Brillet}, {Brinkmann}, {Brockill}, {Brooks}, {Brooks}, {Brown}, {Brunett}, {Bruno}, {Bruntz}, {Bryant}, {Bulik}, {Bulten}, {Buonanno}, {Buscicchio}, {Buskulic}, {Buy}, {Byer}, {Cabourn Davies}, {Cadonati}, {Cagnoli}, {Cahillane}, {Calder{\'o}n Bustillo}, {Callaghan}, {Callister},
  {Calloni}, {Cameron}, {Camp}, {Canepa}, {Canevarolo}, {Cannavacciuolo}, {Cannon}, {Cao}, {Cao}, {Capocasa}, {Capote}, {Carapella}, {Carbognani}, {Carlin}, {Carney}, {Carpinelli}, {Carrillo}, {Carullo}, {Carver}, {Casanueva Diaz}, {Casentini}, {Castaldi}, {Caudill}, {Cavagli{\`a}}, {Cavalier}, {Cavalieri}, {Ceasar}, {Cella}, {Cerd{\'a}-Dur{\'a}n}, {Cesarini}, {Chaibi}, {Chakravarti}, {Chalathadka Subrahmanya}, {Champion}, {Chan}, {Chan}, {Chan}, {Chan}, {Chan}, {Chandra}, {Chanial}, {Chao}, {Chapman-Bird}, {Charlton}, {Chase}, {Chassande-Mottin}, {Chatterjee}, {Chatterjee}, {Chatterjee}, {Chaturvedi}, {Chaty}, {Chatziioannou}, {Chen}, {Chen}, {Chen}, {Chen}, {Chen}, {Chen}, {Chen}, {Chen}, {Cheng}, {Cheong}, {Cheung}, {Chia}, {Chiadini}, {Chiang}, {Chiarini}, {Chierici}, {Chincarini}, {Chiofalo}, {Chiummo}, {Cho}, {Cho}, {Choudhary}, {Choudhary}, {Christensen}, {Chu}, {Chu}, {Chu}, {Chua}, {Chung}, {Ciani}, {Ciecielag}, {Cie{\'s}lar}, {Cifaldi}, {Ciobanu}, {Ciolfi}, {Cipriano}, {Cirone}, {Clara}, {Clark},
  {Clark}, {Clarke}, {Clearwater}, {Clesse}, {Cleva}, {Coccia}, {Codazzo}, {Cohadon}, {Cohen}, {Cohen}, {Colleoni}, {Collette}, {Colombo}, {Colpi}, {Compton}, {Constancio}, {Conti}, {Cooper}, {Corban}, {Corbitt}, {Cordero-Carri{\'o}n}, {Corezzi}, {Corley}, {Cornish}, {Corre}, {Corsi}, {Cortese}, {Costa}, {Cotesta}, {Coughlin}, {Coulon}, {Countryman}, {Cousins}, {Couvares}, {Coward}, {Cowart}, {Coyne}, {Coyne}, {Creighton}, {Creighton}, {Criswell}, {Croquette}, {Crowder}, {Cudell}, {Cullen}, {Cumming}, {Cummings}, {Cunningham}, {Cuoco}, {Cury{\l}o}, {Dabadie}, {Dal Canton}, {Dall'Osso}, {D{\'a}lya}, {Dana}, {DaneshgaranBajastani}, {D'Angelo}, {Danila}, {Danilishin}, {D'Antonio}, {Danzmann}, {Darsow-Fromm}, {Dasgupta}, {Datrier}, {Datta}, {Dattilo}, {Dave}, {Davier}, {Davis}, {Davis}, {Daw}, {de Alarc{\'o}n}, {Dean}, {DeBra}, {Deenadayalan}, {Degallaix}, {De Laurentis}, {Del{\'e}glise}, {Del Favero}, {De Lillo}, {De Lillo}, {Del Pozzo}, {DeMarchi}, {De Matteis}, {D'Emilio}, {Demos}, {Dent}, {Depasse}, {De
  Pietri}, {De Rosa}, {De Rossi}, {DeSalvo}, {De Simone}, {Dhurandhar}, {D{\'\i}az}, {Diaz-Ortiz}, {Didio}, {Dietrich}, {Di Fiore}, {Di Fronzo}, {Di Giorgio}, {Di Giovanni}, {Di Giovanni}, {Di Girolamo}, {Di Lieto}, {Ding}, {Di Pace}, {Di Palma}, {Di Renzo}, {Divakarla}, {Dmitriev}, {Doctor}, {D'Onofrio}, {Donovan}, {Dooley}, {Doravari}, {Dorrington}, {Drago}, {Driggers}, {Drori}, {Ducoin}, {Dupej}, {Durante}, {D'Urso}, {Duverne}, {Dwyer}, {Eassa}, {Easter}, {Ebersold}, {Eckhardt}, {Eddolls}, {Edelman}, {Edo}, {Edy}, {Effler}, {Eguchi}, {Eichholz}, {Eikenberry}, {Eisenmann}, {Eisenstein}, {Ejlli}, {Engelby}, {Enomoto}, {Errico}, {Essick}, {Estell{\'e}s}, {Estevez}, {Etienne}, {Etzel}, {Evans}, {Evans}, {Ewing}, {Fafone}, {Fair}, {Fairhurst}, {Farah}, {Farinon}, {Farr}, {Farr}, {Farrow}, {Fauchon-Jones}, {Favaro}, {Favata}, {Fays}, {Fazio}, {Feicht}, {Fejer}, {Fenyvesi}, {Ferguson}, {Fernandez-Galiana}, {Ferrante}, {Ferreira}, {Fidecaro}, {Figura}, {Fiori}, {Fishbach}, {Fisher}, {Fittipaldi}, {Fiumara},
  {Flaminio}, {Floden}, {Fong}, {Font}, {Fornal}, {Forsyth}, {Franke}, {Frasca}, {Frasconi}, {Frederick}, {Freed}, {Frei}, {Freise}, {Frey}, {Fritschel}, {Frolov}, {Fronz{\'e}}, {Fujii}, {Fujikawa}, {Fukunaga}, {Fukushima}, {Fulda}, {Fyffe}, {Gabbard}, {Gabella}, {Gadre}, {Gair}, {Gais}, {Galaudage}, {Gamba}, {Ganapathy}, {Ganguly}, {Gao}, {Gaonkar}, {Garaventa}, {Garc{\'\i}a}, {Garc{\'\i}a-N{\'u}{\~n}ez}, {Garc{\'\i}a-Quir{\'o}s}, {Garufi}, {Gateley}, {Gaudio}, {Gayathri}, {Ge}, {Gemme}, {Gennai}, {George}, {George}, {Gerberding}, {Gergely}, {Gewecke}, {Ghonge}, {Ghosh}, {Ghosh}, {Ghosh}, {Ghosh}, {Giacomazzo}, {Giacoppo}, {Giaime}, {Giardina}, {Gibson}, {Gier}, {Giesler}, {Giri}, {Gissi}, {Glanzer}, {Gleckl}, {Godwin}, {Goetz}, {Goetz}, {Gohlke}, {Golomb}, {Goncharov}, {Gonz{\'a}lez}, {Gopakumar}, {Gosselin}, {Gouaty}, {Gould}, {Grace}, {Grado}, {Granata}, {Granata}, {Grant}, {Gras}, {Grassia}, {Gray}, {Gray}, {Greco}, {Green}, {Green}, {Gretarsson}, {Gretarsson}, {Griffith}, {Griffiths}, {Griggs},
  {Grignani}, {Grimaldi}, {Grimm}, {Grote}, {Grunewald}, {Gruning}, {Guerra}, {Guidi}, {Guimaraes}, {Guix{\'e}}, {Gulati}, {Guo}, {Guo}, {Gupta}, {Gupta}, {Gupta}, {Gustafson}, {Gustafson}, {Guzman}, {Ha}, {Haegel}, {Hagiwara}, {Haino}, {Halim}, {Hall}, {Hamilton}, {Hammond}, {Han}, {Haney}, {Hanks}, {Hanna}, {Hannam}, {Hannuksela}, {Hansen}, {Hansen}, {Hanson}, {Harder}, {Hardwick}, {Haris}, {Harms}, {Harry}, {Harry}, {Hartwig}, {Hasegawa}, {Haskell}, {Hasskew}, {Haster}, {Hattori}, {Haughian}, {Hayakawa}, {Hayama}, {Hayes}, {Healy}, {Heidmann}, {Heidt}, {Heintze}, {Heinze}, {Heinzel}, {Heitmann}, {Hellman}, {Hello}, {Helmling-Cornell}, {Hemming}, {Hendry}, {Heng}, {Hennes}, {Hennig}, {Hennig}, {Hernandez}, {Hernandez Vivanco}, {Heurs}, {Hild}, {Hill}, {Himemoto}, {Hines}, {Hiranuma}, {Hirata}, {Hirose}, {Hochheim}, {Hofman}, {Hohmann}, {Holcomb}, {Holland}, {Holley-Bockelmann}, {Hollows}, {Holmes}, {Holt}, {Holz}, {Hong}, {Hopkins}, {Hough}, {Hourihane}, {Howell}, {Hoy}, {Hoyland}, {Hreibi}, {Hsieh}, {Hsu},
  {Huang}, {Huang}, {Huang}, {Huang}, {Huang}, {Huang}, {H{\"u}bner}, {Huddart}, {Hughey}, {Hui}, {Hui}, {Husa}, {Huttner}, {Huxford}, {Huynh-Dinh}, {Ide}, {Idzkowski}, {Iess}, {Ikenoue}, {Imam}, {Inayoshi}, {Ingram}, {Inoue}, {Ioka}, {Isi}, {Isleif}, {Ito}, {Itoh}, {Iyer}, {Izumi}, {JaberianHamedan}, {Jacqmin}, {Jadhav}, {Jadhav}, {James}, {Jan}, {Jani}, {Janquart}, {Janssens}, {Janthalur}, {Jaranowski}, {Jariwala}, {Jaume}, {Jenkins}, {Jenner}, {Jeon}, {Jeunon}, {Jia}, {Jin}, {Johns}, {Johnson-McDaniel}, {Jones}, {Jones}, {Jones}, {Jones}, {Jones}, {Jonker}, {Ju}, {Jung}, {Jung}, {Junker}, {Juste}, {Kaihotsu}, {Kajita}, {Kakizaki}, {Kalaghatgi}, {Kalogera}, {Kamai}, {Kamiizumi}, {Kanda}, {Kandhasamy}, {Kang}, {Kanner}, {Kao}, {Kapadia}, {Kapasi}, {Karat}, {Karathanasis}, {Karki}, {Kashyap}, {Kasprzack}, {Kastaun}, {Katsanevas}, {Katsavounidis}, {Katzman}, {Kaur}, {Kawabe}, {Kawaguchi}, {Kawai}, {Kawasaki}, {K{\'e}f{\'e}lian}, {Keitel}, {Key}, {Khadka}, {Khalili}, {Khan}, {Khazanov}, {Khetan}, {Khursheed},
  {Kijbunchoo}, {Kim}, {Kim}, {Kim}, {Kim}, {Kim}, {Kim}, {Kimball}, {Kimura}, {Kinley-Hanlon}, {Kirchhoff}, {Kissel}, {Kita}, {Kitazawa}, {Kleybolte}, {Klimenko}, {Knee}, {Knowles}, {Knyazev}, {Koch}, {Koekoek}, {Kojima}, {Kokeyama}, {Koley}, {Kolitsidou}, {Kolstein}, {Komori}, {Kondrashov}, {Kong}, {Kontos}, {Koper}, {Korobko}, {Kotake}, {Kovalam}, {Kozak}, {Kozakai}, {Kozu}, {Kringel}, {Krishnendu}, {Kr{\'o}lak}, {Kuehn}, {Kuei}, {Kuijer}, {Kulkarni}, {Kumar}, {Kumar}, {Kumar}, {Kumar}, {Kume}, {Kuns}, {Kuo}, {Kuo}, {Kuromiya}, {Kuroyanagi}, {Kusayanagi}, {Kuwahara}, {Kwak}, {Lagabbe}, {Laghi}, {Lalande}, {Lam}, {Lamberts}, {Landry}, {Lane}, {Lang}, {Lange}, {Lantz}, {La Rosa}, {Lartaux-Vollard}, {Lasky}, {Laxen}, {Lazzarini}, {Lazzaro}, {Leaci}, {Leavey}, {Lecoeuche}, {Lee}, {Lee}, {Lee}, {Lee}, {Lee}, {Lee}, {Lehmann}, {Lema{\^\i}tre}, {Leonardi}, {Leroy}, {Letendre}, {Levesque}, {Levin}, {Leviton}, {Leyde}, {Li}, {Li}, {Li}, {Li}, {Li}, {Li}, {Lin}, {Lin}, {Lin}, {Lin}, {Lin}, {Linde}, {Linker},
  {Linley}, {Littenberg}, {Liu}, {Liu}, {Liu}, {Liu}, {Llamas}, {Llorens-Monteagudo}, {Lo}, {Lockwood}, {Loh}, {London}, {Longo}, {Lopez}, {Lopez Portilla}, {Lorenzini}, {Loriette}, {Lormand}, {Losurdo}, {Lott}, {Lough}, {Lousto}, {Lovelace}, {Lucaccioni}, {L{\"u}ck}, {Lumaca}, {Lundgren}, {Luo}, {Lynam}, {Macas}, {MacInnis}, {Macleod}, {MacMillan}, {Macquet}, {Maga{\~n}a Hernandez}, {Magazz{\`u}}, {Magee}, {Maggiore}, {Magnozzi}, {Mahesh}, {Majorana}, {Makarem}, {Maksimovic}, {Maliakal}, {Malik}, {Man}, {Mandic}, {Mangano}, {Mango}, {Mansell}, {Manske}, {Mantovani}, {Mapelli}, {Marchesoni}, {Marchio}, {Marion}, {Mark}, {M{\'a}rka}, {M{\'a}rka}, {Markakis}, {Markosyan}, {Markowitz}, {Maros}, {Marquina}, {Marsat}, {Martelli}, {Martin}, {Martin}, {Martinez}, {Martinez}, {Martinez}, {Martinovic}, {Martynov}, {Marx}, {Masalehdan}, {Mason}, {Massera}, {Masserot}, {Massinger}, {Masso-Reid}, {Mastrogiovanni}, {Matas}, {Mateu-Lucena}, {Matichard}, {Matiushechkina}, {Mavalvala}, {McCann}, {McCarthy}, {McClelland},
  {McClincy}, {McCormick}, {McCuller}, {McGhee}, {McGuire}, {McIsaac}, {McIver}, {McRae}, {McWilliams}, {Meacher}, {Mehmet}, {Mehta}, {Meijer}, {Melatos}, {Melchor}, {Mendell}, {Menendez-Vazquez}, {Menoni}, {Mercer}, {Mereni}, {Merfeld}, {Merilh}, {Merritt}, {Merzougui}, {Meshkov}, {Messenger}, {Messick}, {Meyers}, {Meylahn}, {Mhaske}, {Miani}, {Miao}, {Michaloliakos}, {Michel}, {Michimura}, {Middleton}, {Milano}, {Miller}, {Miller}, {Miller}, {Millhouse}, {Mills}, {Milotti}, {Minazzoli}, {Minenkov}, {Mio}, {Mir}, {Miravet-Ten{\'e}s}, {Mishra}, {Mishra}, {Mistry}, {Mitra}, {Mitrofanov}, {Mitselmakher}, {Mittleman}, {Miyakawa}, {Miyamoto}, {Miyazaki}, {Miyo}, {Miyoki}, {Mo}, {Modafferi}, {Moguel}, {Mogushi}, {Mohapatra}, {Mohite}, {Molina}, {Molina-Ruiz}, {Mondin}, {Montani}, {Moore}, {Moraru}, {Morawski}, {More}, {Moreno}, {Moreno}, {Mori}, {Morisaki}, {Moriwaki}, {Morr{\'a}s}, {Mours}, {Mow-Lowry}, {Mozzon}, {Muciaccia}, {Mukherjee}, {Mukherjee}, {Mukherjee}, {Mukherjee}, {Mukherjee}, {Mukund}, {Mullavey},
  {Munch}, {Mu{\~n}iz}, {Murray}, {Musenich}, {Muusse}, {Nadji}, {Nagano}, {Nagano}, {Nagar}, {Nakamura}, {Nakano}, {Nakano}, {Nakashima}, {Nakayama}, {Napolano}, {Nardecchia}, {Narikawa}, {Naticchioni}, {Nayak}, {Nayak}, {Negishi}, {Neil}, {Neilson}, {Nelemans}, {Nelson}, {Nery}, {Neubauer}, {Neunzert}, {Ng}, {Ng}, {Nguyen}, {Nguyen}, {Nguyen}, {Nguyen Quynh}, {Ni}, {Nichols}, {Nishizawa}, {Nissanke}, {Nitoglia}, {Nocera}, {Norman}, {North}, {Nozaki}, {Nu{\~n}o Siles}, {Nuttall}, {Oberling}, {O'Brien}, {Obuchi}, {O'Dell}, {Oelker}, {Ogaki}, {Oganesyan}, {Oh}, {Oh}, {Oh}, {Ohashi}, {Ohishi}, {Ohkawa}, {Ohme}, {Ohta}, {Okada}, {Okutani}, {Okutomi}, {Olivetto}, {Oohara}, {Ooi}, {Oram}, {O'Reilly}, {Ormiston}, {Ormsby}, {Ortega}, {O'Shaughnessy}, {O'Shea}, {Oshino}, {Ossokine}, {Osthelder}, {Otabe}, {Ottaway}, {Overmier}, {Pace}, {Pagano}, {Page}, {Pagliaroli}, {Pai}, {Pai}, {Palamos}, {Palashov}, {Palomba}, {Pan}, {Pan}, {Panda}, {Pang}, {Pang}, {Pankow}, {Pannarale}, {Pant}, {Panther}, {Paoletti}, {Paoli},
  {Paolone}, {Parisi}, {Park}, {Park}, {Parker}, {Pascucci}, {Pasqualetti}, {Passaquieti}, {Passuello}, {Patel}, {Pathak}, {Patricelli}, {Patron}, {Paul}, {Payne}, {Pedraza}, {Pegoraro}, {Pele}, {Pe{\~n}a Arellano}, {Penn}, {Perego}, {Pereira}, {Pereira}, {Perez}, {P{\'e}rigois}, {Perkins}, {Perreca}, {Perri{\`e}s}, {Petermann}, {Petterson}, {Pfeiffer}, {Pham}, {Phukon}, {Piccinni}, {Pichot}, {Piendibene}, {Piergiovanni}, {Pierini}, {Pierro}, {Pillant}, {Pillas}, {Pilo}, {Pinard}, {Pinto}, {Pinto}, {Piotrzkowski}, {Piotrzkowski}, {Pirello}, {Pitkin}, {Placidi}, {Planas}, {Plastino}, {Pluchar}, {Poggiani}, {Polini}, {Pong}, {Ponrathnam}, {Popolizio}, {Porter}, {Poulton}, {Powell}, {Pracchia}, {Pradier}, {Prajapati}, {Prasai}, {Prasanna}, {Pratten}, {Principe}, {Prodi}, {Prokhorov}, {Prosposito}, {Prudenzi}, {Puecher}, {Punturo}, {Puosi}, {Puppo}, {P{\"u}rrer}, {Qi}, {Quetschke}, {Quitzow-James}, {Qutob}, {Raab}, {Raaijmakers}, {Radkins}, {Radulesco}, {Raffai}, {Rail}, {Raja}, {Rajan}, {Ramirez}, {Ramirez},
  {Ramos-Buades}, {Rana}, {Rapagnani}, {Rapol}, {Ray}, {Raymond}, {Raza}, {Razzano}, {Read}, {Rees}, {Regimbau}, {Rei}, {Reid}, {Reid}, {Reitze}, {Relton}, {Renzini}, {Rettegno}, {Reza}, {Rezac}, {Ricci}, {Richards}, {Richardson}, {Richardson}, {Riemenschneider}, {Riles}, {Rinaldi}, {Rink}, {Rizzo}, {Robertson}, {Robie}, {Robinet}, {Rocchi}, {Rodriguez}, {Rolland}, {Rollins}, {Romanelli}, {Romano}, {Romel}, {Romero-Rodr{\'\i}guez}, {Romero-Shaw}, {Romie}, {Ronchini}, {Rosa}, {Rose}, {Rosi{\'n}ska}, {Ross}, {Rowan}, {Rowlinson}, {Roy}, {Roy}, {Roy}, {Rozza}, {Ruggi}, {Ruiz-Rocha}, {Ryan}, {Sachdev}, {Sadecki}, {Sadiq}, {Sago}, {Saito}, {Saito}, {Sakai}, {Sakai}, {Sakellariadou}, {Sakuno}, {Salafia}, {Salconi}, {Saleem}, {Salemi}, {Samajdar}, {Sanchez}, {Sanchez}, {Sanchez}, {Sanchis-Gual}, {Sanders}, {Sanuy}, {Saravanan}, {Sarin}, {Sassolas}, {Satari}, {Sathyaprakash}, {Sato}, {Sato}, {Sauter}, {Savage}, {Sawada}, {Sawant}, {Sawant}, {Sayah}, {Schaetzl}, {Scheel}, {Scheuer}, {Schiworski}, {Schmidt}, {Schmidt},
  {Schnabel}, {Schneewind}, {Schofield}, {Sch{\"o}nbeck}, {Schulte}, {Schutz}, {Schwartz}, {Scott}, {Scott}, {Seglar-Arroyo}, {Sekiguchi}, {Sekiguchi}, {Sellers}, {Sengupta}, {Sentenac}, {Seo}, {Sequino}, {Sergeev}, {Setyawati}, {Shaffer}, {Shahriar}, {Shams}, {Shao}, {Sharma}, {Sharma}, {Shawhan}, {Shcheblanov}, {Shibagaki}, {Shikauchi}, {Shimizu}, {Shimoda}, {Shimode}, {Shinkai}, {Shishido}, {Shoda}, {Shoemaker}, {Shoemaker}, {ShyamSundar}, {Sieniawska}, {Sigg}, {Singer}, {Singh}, {Singh}, {Singha}, {Sintes}, {Sipala}, {Skliris}, {Slagmolen}, {Slaven-Blair}, {Smetana}, {Smith}, {Smith}, {Soldateschi}, {Somala}, {Somiya}, {Son}, {Soni}, {Soni}, {Sordini}, {Sorrentino}, {Sorrentino}, {Sotani}, {Soulard}, {Souradeep}, {Sowell}, {Spagnuolo}, {Spencer}, {Spera}, {Srinivasan}, {Srivastava}, {Srivastava}, {Staats}, {Stachie}, {Steer}, {Steinhoff}, {Steinlechner}, {Steinlechner}, {Stevenson}, {Stops}, {Stover}, {Strain}, {Strang}, {Stratta}, {Strunk}, {Sturani}, {Stuver}, {Sudhagar}, {Sudhir}, {Sugimoto}, {Suh},
  {Sullivan}, {Sullivan}, {Summerscales}, {Sun}, {Sun}, {Sunil}, {Sur}, {Suresh}, {Sutton}, {Suzuki}, {Suzuki}, {Swinkels}, {Szczepa{\'n}czyk}, {Szewczyk}, {Tacca}, {Tagoshi}, {Tait}, {Takahashi}, {Takahashi}, {Takamori}, {Takano}, {Takeda}, {Takeda}, {Talbot}, {Talbot}, {Tanaka}, {Tanaka}, {Tanaka}, {Tanaka}, {Tanaka}, {Tanasijczuk}, {Tanioka}, {Tanner}, {Tao}, {Tao}, {Tapia San Mart{\'\i}n}, {Taranto}, {Tasson}, {Telada}, {Tenorio}, {Terhune}, {Terkowski}, {Thirugnanasambandam}, {Thomas}, {Thomas}, {Thomas}, {Thompson}, {Thondapu}, {Thorne}, {Thrane}, {Tiwari}, {Tiwari}, {Tiwari}, {Toivonen}, {Toland}, {Tolley}, {Tomaru}, {Tomigami}, {Tomura}, {Tonelli}, {Torres-Forn{\'e}}, {Torrie}, {Tosta e Melo}, {T{\"o}yr{\"a}}, {Trapananti}, {Travasso}, {Traylor}, {Trevor}, {Tringali}, {Tripathee}, {Troiano}, {Trovato}, {Trozzo}, {Trudeau}, {Tsai}, {Tsai}, {Tsang}, {Tsang}, {Tsao}, {Tse}, {Tso}, {Tsubono}, {Tsuchida}, {Tsukada}, {Tsuna}, {Tsutsui}, {Tsuzuki}, {Turbang}, {Turconi}, {Tuyenbayev}, {Ubhi}, {Uchikata},
  {Uchiyama}, {Udall}, {Ueda}, {Uehara}, {Ueno}, {Ueshima}, {Unnikrishnan}, {Uraguchi}, {Urban}, {Ushiba}, {Utina}, {Vahlbruch}, {Vajente}, {Vajpeyi}, {Valdes}, {Valentini}, {Valsan}, {van Bakel}, {van Beuzekom}, {van den Brand}, {Van Den Broeck}, {Vander-Hyde}, {van der Schaaf}, {van Heijningen}, {Vanosky}, {van Putten}, {van Remortel}, {Vardaro}, {Vargas}, {Varma}, {Vas{\'u}th}, {Vecchio}, {Vedovato}, {Veitch}, {Veitch}, {Venneberg}, {Venugopalan}, {Verkindt}, {Verma}, {Verma}, {Veske}, {Vetrano}, {Vicer{\'e}}, {Vidyant}, {Viets}, {Vijaykumar}, {Villa-Ortega}, {Vinet}, {Virtuoso}, {Vitale}, {Vo}, {Vocca}, {von Reis}, {von Wrangel}, {Vorvick}, {Vyatchanin}, {Wade}, {Wade}, {Wagner}, {Walet}, {Walker}, {Wallace}, {Wallace}, {Walsh}, {Wang}, {Wang}, {Wang}, {Ward}, {Warner}, {Was}, {Washimi}, {Washington}, {Watchi}, {Weaver}, {Webster}, {Weinert}, {Weinstein}, {Weiss}, {Weller}, {Weller}, {Wellmann}, {Wen}, {We{\ss}els}, {Wette}, {Whelan}, {White}, {Whiting}, {Whittle}, {Wilken}, {Williams}, {Williams},
  {Williams}, {Williamson}, {Willis}, {Willke}, {Wilson}, {Winkler}, {Wipf}, {Wlodarczyk}, {Woan}, {Woehler}, {Wofford}, {Wong}, {Wu}, {Wu}, {Wu}, {Wu}, {Wysocki}, {Xiao}, {Xu}, {Yamada}, {Yamamoto}, {Yamamoto}, {Yamamoto}, {Yamamoto}, {Yamashita}, {Yamazaki}, {Yang}, {Yang}, {Yang}, {Yang}, {Yang}, {Yap}, {Yeeles}, {Yelikar}, {Ying}, {Yokogawa}, {Yokoyama}, {Yokozawa}, {Yoo}, {Yoshioka}, {Yu}, {Yu}, {Yuzurihara}, {Zadro{\.z}ny}, {Zanolin}, {Zeidler}, {Zelenova}, {Zendri}, {Zevin}, {Zhan}, {Zhang}, {Zhang}, {Zhang}, {Zhang}, {Zhang}, {Zhao}, {Zhao}, {Zhao}, {Zhao}, {Zheng}, {Zhou}, {Zhou}, {Zhu}, {Zhu}, {Zimmerman}, {Zlochower}, {Zucker}, \& {Zweizig}}]{2021arXiv211103606T}
{Abbott}, R., {Abbott}, T.~D., {Acernese}, F., {et~al.} 2021{\natexlab{c}}, arXiv e-prints, arXiv:2111.03606

\bibitem[{{Abbott} {et~al.}(2021{\natexlab{d}}){Abbott}, {Abbott}, {Acernese}, {Ackley}, {Adams}, {Adhikari}, {Adhikari}, {Adya}, {Affeldt}, {Agarwal}, {Agathos}, {Agatsuma}, {Aggarwal}, {Aguiar}, {Aiello}, {Ain}, {Ajith}, {Albanesi}, {Allocca}, {Altin}, {Amato}, {Anand}, {Anand}, {Ananyeva}, {Anderson}, {Anderson}, {Andrade}, {Andres}, {Andri{\'c}}, {Angelova}, {Ansoldi}, {Antelis}, {Antier}, {Appert}, {Arai}, {Araya}, {Areeda}, {Ar{\`e}ne}, {Arnaud}, {Aronson}, {Arun}, {Asali}, {Ashton}, {Assiduo}, {Aston}, {Astone}, {Aubin}, {Austin}, {Babak}, {Badaracco}, {Bader}, {Badger}, {Bae}, {Baer}, {Bagnasco}, {Bai}, {Baird}, {Ball}, {Ballardin}, {Ballmer}, {Balsamo}, {Baltus}, {Banagiri}, {Bankar}, {Barayoga}, {Barbieri}, {Barish}, {Barker}, {Barneo}, {Barone}, {Barr}, {Barsotti}, {Barsuglia}, {Barta}, {Bartlett}, {Barton}, {Bartos}, {Bassiri}, {Basti}, {Bawaj}, {Bayley}, {Baylor}, {Bazzan}, {B{\'e}csy}, {Bedakihale}, {Bejger}, {Belahcene}, {Benedetto}, {Beniwal}, {Bennett}, {Bentley}, {BenYaala}, {Bergamin},
  {Berger}, {Bernuzzi}, {Berry}, {Bersanetti}, {Bertolini}, {Betzwieser}, {Beveridge}, {Bhandare}, {Bhardwaj}, {Bhattacharjee}, {Bhaumik}, {Bilenko}, {Billingsley}, {Bini}, {Birney}, {Birnholtz}, {Biscans}, {Bischi}, {Biscoveanu}, {Bisht}, {Biswas}, {Bitossi}, {Bizouard}, {Blackburn}, {Blair}, {Blair}, {Blair}, {Bobba}, {Bode}, {Boer}, {Bogaert}, {Boldrini}, {Bonavena}, {Bondu}, {Bonilla}, {Bonnand}, {Booker}, {Boom}, {Bork}, {Boschi}, {Bose}, {Bose}, {Bossilkov}, {Boudart}, {Bouffanais}, {Bozzi}, {Bradaschia}, {Brady}, {Bramley}, {Branch}, {Branchesi}, {Brau}, {Breschi}, {Briant}, {Briggs}, {Brillet}, {Brinkmann}, {Brockill}, {Brooks}, {Brooks}, {Brown}, {Brunett}, {Bruno}, {Bruntz}, {Bryant}, {Bulik}, {Bulten}, {Buonanno}, {Buscicchio}, {Buskulic}, {Buy}, {Byer}, {Cadonati}, {Cagnoli}, {Cahillane}, {Calder{\'o}n Bustillo}, {Callaghan}, {Callister}, {Calloni}, {Cameron}, {Camp}, {Canepa}, {Canevarolo}, {Cannavacciuolo}, {Cannon}, {Cao}, {Capote}, {Carapella}, {Carbognani}, {Carlin}, {Carney}, {Carpinelli},
  {Carrillo}, {Carullo}, {Carver}, {Casanueva Diaz}, {Casentini}, {Castaldi}, {Caudill}, {Cavagli{\`a}}, {Cavalier}, {Cavalieri}, {Ceasar}, {Cella}, {Cerd{\'a}-Dur{\'a}n}, {Cesarini}, {Chaibi}, {Chakravarti}, {Chalathadka Subrahmanya}, {Champion}, {Chan}, {Chan}, {Chan}, {Chan}, {Chandra}, {Chanial}, {Chao}, {Charlton}, {Chase}, {Chassande-Mottin}, {Chatterjee}, {Chatterjee}, {Chatterjee}, {Chattopadhyay}, {Chaturvedi}, {Chaty}, {Chatziioannou}, {Chen}, {Chen}, {Chen}, {Chen}, {Chen}, {Cheng}, {Cheong}, {Cheung}, {Chia}, {Chiadini}, {Chiarini}, {Chierici}, {Chincarini}, {Chiofalo}, {Chiummo}, {Cho}, {Cho}, {Choudhary}, {Choudhary}, {Christensen}, {Chu}, {Chua}, {Chung}, {Ciani}, {Ciecielag}, {Cie{\'s}lar}, {Cifaldi}, {Ciobanu}, {Ciolfi}, {Cipriano}, {Cirone}, {Clara}, {Clark}, {Clark}, {Clarke}, {Clearwater}, {Clesse}, {Cleva}, {Coccia}, {Codazzo}, {Cohadon}, {Cohen}, {Cohen}, {Colleoni}, {Collette}, {Colombo}, {Colpi}, {Compton}, {Constancio}, {Conti}, {Cooper}, {Corban}, {Corbitt}, {Cordero-Carri{\'o}n},
  {Corezzi}, {Corley}, {Cornish}, {Corre}, {Corsi}, {Cortese}, {Costa}, {Cotesta}, {Coughlin}, {Coulon}, {Countryman}, {Cousins}, {Couvares}, {Coward}, {Cowart}, {Coyne}, {Coyne}, {Creighton}, {Creighton}, {Criswell}, {Croquette}, {Crowder}, {Cudell}, {Cullen}, {Cumming}, {Cummings}, {Cunningham}, {Cuoco}, {Cury{\l}o}, {Dabadie}, {Dal Canton}, {Dall'Osso}, {D{\'a}lya}, {Dana}, {DaneshgaranBajastani}, {D'Angelo}, {Danila}, {Danilishin}, {D'Antonio}, {Danzmann}, {Darsow-Fromm}, {Dasgupta}, {Datrier}, {Datta}, {Dattilo}, {Dave}, {Davier}, {Davies}, {Davis}, {Davis}, {Daw}, {Dean}, {DeBra}, {Deenadayalan}, {Degallaix}, {De Laurentis}, {Del{\'e}glise}, {Del Favero}, {De Lillo}, {De Lillo}, {Del Pozzo}, {DeMarchi}, {De Matteis}, {D'Emilio}, {Demos}, {Dent}, {Depasse}, {De Pietri}, {De Rosa}, {De Rossi}, {DeSalvo}, {De Simone}, {Dhurandhar}, {D{\'\i}az}, {Diaz-Ortiz}, {Didio}, {Dietrich}, {Di Fiore}, {Di Fronzo}, {Di Giorgio}, {Di Giovanni}, {Di Giovanni}, {Di Girolamo}, {Di Lieto}, {Ding}, {Di Pace}, {Di Palma},
  {Di Renzo}, {Divakarla}, {Divyajyoti}, {Dmitriev}, {Doctor}, {D'Onofrio}, {Donovan}, {Dooley}, {Doravari}, {Dorrington}, {Drago}, {Driggers}, {Drori}, {Ducoin}, {Dupej}, {Durante}, {D'Urso}, {Duverne}, {Dwyer}, {Eassa}, {Easter}, {Ebersold}, {Eckhardt}, {Eddolls}, {Edelman}, {Edo}, {Edy}, {Effler}, {Eichholz}, {Eikenberry}, {Eisenmann}, {Eisenstein}, {Ejlli}, {Engelby}, {Errico}, {Essick}, {Estell{\'e}s}, {Estevez}, {Etienne}, {Etzel}, {Evans}, {Evans}, {Ewing}, {Fafone}, {Fair}, {Fairhurst}, {Fanning}, {Farah}, {Farinon}, {Farr}, {Farr}, {Farrow}, {Fauchon-Jones}, {Favaro}, {Favata}, {Fays}, {Fazio}, {Feicht}, {Fejer}, {Fenyvesi}, {Ferguson}, {Fernandez-Galiana}, {Ferrante}, {Ferreira}, {Fidecaro}, {Figura}, {Fiori}, {Fishbach}, {Fisher}, {Fittipaldi}, {Fiumara}, {Flaminio}, {Floden}, {Fong}, {Font}, {Fornal}, {Forsyth}, {Franke}, {Frasca}, {Frasconi}, {Frederick}, {Freed}, {Frei}, {Freise}, {Frey}, {Fritschel}, {Frolov}, {Fronz{\'e}}, {Fulda}, {Fyffe}, {Gabbard}, {Gabella}, {Gadre}, {Gair}, {Gais},
  {Galaudage}, {Gamba}, {Ganapathy}, {Ganguly}, {Gaonkar}, {Garaventa}, {Garc{\'\i}a}, {Garc{\'\i}a-N{\'u}{\~n}ez}, {Garc{\'\i}a-Quir{\'o}s}, {Garufi}, {Gateley}, {Gaudio}, {Gayathri}, {Gemme}, {Gennai}, {George}, {George}, {Gerberding}, {Gergely}, {Gewecke}, {Ghonge}, {Ghosh}, {Ghosh}, {Ghosh}, {Ghosh}, {Giacomazzo}, {Giacoppo}, {Giaime}, {Giardina}, {Gibson}, {Gier}, {Giesler}, {Giri}, {Gissi}, {Glanzer}, {Gleckl}, {Godwin}, {Goetz}, {Goetz}, {Gohlke}, {Goncharov}, {Gonz{\'a}lez}, {Gopakumar}, {Gosselin}, {Gouaty}, {Gould}, {Grace}, {Grado}, {Granata}, {Granata}, {Grant}, {Gras}, {Grassia}, {Gray}, {Gray}, {Greco}, {Green}, {Green}, {Gretarsson}, {Gretarsson}, {Griffith}, {Griffiths}, {Griggs}, {Grignani}, {Grimaldi}, {Grimm}, {Grote}, {Grunewald}, {Gruning}, {Guerra}, {Guidi}, {Guimaraes}, {Guix{\'e}}, {Gulati}, {Guo}, {Guo}, {Gupta}, {Gupta}, {Gupta}, {Gustafson}, {Gustafson}, {Guzman}, {Haegel}, {Halim}, {Hall}, {Hamilton}, {Hammond}, {Haney}, {Hanks}, {Hanna}, {Hannam}, {Hannuksela}, {Hansen}, {Hansen},
  {Hanson}, {Harder}, {Hardwick}, {Haris}, {Harms}, {Harry}, {Harry}, {Hartwig}, {Haskell}, {Hasskew}, {Haster}, {Haughian}, {Hayes}, {Healy}, {Heidmann}, {Heidt}, {Heintze}, {Heinze}, {Heinzel}, {Heitmann}, {Hellman}, {Hello}, {Helmling-Cornell}, {Hemming}, {Hendry}, {Heng}, {Hennes}, {Hennig}, {Hennig}, {Hernandez}, {Hernandez Vivanco}, {Heurs}, {Hild}, {Hill}, {Hines}, {Hochheim}, {Hofman}, {Hohmann}, {Holcomb}, {Holland}, {Holley-Bockelmann}, {Hollows}, {Holmes}, {Holt}, {Holz}, {Hopkins}, {Hough}, {Hourihane}, {Howell}, {Hoy}, {Hoyland}, {Hreibi}, {Hsu}, {Huang}, {H{\"u}bner}, {Huddart}, {Hughey}, {Hui}, {Husa}, {Huttner}, {Huxford}, {Huynh-Dinh}, {Idzkowski}, {Iess}, {Ingram}, {Isi}, {Isleif}, {Iyer}, {JaberianHamedan}, {Jacqmin}, {Jadhav}, {Jadhav}, {James}, {Jan}, {Jani}, {Janquart}, {Janssens}, {Janthalur}, {Jaranowski}, {Jariwala}, {Jaume}, {Jenkins}, {Jenner}, {Jeunon}, {Jia}, {Johns}, {Johnson-McDaniel}, {Jones}, {Jones}, {Jones}, {Jones}, {Jones}, {Jonker}, {Ju}, {Junker}, {Juste}, {Kalaghatgi},
  {Kalogera}, {Kamai}, {Kandhasamy}, {Kang}, {Kanner}, {Kao}, {Kapadia}, {Kapasi}, {Karat}, {Karathanasis}, {Karki}, {Kashyap}, {Kasprzack}, {Kastaun}, {Katsanevas}, {Katsavounidis}, {Katzman}, {Kaur}, {Kawabe}, {K{\'e}f{\'e}lian}, {Keitel}, {Key}, {Khadka}, {Khalili}, {Khan}, {Khazanov}, {Khetan}, {Khursheed}, {Kijbunchoo}, {Kim}, {Kim}, {Kim}, {Kim}, {Kim}, {Kimball}, {Kinley-Hanlon}, {Kirchhoff}, {Kissel}, {Kleybolte}, {Klimenko}, {Knee}, {Knowles}, {Knyazev}, {Koch}, {Koekoek}, {Koley}, {Kolitsidou}, {Kolstein}, {Komori}, {Kondrashov}, {Kontos}, {Koper}, {Korobko}, {Kovalam}, {Kozak}, {Kringel}, {Krishnendu}, {Kr{\'o}lak}, {Kuehn}, {Kuei}, {Kuijer}, {Kumar}, {Kumar}, {Kumar}, {Kumar}, {Kuns}, {Kuwahara}, {Lagabbe}, {Laghi}, {Lalande}, {Lam}, {Lamberts}, {Landry}, {Lane}, {Lang}, {Lange}, {Lantz}, {La Rosa}, {Lartaux-Vollard}, {Lasky}, {Laxen}, {Lazzarini}, {Lazzaro}, {Leaci}, {Leavey}, {Lecoeuche}, {Lee}, {Lee}, {Lee}, {Lee}, {Lehmann}, {Lema{\^\i}tre}, {Leroy}, {Letendre}, {Levesque}, {Levin}, {Leviton},
  {Leyde}, {Li}, {Li}, {Li}, {Li}, {Li}, {Linde}, {Linker}, {Linley}, {Littenberg}, {Liu}, {Liu}, {Liu}, {Llamas}, {Llorens-Monteagudo}, {Lo}, {Lockwood}, {London}, {Longo}, {Lopez}, {Lopez Portilla}, {Lorenzini}, {Loriette}, {Lormand}, {Losurdo}, {Lott}, {Lough}, {Lousto}, {Lovelace}, {Lucaccioni}, {L{\"u}ck}, {Lumaca}, {Lundgren}, {Lynam}, {Macas}, {MacInnis}, {Macleod}, {MacMillan}, {Macquet}, {Maga{\~n}a Hernandez}, {Magazz{\`u}}, {Magee}, {Maggiore}, {Magnozzi}, {Mahesh}, {Majorana}, {Makarem}, {Maksimovic}, {Maliakal}, {Malik}, {Man}, {Mandic}, {Mangano}, {Mango}, {Mansell}, {Manske}, {Mantovani}, {Mapelli}, {Marchesoni}, {Marion}, {Mark}, {M{\'a}rka}, {M{\'a}rka}, {Markakis}, {Markosyan}, {Markowitz}, {Maros}, {Marquina}, {Marsat}, {Martelli}, {Martin}, {Martin}, {Martinez}, {Martinez}, {Martinez}, {Martinovic}, {Martynov}, {Marx}, {Masalehdan}, {Mason}, {Massera}, {Masserot}, {Massinger}, {Masso-Reid}, {Mastrogiovanni}, {Matas}, {Mateu-Lucena}, {Matichard}, {Matiushechkina}, {Mavalvala}, {McCann},
  {McCarthy}, {McClelland}, {McClincy}, {McCormick}, {McCuller}, {McGhee}, {McGuire}, {McIsaac}, {McIver}, {McRae}, {McWilliams}, {Meacher}, {Mehmet}, {Mehta}, {Meijer}, {Melatos}, {Melchor}, {Mendell}, {Menendez-Vazquez}, {Menoni}, {Mercer}, {Mereni}, {Merfeld}, {Merilh}, {Merritt}, {Merzougui}, {Meshkov}, {Messenger}, {Messick}, {Meyers}, {Meylahn}, {Mhaske}, {Miani}, {Miao}, {Michaloliakos}, {Michel}, {Middleton}, {Milano}, {Miller}, {Miller}, {Miller}, {Millhouse}, {Mills}, {Milotti}, {Minazzoli}, {Minenkov}, {Mir}, {Miravet-Ten{\'e}s}, {Mishra}, {Mishra}, {Mistry}, {Mitra}, {Mitrofanov}, {Mitselmakher}, {Mittleman}, {Mo}, {Moguel}, {Mogushi}, {Mohapatra}, {Mohite}, {Molina}, {Molina-Ruiz}, {Mondin}, {Montani}, {Moore}, {Moraru}, {Morawski}, {More}, {Moreno}, {Moreno}, {Morisaki}, {Mours}, {Mow-Lowry}, {Mozzon}, {Muciaccia}, {Mukherjee}, {Mukherjee}, {Mukherjee}, {Mukherjee}, {Mukherjee}, {Mukund}, {Mullavey}, {Munch}, {Mu{\~n}iz}, {Murray}, {Musenich}, {Muusse}, {Nadji}, {Nagar}, {Napolano},
  {Nardecchia}, {Naticchioni}, {Nayak}, {Nayak}, {Neil}, {Neilson}, {Nelemans}, {Nelson}, {Nery}, {Neubauer}, {Neunzert}, {Ng}, {Ng}, {Nguyen}, {Nguyen}, {Nguyen}, {Nichols}, {Nissanke}, {Nitoglia}, {Nocera}, {Norman}, {North}, {Nuttall}, {Oberling}, {O'Brien}, {O'Dell}, {Oelker}, {Oganesyan}, {Oh}, {Oh}, {Ohme}, {Ohta}, {Okada}, {Olivetto}, {Oram}, {O'Reilly}, {Ormiston}, {Ormsby}, {Ortega}, {O'Shaughnessy}, {O'Shea}, {Ossokine}, {Osthelder}, {Ottaway}, {Overmier}, {Pace}, {Pagano}, {Page}, {Pagliaroli}, {Pai}, {Pai}, {Palamos}, {Palashov}, {Palomba}, {Pan}, {Panda}, {Pang}, {Pankow}, {Pannarale}, {Pant}, {Panther}, {Paoletti}, {Paoli}, {Paolone}, {Park}, {Parker}, {Pascucci}, {Pasqualetti}, {Passaquieti}, {Passuello}, {Patel}, {Pathak}, {Patricelli}, {Patron}, {Patrone}, {Paul}, {Payne}, {Pedraza}, {Pegoraro}, {Pele}, {Penn}, {Perego}, {Pereira}, {Pereira}, {Perez}, {P{\'e}rigois}, {Perkins}, {Perreca}, {Perri{\`e}s}, {Petermann}, {Petterson}, {Pfeiffer}, {Pham}, {Phukon}, {Piccinni}, {Pichot},
  {Piendibene}, {Piergiovanni}, {Pierini}, {Pierro}, {Pillant}, {Pillas}, {Pilo}, {Pinard}, {Pinto}, {Pinto}, {Piotrzkowski}, {Pirello}, {Pitkin}, {Placidi}, {Planas}, {Plastino}, {Pluchar}, {Poggiani}, {Polini}, {Pong}, {Ponrathnam}, {Popolizio}, {Porter}, {Poulton}, {Powell}, {Pracchia}, {Pradier}, {Prajapati}, {Prasai}, {Prasanna}, {Pratten}, {Principe}, {Prodi}, {Prokhorov}, {Prosposito}, {Prudenzi}, {Puecher}, {Punturo}, {Puosi}, {Puppo}, {P{\"u}rrer}, {Qi}, {Quetschke}, {Quitzow-James}, {Raab}, {Raaijmakers}, {Radkins}, {Radulesco}, {Raffai}, {Rail}, {Raja}, {Rajan}, {Ramirez}, {Ramirez}, {Ramos-Buades}, {Rana}, {Rapagnani}, {Rapol}, {Ray}, {Raymond}, {Raza}, {Razzano}, {Read}, {Rees}, {Regimbau}, {Rei}, {Reid}, {Reid}, {Reitze}, {Relton}, {Renzini}, {Rettegno}, {Reza}, {Rezac}, {Ricci}, {Richards}, {Richardson}, {Richardson}, {Riemenschneider}, {Riles}, {Rinaldi}, {Rink}, {Rizzo}, {Robertson}, {Robie}, {Robinet}, {Rocchi}, {Rodriguez}, {Rolland}, {Rollins}, {Romanelli}, {Romano}, {Romel},
  {Romero-Rodr{\'\i}guez}, {Romero-Shaw}, {Romie}, {Ronchini}, {Rosa}, {Rose}, {Rosell}, {Rosi{\'n}ska}, {Ross}, {Rowan}, {Rowlinson}, {Roy}, {Roy}, {Roy}, {Rozza}, {Ruggi}, {Ruiz-Rocha}, {Ryan}, {Sachdev}, {Sadecki}, {Sadiq}, {Sakellariadou}, {Salafia}, {Salconi}, {Saleem}, {Salemi}, {Samajdar}, {Sanchez}, {Sanchez}, {Sanchez}, {Sanchis-Gual}, {Sanders}, {Sanuy}, {Saravanan}, {Sarin}, {Sassolas}, {Satari}, {Sauter}, {Savage}, {Sawant}, {Sawant}, {Sayah}, {Schaetzl}, {Scheel}, {Scheuer}, {Schiworski}, {Schmidt}, {Schmidt}, {Schnabel}, {Schneewind}, {Schofield}, {Sch{\"o}nbeck}, {Schulte}, {Schutz}, {Schwartz}, {Scott}, {Scott}, {Seglar-Arroyo}, {Sellers}, {Sengupta}, {Sentenac}, {Seo}, {Sequino}, {Sergeev}, {Setyawati}, {Shaffer}, {Shahriar}, {Shams}, {Sharma}, {Sharma}, {Shawhan}, {Shcheblanov}, {Shikauchi}, {Shoemaker}, {Shoemaker}, {ShyamSundar}, {Sieniawska}, {Sigg}, {Singer}, {Singh}, {Singh}, {Singha}, {Sintes}, {Sipala}, {Skliris}, {Slagmolen}, {Slaven-Blair}, {Smetana}, {Smith}, {Smith},
  {Soldateschi}, {Somala}, {Son}, {Soni}, {Soni}, {Sordini}, {Sorrentino}, {Sorrentino}, {Soulard}, {Souradeep}, {Sowell}, {Spagnuolo}, {Spencer}, {Spera}, {Srinivasan}, {Srivastava}, {Srivastava}, {Staats}, {Stachie}, {Steer}, {Steinhoff}, {Steinlechner}, {Steinlechner}, {Stevenson}, {Stops}, {Stover}, {Strain}, {Strang}, {Stratta}, {Strunk}, {Sturani}, {Stuver}, {Sudhagar}, {Sudhir}, {Suh}, {Summerscales}, {Sun}, {Sun}, {Sunil}, {Sur}, {Suresh}, {Sutton}, {Swinkels}, {Szczepa{\'n}czyk}, {Szewczyk}, {Tacca}, {Tait}, {Talbot}, {Talbot}, {Tanasijczuk}, {Tanner}, {Tao}, {Tao}, {Tapia San Mart{\'\i}n}, {Taranto}, {Tasson}, {Tenorio}, {Terhune}, {Terkowski}, {Thirugnanasambandam}, {Thomas}, {Thomas}, {Thomas}, {Thompson}, {Thondapu}, {Thorne}, {Thrane}, {Tiwari}, {Tiwari}, {Tiwari}, {Toivonen}, {Toland}, {Tolley}, {Tonelli}, {Torres-Forn{\'e}}, {Torrie}, {Tosta e Melo}, {T{\"o}yr{\"a}}, {Trapananti}, {Travasso}, {Traylor}, {Trevor}, {Tringali}, {Tripathee}, {Troiano}, {Trovato}, {Trozzo}, {Trudeau}, {Tsai},
  {Tsai}, {Tsang}, {Tse}, {Tso}, {Tsukada}, {Tsuna}, {Tsutsui}, {Turbang}, {Turconi}, {Ubhi}, {Udall}, {Ueno}, {Unnikrishnan}, {Urban}, {Utina}, {Vahlbruch}, {Vajente}, {Vajpeyi}, {Valdes}, {Valentini}, {Valsan}, {van Bakel}, {van Beuzekom}, {van den Brand}, {Van Den Broeck}, {Vander-Hyde}, {van der Schaaf}, {van Heijningen}, {Vanosky}, {van Remortel}, {Vardaro}, {Vargas}, {Varma}, {Vas{\'u}th}, {Vecchio}, {Vedovato}, {Veitch}, {Veitch}, {Venneberg}, {Venugopalan}, {Verkindt}, {Verma}, {Verma}, {Veske}, {Vetrano}, {Vicer{\'e}}, {Vidyant}, {Viets}, {Vijaykumar}, {Villa-Ortega}, {Vinet}, {Virtuoso}, {Vitale}, {Vo}, {Vocca}, {von Reis}, {von Wrangel}, {Vorvick}, {Vyatchanin}, {Wade}, {Wade}, {Wagner}, {Walet}, {Walker}, {Wallace}, {Wallace}, {Walsh}, {Wang}, {Wang}, {Ward}, {Warner}, {Was}, {Washington}, {Watchi}, {Weaver}, {Webster}, {Weinert}, {Weinstein}, {Weiss}, {Weller}, {Weller}, {Wellmann}, {Wen}, {We{\ss}els}, {Wette}, {Whelan}, {White}, {Whiting}, {Whittle}, {Wilken}, {Williams}, {Williams},
  {Williamson}, {Willis}, {Willke}, {Wilson}, {Winkler}, {Wipf}, {Wlodarczyk}, {Woan}, {Woehler}, {Wofford}, {Wong}, {Wu}, {Wysocki}, {Xiao}, {Yamamoto}, {Yang}, {Yang}, {Yang}, {Yang}, {Yap}, {Yeeles}, {Yelikar}, {Ying}, {Yoo}, {Yu}, {Yu}, {Zadro{\.z}ny}, {Zanolin}, {Zelenova}, {Zendri}, {Zevin}, {Zhang}, {Zhang}, {Zhang}, {Zhang}, {Zhao}, {Zhao}, {Zhao}, {Zhou}, {Zhou}, {Zhu}, {Zimmerman}, {Zlochower}, {Zucker}, \& {Zweizig}}]{2021arXiv210801045T}
{Abbott}, R., {Abbott}, T.~D., {Acernese}, F., {et~al.} 2021{\natexlab{d}}, arXiv e-prints, arXiv:2108.01045

\bibitem[{{Ablimit} \& {Maeda}(2018)}]{2018ApJ...866..151A}
{Ablimit}, I. \& {Maeda}, K. 2018, \apj, 866, 151

\bibitem[{{Acernese} {et~al.}(2015){Acernese}, {Agathos}, {Agatsuma}, {Aisa}, {Allemandou}, {Allocca}, {Amarni}, {Astone}, {Balestri}, {Ballardin}, {Barone}, {Baronick}, {Barsuglia}, {Basti}, {Basti}, {Bauer}, {Bavigadda}, {Bejger}, {Beker}, {Belczynski}, {Bersanetti}, {Bertolini}, {Bitossi}, {Bizouard}, {Bloemen}, {Blom}, {Boer}, {Bogaert}, {Bondi}, {Bondu}, {Bonelli}, {Bonnand}, {Boschi}, {Bosi}, {Bouedo}, {Bradaschia}, {Branchesi}, {Briant}, {Brillet}, {Brisson}, {Bulik}, {Bulten}, {Buskulic}, {Buy}, {Cagnoli}, {Calloni}, {Campeggi}, {Canuel}, {Carbognani}, {Cavalier}, {Cavalieri}, {Cella}, {Cesarini}, {Mottin}, {Chincarini}, {Chiummo}, {Chua}, {Cleva}, {Coccia}, {Cohadon}, {Colla}, {Colombini}, {Conte}, {Coulon}, {Cuoco}, {Dalmaz}, {D'Antonio}, {Dattilo}, {Davier}, {Day}, {Debreczeni}, {Degallaix}, {Del{\'e}glise}, {Pozzo}, {Dereli}, {Rosa}, {Fiore}, {Lieto}, {Virgilio}, {Doets}, {Dolique}, {Drago}, {Ducrot}, {Endr{\H{o}}czi}, {Fafone}, {Farinon}, {Ferrante}, {Ferrini}, {Fidecaro}, {Fiori}, {Flaminio},
  {Fournier}, {Franco}, {Frasca}, {Frasconi}, {Gammaitoni}, {Garufi}, {Gaspard}, {Gatto}, {Gemme}, {Gendre}, {Genin}, {Gennai}, {Ghosh}, {Giacobone}, {Giazotto}, {Gouaty}, {Granata}, {Greco}, {Groot}, {Guidi}, {Harms}, {Heidmann}, {Heitmann}, {Hello}, {Hemming}, {Hennes}, {Hofman}, {Jaranowski}, {Jonker}, {Kasprzack}, {K{\'e}f{\'e}lian}, {Kowalska}, {Kraan}, {Kr{\'o}lak}, {Kutynia}, {Lazzaro}, {Leonardi}, {Leroy}, {Letendre}, {Li}, {Lieunard}, {Lorenzini}, {Loriette}, {Losurdo}, {Magazz{\`u}}, {Majorana}, {Maksimovic}, {Malvezzi}, {Man}, {Mangano}, {Mantovani}, {Marchesoni}, {Marion}, {Marque}, {Martelli}, {Martellini}, {Masserot}, {Meacher}, {Meidam}, {Mezzani}, {Michel}, {Milano}, {Minenkov}, {Moggi}, {Mohan}, {Montani}, {Morgado}, {Mours}, {Mul}, {Nagy}, {Nardecchia}, {Naticchioni}, {Nelemans}, {Neri}, {Neri}, {Nocera}, {Pacaud}, {Palomba}, {Paoletti}, {Paoli}, {Pasqualetti}, {Passaquieti}, {Passuello}, {Perciballi}, {Petit}, {Pichot}, {Piergiovanni}, {Pillant}, {Piluso}, {Pinard}, {Poggiani}, {Prijatelj},
  {Prodi}, {Punturo}, {Puppo}, {Rabeling}, {R{\'a}cz}, {Rapagnani}, {Razzano}, {Re}, {Regimbau}, {Ricci}, {Robinet}, {Rocchi}, {Rolland}, {Romano}, {Rosi{\'n}ska}, {Ruggi}, {Saracco}, {Sassolas}, {Schimmel}, {Sentenac}, {Sequino}, {Shah}, {Siellez}, {Straniero}, {Swinkels}, {Tacca}, {Tonelli}, {Travasso}, {Turconi}, {Vajente}, {van Bakel}, {van Beuzekom}, {van den Brand}, {Van Den Broeck}, {van der Sluys}, {van Heijningen}, {Vas{\'u}th}, {Vedovato}, {Veitch}, {Verkindt}, {Vetrano}, {Vicer{\'e}}, {Vinet}, {Visser}, {Vocca}, {Ward}, {Was}, {Wei}, {Yvert}, {{\.z}ny}, \& {Zendri}}]{2015CQGra..32b4001A}
{Acernese}, F., {Agathos}, M., {Agatsuma}, K., {et~al.} 2015, Classical and Quantum Gravity, 32, 024001

\bibitem[{{Amaro-Seoane} {et~al.}(2017){Amaro-Seoane}, {Audley}, {Babak}, {Baker}, {Barausse}, {Bender}, {Berti}, {Binetruy}, {Born}, {Bortoluzzi}, {Camp}, {Caprini}, {Cardoso}, {Colpi}, {Conklin}, {Cornish}, {Cutler}, {Danzmann}, {Dolesi}, {Ferraioli}, {Ferroni}, {Fitzsimons}, {Gair}, {Gesa Bote}, {Giardini}, {Gibert}, {Grimani}, {Halloin}, {Heinzel}, {Hertog}, {Hewitson}, {Holley-Bockelmann}, {Hollington}, {Hueller}, {Inchauspe}, {Jetzer}, {Karnesis}, {Killow}, {Klein}, {Klipstein}, {Korsakova}, {Larson}, {Livas}, {Lloro}, {Man}, {Mance}, {Martino}, {Mateos}, {McKenzie}, {McWilliams}, {Miller}, {Mueller}, {Nardini}, {Nelemans}, {Nofrarias}, {Petiteau}, {Pivato}, {Plagnol}, {Porter}, {Reiche}, {Robertson}, {Robertson}, {Rossi}, {Russano}, {Schutz}, {Sesana}, {Shoemaker}, {Slutsky}, {Sopuerta}, {Sumner}, {Tamanini}, {Thorpe}, {Troebs}, {Vallisneri}, {Vecchio}, {Vetrugno}, {Vitale}, {Volonteri}, {Wanner}, {Ward}, {Wass}, {Weber}, {Ziemer}, \& {Zweifel}}]{2017arXiv170200786A}
{Amaro-Seoane}, P., {Audley}, H., {Babak}, S., {et~al.} 2017, arXiv e-prints, arXiv:1702.00786

\bibitem[{{Arca Sedda}(2020)}]{2020CmPhy...3...43A}
{Arca Sedda}, M. 2020, Communications Physics, 3, 43

\bibitem[{{Arca Sedda}(2021)}]{2021ApJ...908L..38A}
{Arca Sedda}, M. 2021, \apjl, 908, L38

\bibitem[{{Baker} {et~al.}(2019){Baker}, {Bellovary}, {Bender}, {Berti}, {Caldwell}, {Camp}, {Conklin}, {Cornish}, {Cutler}, {DeRosa}, {Eracleous}, {Ferrara}, {Francis}, {Hewitson}, {Holley-Bockelmann}, {Hornschemeier}, {Hogan}, {Kamai}, {Kelly}, {Shapiro Key}, {Larson}, {Livas}, {Manthripragada}, {McKenzie}, {McWilliams}, {Mueller}, {Natarajan}, {Numata}, {Rioux}, {Sankar}, {Schnittman}, {Shoemaker}, {Shoemaker}, {Slutsky}, {Spero}, {Stebbins}, {Thorpe}, {Vallisneri}, {Ware}, {Wass}, {Yu}, \& {Ziemer}}]{2019arXiv190706482B}
{Baker}, J., {Bellovary}, J., {Bender}, P.~L., {et~al.} 2019, arXiv e-prints, arXiv:1907.06482

\bibitem[{{Bavera} {et~al.}(2023){Bavera}, {Fragos}, {Zapartas}, {Andrews}, {Kalogera}, {Berry}, {Kruckow}, {Dotter}, {Kovlakas}, {Misra}, {Rocha}, {Srivastava}, {Sun}, \& {Xing}}]{2023NatAs.tmp..142B}
{Bavera}, S.~S., {Fragos}, T., {Zapartas}, E., {et~al.} 2023, Nature Astronomy [\eprint[arXiv]{2212.10924}]

\bibitem[{{Bavera} {et~al.}(2021){Bavera}, {Fragos}, {Zevin}, {Berry}, {Marchant}, {Andrews}, {Coughlin}, {Dotter}, {Kovlakas}, {Misra}, {Serra-Perez}, {Qin}, {Rocha}, {Rom{\'a}n-Garza}, {Tran}, \& {Zapartas}}]{2021A&A...647A.153B}
{Bavera}, S.~S., {Fragos}, T., {Zevin}, M., {et~al.} 2021, \aap, 647, A153

\bibitem[{{Begelman}(1979)}]{1979MNRAS.187..237B}
{Begelman}, M.~C. 1979, \mnras, 187, 237

\bibitem[{{Belczynski} {et~al.}(2020){Belczynski}, {Klencki}, {Fields}, {Olejak}, {Berti}, {Meynet}, {Fryer}, {Holz}, {O'Shaughnessy}, {Brown}, {Bulik}, {Leung}, {Nomoto}, {Madau}, {Hirschi}, {Kaiser}, {Jones}, {Mondal}, {Chruslinska}, {Drozda}, {Gerosa}, {Doctor}, {Giersz}, {Ekstrom}, {Georgy}, {Askar}, {Baibhav}, {Wysocki}, {Natan}, {Farr}, {Wiktorowicz}, {Coleman Miller}, {Farr}, \& {Lasota}}]{2020A&A...636A.104B}
{Belczynski}, K., {Klencki}, J., {Fields}, C.~E., {et~al.} 2020, \aap, 636, A104

\bibitem[{{Bhattacharya} {et~al.}(2019){Bhattacharya}, {Kumar}, \& {Smoot}}]{2019MNRAS.486.5289B}
{Bhattacharya}, M., {Kumar}, P., \& {Smoot}, G. 2019, \mnras, 486, 5289

\bibitem[{{Bloecker}(1995)}]{1995A&A...297..727B}
{Bloecker}, T. 1995, \aap, 297, 727

\bibitem[{{B{\"o}hm-Vitense}(1958)}]{1958ZA.....46..108B}
{B{\"o}hm-Vitense}, E. 1958, \zap, 46, 108

\bibitem[{{Booth} {et~al.}(2009){Booth}, {de Blok}, {Jonas}, \& {Fanaroff}}]{2009arXiv0910.2935B}
{Booth}, R.~S., {de Blok}, W.~J.~G., {Jonas}, J.~L., \& {Fanaroff}, B. 2009, arXiv e-prints, arXiv:0910.2935

\bibitem[{{Breivik} {et~al.}(2020){Breivik}, {Coughlin}, {Zevin}, {Rodriguez}, {Kremer}, {Ye}, {Andrews}, {Kurkowski}, {Digman}, {Larson}, \& {Rasio}}]{2020ApJ...898...71B}
{Breivik}, K., {Coughlin}, S., {Zevin}, M., {et~al.} 2020, \apj, 898, 71

\bibitem[{{Briel} {et~al.}(2023){Briel}, {Stevance}, \& {Eldridge}}]{2023MNRAS.520.5724B}
{Briel}, M.~M., {Stevance}, H.~F., \& {Eldridge}, J.~J. 2023, \mnras, 520, 5724

\bibitem[{{Broekgaarden} {et~al.}(2021){Broekgaarden}, {Berger}, {Neijssel}, {Vigna-G{\'o}mez}, {Chattopadhyay}, {Stevenson}, {Chruslinska}, {Justham}, {de Mink}, \& {Mandel}}]{2021MNRAS.508.5028B}
{Broekgaarden}, F.~S., {Berger}, E., {Neijssel}, C.~J., {et~al.} 2021, \mnras, 508, 5028

\bibitem[{{Broekgaarden} {et~al.}(2022){Broekgaarden}, {Berger}, {Stevenson}, {Justham}, {Mandel}, {Chru{\'s}li{\'n}ska}, {van Son}, {Wagg}, {Vigna-G{\'o}mez}, {de Mink}, {Chattopadhyay}, \& {Neijssel}}]{2022MNRAS.tmp.1776B}
{Broekgaarden}, F.~S., {Berger}, E., {Stevenson}, S., {et~al.} 2022, \mnras [\eprint[arXiv]{2112.05763}]

\bibitem[{{Brott} {et~al.}(2011){Brott}, {de Mink}, {Cantiello}, {Langer}, {de Koter}, {Evans}, {Hunter}, {Trundle}, \& {Vink}}]{2011A&A...530A.115B}
{Brott}, I., {de Mink}, S.~E., {Cantiello}, M., {et~al.} 2011, \aap, 530, A115

\bibitem[{{Chattopadhyay} {et~al.}(2022){Chattopadhyay}, {Stevenson}, {Broekgaarden}, {Antonini}, \& {Belczynski}}]{2022MNRAS.513.5780C}
{Chattopadhyay}, D., {Stevenson}, S., {Broekgaarden}, F., {Antonini}, F., \& {Belczynski}, K. 2022, \mnras, 513, 5780

\bibitem[{{Chattopadhyay} {et~al.}(2021){Chattopadhyay}, {Stevenson}, {Hurley}, {Bailes}, \& {Broekgaarden}}]{2021MNRAS.504.3682C}
{Chattopadhyay}, D., {Stevenson}, S., {Hurley}, J.~R., {Bailes}, M., \& {Broekgaarden}, F. 2021, \mnras, 504, 3682

\bibitem[{{Choi} {et~al.}(2016){Choi}, {Dotter}, {Conroy}, {Cantiello}, {Paxton}, \& {Johnson}}]{2016ApJ...823..102C}
{Choi}, J., {Dotter}, A., {Conroy}, C., {et~al.} 2016, \apj, 823, 102

\bibitem[{{Claret} \& {Torres}(2017)}]{2017ApJ...849...18C}
{Claret}, A. \& {Torres}, G. 2017, \apj, 849, 18

\bibitem[{{Clausen} {et~al.}(2013){Clausen}, {Sigurdsson}, \& {Chernoff}}]{2013MNRAS.428.3618C}
{Clausen}, D., {Sigurdsson}, S., \& {Chernoff}, D.~F. 2013, \mnras, 428, 3618

\bibitem[{{Cordes} {et~al.}(2006){Cordes}, {Freire}, {Lorimer}, {Camilo}, {Champion}, {Nice}, {Ramachandran}, {Hessels}, {Vlemmings}, {van Leeuwen}, {Ransom}, {Bhat}, {Arzoumanian}, {McLaughlin}, {Kaspi}, {Kasian}, {Deneva}, {Reid}, {Chatterjee}, {Han}, {Backer}, {Stairs}, {Deshpande}, \& {Faucher-Gigu{\`e}re}}]{2006ApJ...637..446C}
{Cordes}, J.~M., {Freire}, P.~C.~C., {Lorimer}, D.~R., {et~al.} 2006, \apj, 637, 446

\bibitem[{{Darbha} {et~al.}(2021){Darbha}, {Kasen}, {Foucart}, \& {Price}}]{2021ApJ...915...69D}
{Darbha}, S., {Kasen}, D., {Foucart}, F., \& {Price}, D.~J. 2021, \apj, 915, 69

\bibitem[{{de Jager} {et~al.}(1988){de Jager}, {Nieuwenhuijzen}, \& {van der Hucht}}]{1988A&AS...72..259D}
{de Jager}, C., {Nieuwenhuijzen}, H., \& {van der Hucht}, K.~A. 1988, \aaps, 72, 259

\bibitem[{{de Kool}(1990)}]{1990ApJ...358..189D}
{de Kool}, M. 1990, \apj, 358, 189

\bibitem[{{de Mink} \& {Belczynski}(2015)}]{2015ApJ...814...58D}
{de Mink}, S.~E. \& {Belczynski}, K. 2015, \apj, 814, 58

\bibitem[{{de Mink} {et~al.}(2013){de Mink}, {Langer}, {Izzard}, {Sana}, \& {de Koter}}]{2013ApJ...764..166D}
{de Mink}, S.~E., {Langer}, N., {Izzard}, R.~G., {Sana}, H., \& {de Koter}, A. 2013, \apj, 764, 166

\bibitem[{{Dewi} \& {Tauris}(2000)}]{2000A&A...360.1043D}
{Dewi}, J.~D.~M. \& {Tauris}, T.~M. 2000, \aap, 360, 1043

\bibitem[{{Dominik} {et~al.}(2015){Dominik}, {Berti}, {O'Shaughnessy}, {Mandel}, {Belczynski}, {Fryer}, {Holz}, {Bulik}, \& {Pannarale}}]{2015ApJ...806..263D}
{Dominik}, M., {Berti}, E., {O'Shaughnessy}, R., {et~al.} 2015, \apj, 806, 263

\bibitem[{{Dorozsmai} \& {Toonen}(2022)}]{2022arXiv220708837D}
{Dorozsmai}, A. \& {Toonen}, S. 2022, arXiv e-prints, arXiv:2207.08837

\bibitem[{{Foucart} {et~al.}(2013){Foucart}, {Deaton}, {Duez}, {Kidder}, {MacDonald}, {Ott}, {Pfeiffer}, {Scheel}, {Szilagyi}, \& {Teukolsky}}]{2013PhRvD..87h4006F}
{Foucart}, F., {Deaton}, M.~B., {Duez}, M.~D., {et~al.} 2013, \prd, 87, 084006

\bibitem[{{Foucart} {et~al.}(2018){Foucart}, {Hinderer}, \& {Nissanke}}]{2018PhRvD..98h1501F}
{Foucart}, F., {Hinderer}, T., \& {Nissanke}, S. 2018, \prd, 98, 081501

\bibitem[{{Fragione}(2021)}]{2021ApJ...923L...2F}
{Fragione}, G. 2021, \apjl, 923, L2

\bibitem[{{Fragione} \& {Banerjee}(2020)}]{2020ApJ...901L..16F}
{Fragione}, G. \& {Banerjee}, S. 2020, \apjl, 901, L16

\bibitem[{{Fragione} \& {Loeb}(2019)}]{2019MNRAS.490.4991F}
{Fragione}, G. \& {Loeb}, A. 2019, \mnras, 490, 4991

\bibitem[{{Fragione} {et~al.}(2021){Fragione}, {Loeb}, \& {Rasio}}]{2021ApJ...918L..38F}
{Fragione}, G., {Loeb}, A., \& {Rasio}, F.~A. 2021, \apjl, 918, L38

\bibitem[{{Fragos} {et~al.}(2023){Fragos}, {Andrews}, {Bavera}, {Berry}, {Coughlin}, {Dotter}, {Giri}, {Kalogera}, {Katsaggelos}, {Kovlakas}, {Lalvani}, {Misra}, {Srivastava}, {Qin}, {Rocha}, {Rom{\'a}n-Garza}, {Serra}, {Stahle}, {Sun}, {Teng}, {Trajcevski}, {Tran}, {Xing}, {Zapartas}, \& {Zevin}}]{2023ApJS..264...45F}
{Fragos}, T., {Andrews}, J.~J., {Bavera}, S.~S., {et~al.} 2023, \apjs, 264, 45

\bibitem[{{Fragos} \& {McClintock}(2015)}]{2015ApJ...800...17F}
{Fragos}, T. \& {McClintock}, J.~E. 2015, \apj, 800, 17

\bibitem[{{Fragos} {et~al.}(2010){Fragos}, {Tremmel}, {Rantsiou}, \& {Belczynski}}]{2010ApJ...719L..79F}
{Fragos}, T., {Tremmel}, M., {Rantsiou}, E., \& {Belczynski}, K. 2010, \apjl, 719, L79

\bibitem[{{Fryer} {et~al.}(2012){Fryer}, {Belczynski}, {Wiktorowicz}, {Dominik}, {Kalogera}, \& {Holz}}]{2012ApJ...749...91F}
{Fryer}, C.~L., {Belczynski}, K., {Wiktorowicz}, G., {et~al.} 2012, \apj, 749, 91

\bibitem[{{Fryer} {et~al.}(1999){Fryer}, {Woosley}, \& {Hartmann}}]{1999ApJ...526..152F}
{Fryer}, C.~L., {Woosley}, S.~E., \& {Hartmann}, D.~H. 1999, \apj, 526, 152

\bibitem[{{Fuller} \& {Ma}(2019)}]{2019ApJ...881L...1F}
{Fuller}, J. \& {Ma}, L. 2019, \apjl, 881, L1

\bibitem[{{Giacobbo} \& {Mapelli}(2018)}]{2018MNRAS.480.2011G}
{Giacobbo}, N. \& {Mapelli}, M. 2018, \mnras, 480, 2011

\bibitem[{{Giacobbo} \& {Mapelli}(2019)}]{2019MNRAS.482.2234G}
{Giacobbo}, N. \& {Mapelli}, M. 2019, \mnras, 482, 2234

\bibitem[{{Gompertz} {et~al.}(2020){Gompertz}, {Levan}, \& {Tanvir}}]{2020ApJ...895...58G}
{Gompertz}, B.~P., {Levan}, A.~J., \& {Tanvir}, N.~R. 2020, \apj, 895, 58

\bibitem[{{Gottlieb} {et~al.}(2023){Gottlieb}, {Metzger}, {Quataert}, {Issa}, \& {Foucart}}]{2023arXiv230900038G}
{Gottlieb}, O., {Metzger}, B.~D., {Quataert}, E., {Issa}, D., \& {Foucart}, F. 2023, arXiv e-prints, arXiv:2309.00038

\bibitem[{{Herwig}(2000)}]{2000A&A...360..952H}
{Herwig}, F. 2000, \aap, 360, 952

\bibitem[{{Hild} {et~al.}(2011){Hild}, {Abernathy}, {Acernese}, {Amaro-Seoane}, {Andersson}, {Arun}, {Barone}, {Barr}, {Barsuglia}, {Beker}, {Beveridge}, {Birindelli}, {Bose}, {Bosi}, {Braccini}, {Bradaschia}, {Bulik}, {Calloni}, {Cella}, {Chassande Mottin}, {Chelkowski}, {Chincarini}, {Clark}, {Coccia}, {Colacino}, {Colas}, {Cumming}, {Cunningham}, {Cuoco}, {Danilishin}, {Danzmann}, {De Salvo}, {Dent}, {De Rosa}, {Di Fiore}, {Di Virgilio}, {Doets}, {Fafone}, {Falferi}, {Flaminio}, {Franc}, {Frasconi}, {Freise}, {Friedrich}, {Fulda}, {Gair}, {Gemme}, {Genin}, {Gennai}, {Giazotto}, {Glampedakis}, {Gr{\"a}f}, {Granata}, {Grote}, {Guidi}, {Gurkovsky}, {Hammond}, {Hannam}, {Harms}, {Heinert}, {Hendry}, {Heng}, {Hennes}, {Hough}, {Husa}, {Huttner}, {Jones}, {Khalili}, {Kokeyama}, {Kokkotas}, {Krishnan}, {Li}, {Lorenzini}, {L{\"u}ck}, {Majorana}, {Mandel}, {Mandic}, {Mantovani}, {Martin}, {Michel}, {Minenkov}, {Morgado}, {Mosca}, {Mours}, {M{\"u}ller{\textendash}Ebhardt}, {Murray}, {Nawrodt}, {Nelson},
  {Oshaughnessy}, {Ott}, {Palomba}, {Paoli}, {Parguez}, {Pasqualetti}, {Passaquieti}, {Passuello}, {Pinard}, {Plastino}, {Poggiani}, {Popolizio}, {Prato}, {Punturo}, {Puppo}, {Rabeling}, {Rapagnani}, {Read}, {Regimbau}, {Rehbein}, {Reid}, {Ricci}, {Richard}, {Rocchi}, {Rowan}, {R{\"u}diger}, {Santamar{\'\i}a}, {Sassolas}, {Sathyaprakash}, {Schnabel}, {Schwarz}, {Seidel}, {Sintes}, {Somiya}, {Speirits}, {Strain}, {Strigin}, {Sutton}, {Tarabrin}, {Th{\"u}ring}, {van den Brand}, {van Veggel}, {van den Broeck}, {Vecchio}, {Veitch}, {Vetrano}, {Vicere}, {Vyatchanin}, {Willke}, {Woan}, \& {Yamamoto}}]{2011CQGra..28i4013H}
{Hild}, S., {Abernathy}, M., {Acernese}, F., {et~al.} 2011, Classical and Quantum Gravity, 28, 094013

\bibitem[{{Hobbs} {et~al.}(2005){Hobbs}, {Lorimer}, {Lyne}, \& {Kramer}}]{2005MNRAS.360..974H}
{Hobbs}, G., {Lorimer}, D.~R., {Lyne}, A.~G., \& {Kramer}, M. 2005, \mnras, 360, 974

\bibitem[{{Hotokezaka} {et~al.}(2016){Hotokezaka}, {Nissanke}, {Hallinan}, {Lazio}, {Nakar}, \& {Piran}}]{2016ApJ...831..190H}
{Hotokezaka}, K., {Nissanke}, S., {Hallinan}, G., {et~al.} 2016, \apj, 831, 190

\bibitem[{{Hu} {et~al.}(2022){Hu}, {Zhu}, {Qin}, {Zhang}, {Liang}, \& {Shao}}]{2022ApJ...928..163H}
{Hu}, R.-C., {Zhu}, J.-P., {Qin}, Y., {et~al.} 2022, \apj, 928, 163

\bibitem[{{Huang} {et~al.}(2020){Huang}, {Hu}, {Zhang}, \& {Shen}}]{2020ApJ...904...39H}
{Huang}, K., {Hu}, J., {Zhang}, Y., \& {Shen}, H. 2020, \apj, 904, 39

\bibitem[{{Hurley} {et~al.}(2000){Hurley}, {Pols}, \& {Tout}}]{2000MNRAS.315..543H}
{Hurley}, J.~R., {Pols}, O.~R., \& {Tout}, C.~A. 2000, \mnras, 315, 543

\bibitem[{{Hurley} {et~al.}(2002){Hurley}, {Tout}, \& {Pols}}]{2002MNRAS.329..897H}
{Hurley}, J.~R., {Tout}, C.~A., \& {Pols}, O.~R. 2002, \mnras, 329, 897

\bibitem[{{Hut}(1981)}]{1981A&A....99..126H}
{Hut}, P. 1981, \aap, 99, 126

\bibitem[{{Jermyn} {et~al.}(2023){Jermyn}, {Bauer}, {Schwab}, {Farmer}, {Ball}, {Bellinger}, {Dotter}, {Joyce}, {Marchant}, {Mombarg}, {Wolf}, {Sunny Wong}, {Cinquegrana}, {Farrell}, {Smolec}, {Thoul}, {Cantiello}, {Herwig}, {Toloza}, {Bildsten}, {Townsend}, \& {Timmes}}]{2023ApJS..265...15J}
{Jermyn}, A.~S., {Bauer}, E.~B., {Schwab}, J., {et~al.} 2023, \apjs, 265, 15

\bibitem[{{Kalogera}(1996)}]{1996ApJ...471..352K}
{Kalogera}, V. 1996, \apj, 471, 352

\bibitem[{{Kawaguchi} {et~al.}(2020){Kawaguchi}, {Shibata}, \& {Tanaka}}]{2020ApJ...893..153K}
{Kawaguchi}, K., {Shibata}, M., \& {Tanaka}, M. 2020, \apj, 893, 153

\bibitem[{{King} {et~al.}(2005){King}, {Lubow}, {Ogilvie}, \& {Pringle}}]{2005MNRAS.363...49K}
{King}, A.~R., {Lubow}, S.~H., {Ogilvie}, G.~I., \& {Pringle}, J.~E. 2005, \mnras, 363, 49

\bibitem[{{Klencki} {et~al.}(2022){Klencki}, {Istrate}, {Nelemans}, \& {Pols}}]{2022A&A...662A..56K}
{Klencki}, J., {Istrate}, A., {Nelemans}, G., \& {Pols}, O. 2022, \aap, 662, A56

\bibitem[{{Kobulnicky} \& {Fryer}(2007)}]{2007ApJ...670..747K}
{Kobulnicky}, H.~A. \& {Fryer}, C.~L. 2007, \apj, 670, 747

\bibitem[{{Kolb} \& {Ritter}(1990)}]{1990A&A...236..385K}
{Kolb}, U. \& {Ritter}, H. 1990, \aap, 236, 385

\bibitem[{{Kramer} {et~al.}(2004){Kramer}, {Backer}, {Cordes}, {Lazio}, {Stappers}, \& {Johnston}}]{2004NewAR..48..993K}
{Kramer}, M., {Backer}, D.~C., {Cordes}, J.~M., {et~al.} 2004, \nar, 48, 993

\bibitem[{{Kroupa}(2001)}]{2001MNRAS.322..231K}
{Kroupa}, P. 2001, \mnras, 322, 231

\bibitem[{{Kruckow} {et~al.}(2018){Kruckow}, {Tauris}, {Langer}, {Kramer}, \& {Izzard}}]{2018MNRAS.481.1908K}
{Kruckow}, M.~U., {Tauris}, T.~M., {Langer}, N., {Kramer}, M., \& {Izzard}, R.~G. 2018, \mnras, 481, 1908

\bibitem[{{Kyutoku} {et~al.}(2011){Kyutoku}, {Okawa}, {Shibata}, \& {Taniguchi}}]{2011PhRvD..84f4018K}
{Kyutoku}, K., {Okawa}, H., {Shibata}, M., \& {Taniguchi}, K. 2011, \prd, 84, 064018

\bibitem[{{Kyutoku} {et~al.}(2010){Kyutoku}, {Shibata}, \& {Taniguchi}}]{2010PhRvD..82d4049K}
{Kyutoku}, K., {Shibata}, M., \& {Taniguchi}, K. 2010, \prd, 82, 044049

\bibitem[{{Langer}(1998)}]{1998A&A...329..551L}
{Langer}, N. 1998, \aap, 329, 551

\bibitem[{{Langer} {et~al.}(2020){Langer}, {Sch{\"u}rmann}, {Stoll}, {Marchant}, {Lennon}, {Mahy}, {de Mink}, {Quast}, {Riedel}, {Sana}, {Schneider}, {Schootemeijer}, {Wang}, {Almeida}, {Bestenlehner}, {Bodensteiner}, {Castro}, {Clark}, {Crowther}, {Dufton}, {Evans}, {Fossati}, {Gr{\"a}fener}, {Grassitelli}, {Grin}, {Hastings}, {Herrero}, {de Koter}, {Menon}, {Patrick}, {Puls}, {Renzo}, {Sander}, {Schneider}, {Sen}, {Shenar}, {Sim{\'o}n-D{\'\i}as}, {Tauris}, {Tramper}, {Vink}, \& {Xu}}]{2020A&A...638A..39L}
{Langer}, N., {Sch{\"u}rmann}, C., {Stoll}, K., {et~al.} 2020, \aap, 638, A39

\bibitem[{{Li} \& {Paczy{\'n}ski}(1998)}]{1998ApJ...507L..59L}
{Li}, L.-X. \& {Paczy{\'n}ski}, B. 1998, \apjl, 507, L59

\bibitem[{{LIGO Scientific Collaboration} {et~al.}(2015){LIGO Scientific Collaboration}, {Aasi}, {Abbott}, {Abbott}, {Abbott}, {Abernathy}, {Ackley}, {Adams}, {Adams}, {Addesso}, {Adhikari}, {Adya}, {Affeldt}, {Aggarwal}, {Aguiar}, {Ain}, {Ajith}, {Alemic}, {Allen}, {Amariutei}, {Anderson}, {Anderson}, {Arai}, {Araya}, {Arceneaux}, {Areeda}, {Ashton}, {Ast}, {Aston}, {Aufmuth}, {Aulbert}, {Aylott}, {Babak}, {Baker}, {Ballmer}, {Barayoga}, {Barbet}, {Barclay}, {Barish}, {Barker}, {Barr}, {Barsotti}, {Bartlett}, {Barton}, {Bartos}, {Bassiri}, {Batch}, {Baune}, {Behnke}, {Bell}, {Bell}, {Benacquista}, {Bergman}, {Bergmann}, {Berry}, {Betzwieser}, {Bhagwat}, {Bhandare}, {Bilenko}, {Billingsley}, {Birch}, {Biscans}, {Biwer}, {Blackburn}, {Blackburn}, {Blair}, {Blair}, {Bock}, {Bodiya}, {Bojtos}, {Bond}, {Bork}, {Born}, {Bose}, {Brady}, {Braginsky}, {Brau}, {Bridges}, {Brinkmann}, {Brooks}, {Brown}, {Brown}, {Brown}, {Buchman}, {Buikema}, {Buonanno}, {Cadonati}, {Calder{\'o}n Bustillo}, {Camp}, {Cannon}, {Cao},
  {Capano}, {Caride}, {Caudill}, {Cavagli{\`a}}, {Cepeda}, {Chakraborty}, {Chalermsongsak}, {Chamberlin}, {Chao}, {Charlton}, {Chen}, {Cho}, {Cho}, {Chow}, {Christensen}, {Chu}, {Chung}, {Ciani}, {Clara}, {Clark}, {Collette}, {Cominsky}, {Constancio}, {Cook}, {Corbitt}, {Cornish}, {Corsi}, {Costa}, {Coughlin}, {Countryman}, {Couvares}, {Coward}, {Cowart}, {Coyne}, {Coyne}, {Craig}, {Creighton}, {Creighton}, {Cripe}, {Crowder}, {Cumming}, {Cunningham}, {Cutler}, {Dahl}, {Dal Canton}, {Damjanic}, {Danilishin}, {Danzmann}, {Dartez}, {Dave}, {Daveloza}, {Davies}, {Daw}, {DeBra}, {Del Pozzo}, {Denker}, {Dent}, {Dergachev}, {DeRosa}, {DeSalvo}, {Dhurandhar}, {D{\textasciiacute}{\i}az}, {Di Palma}, {Dojcinoski}, {Dominguez}, {Donovan}, {Dooley}, {Doravari}, {Douglas}, {Downes}, {Driggers}, {Du}, {Dwyer}, {Eberle}, {Edo}, {Edwards}, {Edwards}, {Effler}, {Eggenstein}, {Ehrens}, {Eichholz}, {Eikenberry}, {Essick}, {Etzel}, {Evans}, {Evans}, {Factourovich}, {Fairhurst}, {Fan}, {Fang}, {Farr}, {Farr}, {Favata}, {Fays},
  {Fehrmann}, {Fejer}, {Feldbaum}, {Ferreira}, {Fisher}, {Frei}, {Freise}, {Frey}, {Fricke}, {Fritschel}, {Frolov}, {Fuentes-Tapia}, {Fulda}, {Fyffe}, {Gair}, {Gaonkar}, {Gehrels}, {Gergely}, {Giaime}, {Giardina}, {Gleason}, {Goetz}, {Goetz}, {Gondan}, {Gonz{\'a}lez}, {Gordon}, {Gorodetsky}, {Gossan}, {Go{\ss}ler}, {Gr{\"a}f}, {Graff}, {Grant}, {Gras}, {Gray}, {Greenhalgh}, {Gretarsson}, {Grote}, {Grunewald}, {Guido}, {Guo}, {Gushwa}, {Gustafson}, {Gustafson}, {Hacker}, {Hall}, {Hammond}, {Hanke}, {Hanks}, {Hanna}, {Hannam}, {Hanson}, {Hardwick}, {Harry}, {Harry}, {Hart}, {Hartman}, {Haster}, {Haughian}, {Hee}, {Heintze}, {Heinzel}, {Hendry}, {Heng}, {Heptonstall}, {Heurs}, {Hewitson}, {Hild}, {Hoak}, {Hodge}, {Hollitt}, {Holt}, {Hopkins}, {Hosken}, {Hough}, {Houston}, {Howell}, {Hu}, {Huerta}, {Hughey}, {Husa}, {Huttner}, {Huynh}, {Huynh-Dinh}, {Idrisy}, {Indik}, {Ingram}, {Inta}, {Islas}, {Isler}, {Isogai}, {Iyer}, {Izumi}, {Jacobson}, {Jang}, {Jawahar}, {Ji}, {Jim{\'e}nez-Forteza}, {Johnson}, {Jones},
  {Jones}, {Ju}, {Haris}, {Kalogera}, {Kandhasamy}, {Kang}, {Kanner}, {Katsavounidis}, {Katzman}, {Kaufer}, {Kaufer}, {Kaur}, {Kawabe}, {Kawazoe}, {Keiser}, {Keitel}, {Kelley}, {Kells}, {Keppel}, {Key}, {Khalaidovski}, {Khalili}, {Khazanov}, {Kim}, {Kim}, {Kim}, {Kim}, {Kim}, {King}, {King}, {Kinzel}, {Kissel}, {Klimenko}, {Kline}, {Koehlenbeck}, {Kokeyama}, {Kondrashov}, {Korobko}, {Korth}, {Kozak}, {Kringel}, {Krishnan}, {Krueger}, {Kuehn}, {Kumar}, {Kumar}, {Kuo}, {Landry}, {Lantz}, {Larson}, {Lasky}, {Lazzarini}, {Lazzaro}, {Le}, {Leaci}, {Leavey}, {Lebigot}, {Lee}, {Lee}, {Lee}, {Leong}, {Levin}, {Levine}, {Lewis}, {Li}, {Libbrecht}, {Libson}, {Lin}, {Littenberg}, {Lockerbie}, {Lockett}, {Logue}, {Lombardi}, {Lormand}, {Lough}, {Lubinski}, {L{\"u}ck}, {Lundgren}, {Lynch}, {Ma}, {Macarthur}, {MacDonald}, {Machenschalk}, {MacInnis}, {Macleod}, {Maga{\~n}a-Sandoval}, {Magee}, {Mageswaran}, {Maglione}, {Mailand}, {Mandel}, {Mandic}, {Mangano}, {Mansell}, {M{\'a}rka}, {M{\'a}rka}, {Markosyan}, {Maros},
  {Martin}, {Martin}, {Martynov}, {Marx}, {Mason}, {Massinger}, {Matichard}, {Matone}, {Mavalvala}, {Mazumder}, {Mazzolo}, {McCarthy}, {McClelland}, {McCormick}, {McGuire}, {McIntyre}, {McIver}, {McLin}, {McWilliams}, {Meadors}, {Meinders}, {Melatos}, {Mendell}, {Mercer}, {Meshkov}, {Messenger}, {Meyers}, {Miao}, {Middleton}, {Mikhailov}, {Miller}, {Miller}, {Millhouse}, {Ming}, {Mirshekari}, {Mishra}, {Mitra}, {Mitrofanov}, {Mitselmakher}, {Mittleman}, {Moe}, {Mohanty}, {Mohapatra}, {Moore}, {Moraru}, {Moreno}, {Morriss}, {Mossavi}, {Mow-Lowry}, {Mueller}, {Mueller}, {Mukherjee}, {Mullavey}, {Munch}, {Murphy}, {Murray}, {Mytidis}, {Nash}, {Nayak}, {Necula}, {Nedkova}, {Newton}, {Nguyen}, {Nielsen}, {Nissanke}, {Nitz}, {Nolting}, {Normandin}, {Nuttall}, {Ochsner}, {O'Dell}, {Oelker}, {Ogin}, {Oh}, {Oh}, {Ohme}, {Oppermann}, {Oram}, {O'Reilly}, {Ortega}, {O'Shaughnessy}, {Osthelder}, {Ott}, {Ottaway}, {Ottens}, {Overmier}, {Owen}, {Padilla}, {Pai}, {Pai}, {Palashov}, {Pal-Singh}, {Pan}, {Pankow}, {Pannarale},
  {Pant}, {Papa}, {Paris}, {Patrick}, {Pedraza}, {Pekowsky}, {Pele}, {Penn}, {Perreca}, {Phelps}, {Pierro}, {Pinto}, {Pitkin}, {Poeld}, {Post}, {Poteomkin}, {Powell}, {Prasad}, {Predoi}, {Premachandra}, {Prestegard}, {Price}, {Principe}, {Privitera}, {Prix}, {Prokhorov}, {Puncken}, {P{\"u}rrer}, {Qin}, {Quetschke}, {Quintero}, {Quiroga}, {Quitzow-James}, {Raab}, {Rabeling}, {Radkins}, {Raffai}, {Raja}, {Rajalakshmi}, {Rakhmanov}, {Ramirez}, {Raymond}, {Reed}, {Reid}, {Reitze}, {Reula}, {Riles}, {Robertson}, {Robie}, {Rollins}, {Roma}, {Romano}, {Romanov}, {Romie}, {Rowan}, {R{\"u}diger}, {Ryan}, {Sachdev}, {Sadecki}, {Sadeghian}, {Saleem}, {Salemi}, {Sammut}, {Sandberg}, {Sanders}, {Sannibale}, {Santiago-Prieto}, {Sathyaprakash}, {Saulson}, {Savage}, {Sawadsky}, {Scheuer}, {Schilling}, {Schmidt}, {Schnabel}, {Schofield}, {Schreiber}, {Schuette}, {Schutz}, {Scott}, {Scott}, {Sellers}, {Sengupta}, {Sergeev}, {Serna}, {Sevigny}, {Shaddock}, {Shahriar}, {Shaltev}, {Shao}, {Shapiro}, {Shawhan}, {Shoemaker},
  {Sidery}, {Siemens}, {Sigg}, {Silva}, {Simakov}, {Singer}, {Singer}, {Singh}, {Sintes}, {Slagmolen}, {Smith}, {Smith}, {Smith}, {Smith-Lefebvre}, {Son}, {Sorazu}, {Souradeep}, {Staley}, {Stebbins}, {Steinke}, {Steinlechner}, {Steinlechner}, {Steinmeyer}, {Stephens}, {Steplewski}, {Stevenson}, {Stone}, {Strain}, {Strigin}, {Sturani}, {Stuver}, {Summerscales}, {Sutton}, {Szczepanczyk}, {Szeifert}, {Talukder}, {Tanner}, {T{\'a}pai}, {Tarabrin}, {Taracchini}, {Taylor}, {Tellez}, {Theeg}, {Thirugnanasambandam}, {Thomas}, {Thomas}, {Thorne}, {Thorne}, {Thrane}, {Tiwari}, {Tomlinson}, {Torres}, {Torrie}, {Traylor}, {Tse}, {Tshilumba}, {Ugolini}, {Unnikrishnan}, {Urban}, {Usman}, {Vahlbruch}, {Vajente}, {Valdes}, {Vallisneri}, {van Veggel}, {Vass}, {Vaulin}, {Vecchio}, {Veitch}, {Veitch}, {Venkateswara}, {Vincent-Finley}, {Vitale}, {Vo}, {Vorvick}, {Vousden}, {Vyatchanin}, {Wade}, {Wade}, {Wade}, {Walker}, {Wallace}, {Walsh}, {Wang}, {Wang}, {Wang}, {Ward}, {Warner}, {Was}, {Weaver}, {Weinert}, {Weinstein},
  {Weiss}, {Welborn}, {Wen}, {Wessels}, {Westphal}, {Wette}, {Whelan}, {Whitcomb}, {White}, {Whiting}, {Wilkinson}, {Williams}, {Williams}, {Williamson}, {Willis}, {Willke}, {Wimmer}, {Winkler}, {Wipf}, {Wittel}, {Woan}, {Worden}, {Xie}, {Yablon}, {Yakushin}, {Yam}, {Yamamoto}, {Yancey}, {Yang}, {Zanolin}, {Zhang}, {Zhang}, {Zhang}, {Zhang}, {Zhao}, {Zhou}, {Zhu}, {Zucker}, {Zuraw}, \& {Zweizig}}]{2015CQGra..32g4001L}
{LIGO Scientific Collaboration}, {Aasi}, J., {Abbott}, B.~P., {et~al.} 2015, Classical and Quantum Gravity, 32, 074001

\bibitem[{{Livio} \& {Soker}(1988)}]{1988ApJ...329..764L}
{Livio}, M. \& {Soker}, N. 1988, \apj, 329, 764

\bibitem[{{Luo} {et~al.}(2016){Luo}, {Chen}, {Duan}, {Gong}, {Hu}, {Ji}, {Liu}, {Mei}, {Milyukov}, {Sazhin}, {Shao}, {Toth}, {Tu}, {Wang}, {Wang}, {Yeh}, {Zhan}, {Zhang}, {Zharov}, \& {Zhou}}]{2016CQGra..33c5010L}
{Luo}, J., {Chen}, L.-S., {Duan}, H.-Z., {et~al.} 2016, Classical and Quantum Gravity, 33, 035010

\bibitem[{{Mandel} \& {Smith}(2021)}]{2021ApJ...922L..14M}
{Mandel}, I. \& {Smith}, R. J.~E. 2021, \apjl, 922, L14

\bibitem[{{Marchant} {et~al.}(2017){Marchant}, {Langer}, {Podsiadlowski}, {Tauris}, {de Mink}, {Mandel}, \& {Moriya}}]{2017A&A...604A..55M}
{Marchant}, P., {Langer}, N., {Podsiadlowski}, P., {et~al.} 2017, \aap, 604, A55

\bibitem[{{Marchant} {et~al.}(2016){Marchant}, {Langer}, {Podsiadlowski}, {Tauris}, \& {Moriya}}]{2016A&A...588A..50M}
{Marchant}, P., {Langer}, N., {Podsiadlowski}, P., {Tauris}, T.~M., \& {Moriya}, T.~J. 2016, \aap, 588, A50

\bibitem[{{Mei} {et~al.}(2021){Mei}, {Bai}, {Bao}, {Barausse}, {Cai}, {Canuto}, {Cao}, {Chen}, {Chen}, {Ding}, {Duan}, {Fan}, {Feng}, {Fu}, {Gao}, {Gao}, {Gong}, {Gou}, {Gu}, {Gu}, {He}, {Hendry}, {Hong}, {Hu}, {Hu}, {Hu}, {Huang}, {Huang}, {Jiang}, {Jiang}, {Jiang}, {Jiang}, {Jin}, {Korol}, {Li}, {Li}, {Li}, {Li}, {Li}, {Li}, {Li}, {Li}, {Li}, {Liang}, {Liang}, {Liao}, {Liu}, {Liu}, {Liu}, {Liu}, {Liu}, {Liu}, {Liu}, {Lu}, {Lu}, {Lu}, {Luo}, {Luo}, {Milyukov}, {Ming}, {Pi}, {Qin}, {Qu}, {Sesana}, {Shao}, {Shi}, {Su}, {Tan}, {Tan}, {Tan}, {Tu}, {Wang}, {Wang}, {Wang}, {Wang}, {Wang}, {Wang}, {Wang}, {Wang}, {Wang}, {Wang}, {Wang}, {Wei}, {Wu}, {Xiao}, {Xu}, {Xue}, {Yang}, {Yang}, {Yang}, {Yang}, {Ye}, {Yeh}, {Yu}, {Zhai}, {Zhang}, {Zhang}, {Zhang}, {Zhang}, {Zhang}, {Zhang}, {Zhang}, {Zhou}, {Zhou}, {Zhou}, {Zhu}, {Zi}, \& {Luo}}]{2021PTEP.2021eA107M}
{Mei}, J., {Bai}, Y.-Z., {Bao}, J., {et~al.} 2021, Progress of Theoretical and Experimental Physics, 2021, 05A107

\bibitem[{{Misra} {et~al.}(2020){Misra}, {Fragos}, {Tauris}, {Zapartas}, \& {Aguilera-Dena}}]{2020A&A...642A.174M}
{Misra}, D., {Fragos}, T., {Tauris}, T.~M., {Zapartas}, E., \& {Aguilera-Dena}, D.~R. 2020, \aap, 642, A174

\bibitem[{{Moreno M{\'e}ndez} {et~al.}(2008){Moreno M{\'e}ndez}, {Brown}, {Lee}, \& {Park}}]{2008ApJ...689L...9M}
{Moreno M{\'e}ndez}, E., {Brown}, G.~E., {Lee}, C.-H., \& {Park}, I.~H. 2008, \apjl, 689, L9

\bibitem[{{Nakar}(2007)}]{2007PhR...442..166N}
{Nakar}, E. 2007, \physrep, 442, 166

\bibitem[{{Nan} {et~al.}(2011){Nan}, {Li}, {Jin}, {Wang}, {Zhu}, {Zhu}, {Zhang}, {Yue}, \& {Qian}}]{2011IJMPD..20..989N}
{Nan}, R., {Li}, D., {Jin}, C., {et~al.} 2011, International Journal of Modern Physics D, 20, 989

\bibitem[{{Nandez} {et~al.}(2014){Nandez}, {Ivanova}, \& {Lombardi}}]{2014ApJ...786...39N}
{Nandez}, J.~L.~A., {Ivanova}, N., \& {Lombardi}, J.~C., J. 2014, \apj, 786, 39

\bibitem[{{Neijssel} {et~al.}(2019){Neijssel}, {Vigna-G{\'o}mez}, {Stevenson}, {Barrett}, {Gaebel}, {Broekgaarden}, {de Mink}, {Sz{\'e}csi}, {Vinciguerra}, \& {Mandel}}]{2019MNRAS.490.3740N}
{Neijssel}, C.~J., {Vigna-G{\'o}mez}, A., {Stevenson}, S., {et~al.} 2019, \mnras, 490, 3740

\bibitem[{{Nugis} \& {Lamers}(2000)}]{2000A&A...360..227N}
{Nugis}, T. \& {Lamers}, H.~J.~G.~L.~M. 2000, \aap, 360, 227

\bibitem[{{Paczynski}(1991)}]{1991ApJ...370..597P}
{Paczynski}, B. 1991, \apj, 370, 597

\bibitem[{{Pannarale} \& {Ohme}(2014)}]{2014ApJ...791L...7P}
{Pannarale}, F. \& {Ohme}, F. 2014, \apjl, 791, L7

\bibitem[{{Pannarale} {et~al.}(2011){Pannarale}, {Tonita}, \& {Rezzolla}}]{2011ApJ...727...95P}
{Pannarale}, F., {Tonita}, A., \& {Rezzolla}, L. 2011, \apj, 727, 95

\bibitem[{{Patton} \& {Sukhbold}(2020)}]{2020MNRAS.499.2803P}
{Patton}, R.~A. \& {Sukhbold}, T. 2020, \mnras, 499, 2803

\bibitem[{{Paxton} {et~al.}(2011){Paxton}, {Bildsten}, {Dotter}, {Herwig}, {Lesaffre}, \& {Timmes}}]{2011ApJS..192....3P}
{Paxton}, B., {Bildsten}, L., {Dotter}, A., {et~al.} 2011, \apjs, 192, 3

\bibitem[{{Paxton} {et~al.}(2013){Paxton}, {Cantiello}, {Arras}, {Bildsten}, {Brown}, {Dotter}, {Mankovich}, {Montgomery}, {Stello}, {Timmes}, \& {Townsend}}]{2013ApJS..208....4P}
{Paxton}, B., {Cantiello}, M., {Arras}, P., {et~al.} 2013, \apjs, 208, 4

\bibitem[{{Paxton} {et~al.}(2015){Paxton}, {Marchant}, {Schwab}, {Bauer}, {Bildsten}, {Cantiello}, {Dessart}, {Farmer}, {Hu}, {Langer}, {Townsend}, {Townsley}, \& {Timmes}}]{2015ApJS..220...15P}
{Paxton}, B., {Marchant}, P., {Schwab}, J., {et~al.} 2015, \apjs, 220, 15

\bibitem[{{Paxton} {et~al.}(2018){Paxton}, {Schwab}, {Bauer}, {Bildsten}, {Blinnikov}, {Duffell}, {Farmer}, {Goldberg}, {Marchant}, {Sorokina}, {Thoul}, {Townsend}, \& {Timmes}}]{2018ApJS..234...34P}
{Paxton}, B., {Schwab}, J., {Bauer}, E.~B., {et~al.} 2018, \apjs, 234, 34

\bibitem[{{Paxton} {et~al.}(2019){Paxton}, {Smolec}, {Schwab}, {Gautschy}, {Bildsten}, {Cantiello}, {Dotter}, {Farmer}, {Goldberg}, {Jermyn}, {Kanbur}, {Marchant}, {Thoul}, {Townsend}, {Wolf}, {Zhang}, \& {Timmes}}]{2019ApJS..243...10P}
{Paxton}, B., {Smolec}, R., {Schwab}, J., {et~al.} 2019, \apjs, 243, 10

\bibitem[{{Petrovich} \& {Antonini}(2017)}]{2017ApJ...846..146P}
{Petrovich}, C. \& {Antonini}, F. 2017, \apj, 846, 146

\bibitem[{{Piran} {et~al.}(2013){Piran}, {Nakar}, \& {Rosswog}}]{2013MNRAS.430.2121P}
{Piran}, T., {Nakar}, E., \& {Rosswog}, S. 2013, \mnras, 430, 2121

\bibitem[{{Podsiadlowski} {et~al.}(2004){Podsiadlowski}, {Langer}, {Poelarends}, {Rappaport}, {Heger}, \& {Pfahl}}]{2004ApJ...612.1044P}
{Podsiadlowski}, P., {Langer}, N., {Poelarends}, A.~J.~T., {et~al.} 2004, \apj, 612, 1044

\bibitem[{{Pol} {et~al.}(2021){Pol}, {McLaughlin}, \& {Lorimer}}]{2021arXiv210904512P}
{Pol}, N., {McLaughlin}, M., \& {Lorimer}, D. 2021, arXiv e-prints, arXiv:2109.04512

\bibitem[{{Popham} \& {Narayan}(1991)}]{1991ApJ...370..604P}
{Popham}, R. \& {Narayan}, R. 1991, \apj, 370, 604

\bibitem[{{Punturo} {et~al.}(2010){Punturo}, {Abernathy}, {Acernese}, {Allen}, {Andersson}, {Arun}, {Barone}, {Barr}, {Barsuglia}, {Beker}, {Beveridge}, {Birindelli}, {Bose}, {Bosi}, {Braccini}, {Bradaschia}, {Bulik}, {Calloni}, {Cella}, {Chassande Mottin}, {Chelkowski}, {Chincarini}, {Clark}, {Coccia}, {Colacino}, {Colas}, {Cumming}, {Cunningham}, {Cuoco}, {Danilishin}, {Danzmann}, {De Luca}, {De Salvo}, {Dent}, {De Rosa}, {Di Fiore}, {Di Virgilio}, {Doets}, {Fafone}, {Falferi}, {Flaminio}, {Franc}, {Frasconi}, {Freise}, {Fulda}, {Gair}, {Gemme}, {Gennai}, {Giazotto}, {Glampedakis}, {Granata}, {Grote}, {Guidi}, {Hammond}, {Hannam}, {Harms}, {Heinert}, {Hendry}, {Heng}, {Hennes}, {Hild}, {Hough}, {Husa}, {Huttner}, {Jones}, {Khalili}, {Kokeyama}, {Kokkotas}, {Krishnan}, {Lorenzini}, {L{\"u}ck}, {Majorana}, {Mandel}, {Mandic}, {Martin}, {Michel}, {Minenkov}, {Morgado}, {Mosca}, {Mours}, {M{\"u}ller{\textendash}Ebhardt}, {Murray}, {Nawrodt}, {Nelson}, {Oshaughnessy}, {Ott}, {Palomba}, {Paoli}, {Parguez},
  {Pasqualetti}, {Passaquieti}, {Passuello}, {Pinard}, {Poggiani}, {Popolizio}, {Prato}, {Puppo}, {Rabeling}, {Rapagnani}, {Read}, {Regimbau}, {Rehbein}, {Reid}, {Rezzolla}, {Ricci}, {Richard}, {Rocchi}, {Rowan}, {R{\"u}diger}, {Sassolas}, {Sathyaprakash}, {Schnabel}, {Schwarz}, {Seidel}, {Sintes}, {Somiya}, {Speirits}, {Strain}, {Strigin}, {Sutton}, {Tarabrin}, {Th{\"u}ring}, {van den Brand}, {van Leewen}, {van Veggel}, {van den Broeck}, {Vecchio}, {Veitch}, {Vetrano}, {Vicere}, {Vyatchanin}, {Willke}, {Woan}, {Wolfango}, \& {Yamamoto}}]{2010CQGra..27s4002P}
{Punturo}, M., {Abernathy}, M., {Acernese}, F., {et~al.} 2010, Classical and Quantum Gravity, 27, 194002

\bibitem[{{Qin} {et~al.}(2018){Qin}, {Fragos}, {Meynet}, {Andrews}, {S{\o}rensen}, \& {Song}}]{2018A&A...616A..28Q}
{Qin}, Y., {Fragos}, T., {Meynet}, G., {et~al.} 2018, \aap, 616, A28

\bibitem[{{Qin} {et~al.}(2022){Qin}, {Shu}, {Yi}, \& {Wang}}]{2022RAA....22c5023Q}
{Qin}, Y., {Shu}, X., {Yi}, S., \& {Wang}, Y.-Z. 2022, Research in Astronomy and Astrophysics, 22, 035023

\bibitem[{{Rastello} {et~al.}(2020){Rastello}, {Mapelli}, {Di Carlo}, {Giacobbo}, {Santoliquido}, {Spera}, {Ballone}, \& {Iorio}}]{2020MNRAS.497.1563R}
{Rastello}, S., {Mapelli}, M., {Di Carlo}, U.~N., {et~al.} 2020, \mnras, 497, 1563

\bibitem[{{Reimers}(1975)}]{1975psae.book..229R}
{Reimers}, D. 1975, in Problems in stellar atmospheres and envelopes., 229--256

\bibitem[{{Reitze} {et~al.}(2019){Reitze}, {Adhikari}, {Ballmer}, {Barish}, {Barsotti}, {Billingsley}, {Brown}, {Chen}, {Coyne}, {Eisenstein}, {Evans}, {Fritschel}, {Hall}, {Lazzarini}, {Lovelace}, {Read}, {Sathyaprakash}, {Shoemaker}, {Smith}, {Torrie}, {Vitale}, {Weiss}, {Wipf}, \& {Zucker}}]{2019BAAS...51g..35R}
{Reitze}, D., {Adhikari}, R.~X., {Ballmer}, S., {et~al.} 2019, in Bulletin of the American Astronomical Society, Vol.~51, 35

\bibitem[{{Rom{\'a}n-Garza} {et~al.}(2021){Rom{\'a}n-Garza}, {Bavera}, {Fragos}, {Zapartas}, {Misra}, {Andrews}, {Coughlin}, {Dotter}, {Kovlakas}, {Serra}, {Qin}, {Rocha}, \& {Tran}}]{2021ApJ...912L..23R}
{Rom{\'a}n-Garza}, J., {Bavera}, S.~S., {Fragos}, T., {et~al.} 2021, \apjl, 912, L23

\bibitem[{{Ruan} {et~al.}(2020){Ruan}, {Guo}, {Cai}, \& {Zhang}}]{2020IJMPA..3550075R}
{Ruan}, W.-H., {Guo}, Z.-K., {Cai}, R.-G., \& {Zhang}, Y.-Z. 2020, International Journal of Modern Physics A, 35, 2050075

\bibitem[{{Santoliquido} {et~al.}(2020){Santoliquido}, {Mapelli}, {Bouffanais}, {Giacobbo}, {Di Carlo}, {Rastello}, {Artale}, \& {Ballone}}]{2020ApJ...898..152S}
{Santoliquido}, F., {Mapelli}, M., {Bouffanais}, Y., {et~al.} 2020, \apj, 898, 152

\bibitem[{{Scheuer} \& {Feiler}(1996)}]{1996MNRAS.282..291S}
{Scheuer}, P.~A.~G. \& {Feiler}, R. 1996, \mnras, 282, 291

\bibitem[{{Shao} \& {Li}(2021)}]{2021ApJ...920...81S}
{Shao}, Y. \& {Li}, X.-D. 2021, \apj, 920, 81

\bibitem[{{Sipior} {et~al.}(2004){Sipior}, {Portegies Zwart}, \& {Nelemans}}]{2004MNRAS.354L..49S}
{Sipior}, M.~S., {Portegies Zwart}, S., \& {Nelemans}, G. 2004, \mnras, 354, L49

\bibitem[{{S{\k{a}}dowski} \& {Narayan}(2016)}]{2016MNRAS.456.3929S}
{S{\k{a}}dowski}, A. \& {Narayan}, R. 2016, \mnras, 456, 3929

\bibitem[{{Spruit}(2002)}]{2002A&A...381..923S}
{Spruit}, H.~C. 2002, \aap, 381, 923

\bibitem[{{Tang} {et~al.}(2020){Tang}, {Eldridge}, {Stanway}, \& {Bray}}]{2020MNRAS.493L...6T}
{Tang}, P.~N., {Eldridge}, J.~J., {Stanway}, E.~R., \& {Bray}, J.~C. 2020, \mnras, 493, L6

\bibitem[{{Tiwari} {et~al.}(2021){Tiwari}, {Ebersold}, \& {Hamilton}}]{2021PhRvD.104l3024T}
{Tiwari}, S., {Ebersold}, M., \& {Hamilton}, E.~Z. 2021, \prd, 104, 123024

\bibitem[{{Trani} {et~al.}(2022){Trani}, {Rastello}, {Di Carlo}, {Santoliquido}, {Tanikawa}, \& {Mapelli}}]{2022MNRAS.511.1362T}
{Trani}, A.~A., {Rastello}, S., {Di Carlo}, U.~N., {et~al.} 2022, \mnras, 511, 1362

\bibitem[{{Tutukov} \& {Yungelson}(1993)}]{1993MNRAS.260..675T}
{Tutukov}, A.~V. \& {Yungelson}, L.~R. 1993, \mnras, 260, 675

\bibitem[{{Tylenda} {et~al.}(2011){Tylenda}, {Hajduk}, {Kami{\'n}ski}, {Udalski}, {Soszy{\'n}ski}, {Szyma{\'n}ski}, {Kubiak}, {Pietrzy{\'n}ski}, {Poleski}, {Wyrzykowski}, \& {Ulaczyk}}]{2011A&A...528A.114T}
{Tylenda}, R., {Hajduk}, M., {Kami{\'n}ski}, T., {et~al.} 2011, \aap, 528, A114

\bibitem[{{Vink} {et~al.}(2001){Vink}, {de Koter}, \& {Lamers}}]{2001A&A...369..574V}
{Vink}, J.~S., {de Koter}, A., \& {Lamers}, H.~J.~G.~L.~M. 2001, \aap, 369, 574

\bibitem[{{Voss} \& {Tauris}(2003)}]{2003MNRAS.342.1169V}
{Voss}, R. \& {Tauris}, T.~M. 2003, \mnras, 342, 1169

\bibitem[{{Wang} {et~al.}(2021){Wang}, {Stephan}, {Naoz}, {Hoang}, \& {Breivik}}]{2021ApJ...917...76W}
{Wang}, H., {Stephan}, A.~P., {Naoz}, S., {Hoang}, B.-M., \& {Breivik}, K. 2021, \apj, 917, 76

\bibitem[{{Webbink}(1984)}]{1984ApJ...277..355W}
{Webbink}, R.~F. 1984, \apj, 277, 355

\bibitem[{{Ye} {et~al.}(2020){Ye}, {Fong}, {Kremer}, {Rodriguez}, {Chatterjee}, {Fragione}, \& {Rasio}}]{2020ApJ...888L..10Y}
{Ye}, C.~S., {Fong}, W.-f., {Kremer}, K., {et~al.} 2020, \apjl, 888, L10

\bibitem[{{Zevin} {et~al.}(2020){Zevin}, {Spera}, {Berry}, \& {Kalogera}}]{2020ApJ...899L...1Z}
{Zevin}, M., {Spera}, M., {Berry}, C. P.~L., \& {Kalogera}, V. 2020, \apjl, 899, L1

\bibitem[{{Zhou} {et~al.}(2021){Zhou}, {Li}, \& {Li}}]{2021ApJ...910...62Z}
{Zhou}, X., {Li}, A., \& {Li}, B.-A. 2021, \apj, 910, 62

\bibitem[{{Zhu} {et~al.}(2021){Zhu}, {Wu}, {Yang}, {Zhang}, {Gao}, {Yu}, {Li}, {Cao}, {Liu}, {Huang}, \& {Zhang}}]{2021ApJ...917...24Z}
{Zhu}, J.-P., {Wu}, S., {Yang}, Y.-P., {et~al.} 2021, \apj, 917, 24

\bibitem[{{Ziosi} {et~al.}(2014){Ziosi}, {Mapelli}, {Branchesi}, \& {Tormen}}]{2014MNRAS.441.3703Z}
{Ziosi}, B.~M., {Mapelli}, M., {Branchesi}, M., \& {Tormen}, G. 2014, \mnras, 441, 3703

\end{thebibliography}
\end{document}